\documentclass[10pt,draftcls,onecolumn]{IEEEtran}

\usepackage{cite}
\usepackage{amsmath}
\usepackage{amsmath,amssymb,latexsym,mathrsfs,marvosym}
\usepackage{graphicx,color}
\usepackage{psfig}
\usepackage{psfrag}
\usepackage{subfigure}
\usepackage{url}
\usepackage{array,multirow}
\usepackage{fontenc,times}
\usepackage{pxfonts}
\usepackage{dsfont}
\makeatletter

\graphicspath{{./}}

\DeclareMathAlphabet{\eurm}{U}{eur}{m}{n}
\DeclareMathAlphabet{\mathbsf}{OT1}{cmss}{bx}{n}
\DeclareMathAlphabet{\mathssf}{OT1}{cmss}{m}{sl}
\DeclareMathAlphabet{\mathcsf}{OT1}{cmss}{sbc}{n}

\DeclareSymbolFont{bsfletters}{OT1}{cmss}{bx}{n}  
\DeclareSymbolFont{ssfletters}{OT1}{cmss}{m}{n}
\DeclareMathSymbol{\bsfGamma}{0}{bsfletters}{'000}
\DeclareMathSymbol{\ssfGamma}{0}{ssfletters}{'000}
\DeclareMathSymbol{\bsfDelta}{0}{bsfletters}{'001}
\DeclareMathSymbol{\ssfDelta}{0}{ssfletters}{'001}
\DeclareMathSymbol{\bsfTheta}{0}{bsfletters}{'002}
\DeclareMathSymbol{\ssfTheta}{0}{ssfletters}{'002}
\DeclareMathSymbol{\bsfLambda}{0}{bsfletters}{'003}
\DeclareMathSymbol{\ssfLambda}{0}{ssfletters}{'003}
\DeclareMathSymbol{\bsfXi}{0}{bsfletters}{'004}
\DeclareMathSymbol{\ssfXi}{0}{ssfletters}{'004}
\DeclareMathSymbol{\bsfPi}{0}{bsfletters}{'005}
\DeclareMathSymbol{\ssfPi}{0}{ssfletters}{'005}
\DeclareMathSymbol{\bsfSigma}{0}{bsfletters}{'006}
\DeclareMathSymbol{\ssfSigma}{0}{ssfletters}{'006}
\DeclareMathSymbol{\bsfUpsilon}{0}{bsfletters}{'007}
\DeclareMathSymbol{\ssfUpsilon}{0}{ssfletters}{'007}
\DeclareMathSymbol{\bsfPhi}{0}{bsfletters}{'010}
\DeclareMathSymbol{\ssfPhi}{0}{ssfletters}{'010}
\DeclareMathSymbol{\bsfPsi}{0}{bsfletters}{'011}
\DeclareMathSymbol{\ssfPsi}{0}{ssfletters}{'011}
\DeclareMathSymbol{\bsfOmega}{0}{bsfletters}{'012}
\DeclareMathSymbol{\ssfOmega}{0}{ssfletters}{'012}

\newtheorem{theorem}{\textbf{Theorem}}

\newtheorem{proposition}{\textbf{Proposition}}
\newtheorem{example}{\textbf{Example}}
\newtheorem{definition}{\textbf{Definition}}

\newtheorem{remark}{Remark}
\newcommand{\dv}{\mathbf} 
\newcommand{\mc}{\mathcal} 
\newcommand{\qed}{\hfill \ensuremath{\Box}}

\begin{document}
\title{On Cooperative Multiple Access Channels with Delayed CSI at Transmitters\\}

\author{\vspace{0cm}
\authorblockN{ \small Abdellatif Zaidi \qquad \qquad Shlomo Shamai (Shitz)\thanks{The material in this paper was presented in part at the IEEE International Symposium on Information Theory, Istanbul, Turkey, July 2013, and the International Symposium on Information Theory, Honolulu, Hawai, July 2014. This work has been supported by the European Commission in the framework of the FP7 Network of Excellence in Wireless Communications (NEWCOM\#).}
\thanks{Abdellatif Zaidi is with Universit\'e Paris-Est Marne La Vall\'ee, 77454 Marne la Vall\'ee Cedex 2, France. Email: abdellatif.zaidi@univ-mlv.fr}
\thanks{Shlomo Shamai is with the Department of Electrical Engineering, Technion Institute of Technology, Technion City, Haifa 32000, Israel. Email: sshlomo@ee.technion.ac.il}}}

\vspace{1cm}

\maketitle

\begin{abstract}

We consider a cooperative two-user multiaccess channel in which the transmission is controlled by a random state. Both encoders transmit a common message and, one of the encoders also transmits an individual message. We study the capacity region of this communication model for different degrees of availability of the states at the encoders, causally or strictly causally. In the case in which the states are revealed causally to both encoders but not to the decoder we find an explicit characterization of the capacity region in the discrete memoryless case. In the case in which the states are revealed only strictly causally to both encoders, we establish inner and outer bounds on the capacity region. The outer bound is non-trivial, and has a relatively simple form. It has the advantage of incorporating \textit{only one auxiliary random variable}. In particular, it \textit{suggests} that there is none, or at best only little, to gain from having the encoder that transmits both messages also sending an individual description of the state to the receiver, in addition to the compressed version that is sent cooperatively with the other encoder. We then introduce a class of cooperative multiaccess channels with states known strictly causally at both encoders for which the inner and outer bounds agree; and so we characterize the capacity region for this class. In this class of channels, the state can be obtained as a deterministic function of the channel inputs and output. We also study the model in which the states are revealed, strictly causally, in an asymmetric manner, to only one encoder. Throughout the paper, we discuss a number of examples; and compute the capacity region of some of these examples. The results shed more light on the utility of delayed channel state information for increasing the capacity region of state-dependent cooperative multiaccess channels; and tie with recent progress in this framework.

\end{abstract}

\section{Introduction}\label{secI}

In this paper, we study a two-user state-dependent multiple access channel with the channel states revealed -- depending on the scenario, only strictly-causally or causally, to both or only one of the encoders. Both encoders transmit a common message and, in addition, one of the encoders also transmits an individual message. More precisely, let $W_c$ and $W_1$ denote the common message and the individual message to be transmitted in, say, $n$ uses of the channel; and $S^n=(S_1,\hdots,S_n)$ denote the state sequence affecting the channel during the transmission. In the causal setting, at time $i$ both encoders know the channel states up to and including time $i$, i.e., the sequence $S^i=(S_1,\hdots,S_{i-1},S_i)$. In the strictly causal setting, at time $i$ the encoders know the channel states only up to time $i-1$, i.e., the sequence $S^{i-1}=(S_1,\hdots,S_{i-1})$. We study the capacity region of this state-dependent MAC model under both causal and strictly causal settings. 

For the model with causal states, we characterize the capacity region in the discrete memoryless case. We show that a cooperative scheme that is based on Shannon strategies \cite{Sh58} is optimal. This is to be opposed to the case of MAC with independent inputs in which it has been shown in \cite[Section III]{LS13a} that Shannon strategies are suboptimal in general. 

For the model with strictly causal states at both encoders, while building on the recent related work \cite{LS13a} (see also \cite{LS13b,LSY13,ZPS13}), it can be shown that the knowledge of the states strictly causally at the encoders is generally helpful, characterizing the capacity region of this model does not seem to be easy to obtain, even though one of the encoders knows both messages. In particular, while it can be expected that gains can be obtained by having the encoders cooperate in sending a description of the state to the receiver through a block Markov coding scheme, it is not easy to see how the compression of the state should be performed optimally. For instance, it is not clear whether sending an individual layer of state compression by the encoder that transmits both messages increases the transmission rates beyond what is possible with only the cooperative layer. Note that for the non-cooperative MAC of \cite{LS13a} it is beneficial that each encoder sends also an individual description of the state to the receiver, in addition to the description of the state that is sent cooperatively by both encoders; and this is reflected therein through that the inner bound of \cite[Theorem 2]{LS13a} strictly outperforms that of \cite[Theorem 1]{LS13a} -- the improvement comes precisely from the fact that, for both encoders, in each block a part of the input is composed of an individual compression of the state and the input in the previous block.

In this paper, for the model with states known strictly causally at both encoders we establish inner and outer bounds on the capacity region. The outer bound is non trivial, and has the advantage of having a relatively simple form that incorporates directly the channel inputs $X_1$ and $X_2$ from the encoders and \textit{only one auxiliary random variable}. To establish this outer bound, we first derive another outer bound on the capacity region whose expression involves two auxiliary random variables. We then show that this outer bound can be recast into a simpler form which is more insightful, and whose expression depends on only one auxiliary random variable. This is obtained by showing that the second auxiliary random variable can be chosen optimally to be a constant. In addition to its simplicity, the resulting expression of the outer bound has the advantage of suggesting that, by opposition to the MAC with independent inputs of \cite{LS13a}, for the model that we study there is no gain, or at best only little, to expect from having the encoder that transmits both messages also sending an individual compression of the state to the receiver, in addition to the cooperative compression. Note, however, that optimal forms of compressions are still to be found, since the tightness of the outer bound is still to be shown in general. Next, using the insights that we gain from the obtained outer bound, we establish an inner bound on the capacity region. This inner bound is based on a Block-Markov coding scheme in which the two encoders collaborate in both transmitting the common message and also conveying a lossy version of the state to the decoder. In this coding scheme, the encoder that transmits both messages does \textit{not} send any individual compression of the state beyond what is performed cooperatively with the other encoder.

The inner and outer bounds differ only through the associated joint measures; and, for instance, a Markov-chain relation that holds for the inner bound and not for the outer bound. Next, by investigating a class of channels for which the state can be obtained as a deterministic function of the channel inputs and output, we show that the inner and outer bounds agree; and, so, we characterize the capacity region in this case. 

Furthermore, we also study the case in which the state is revealed (strictly causally) to only one encoder. In this case, we show that revealing the state to the encoder that sends only the common message can increase the capacity region, whereas revealing it to the encoder that sends both messages does not increase the capacity region. In the former case, we show that there is dilemma at the informed encoder among exploiting the available state and creating message-cooperation with the other encoder. We develop a coding scheme that resolves this tension by splitting the codeword of the informed encoder into two parts, one that is meant to carry only the description of the state and is independent of the other encoder's input and one which is sent cooperatively with the other encoder and is generated independently of the state. We also show that this scheme is optimal in some special cases. Throughout the paper, we also discuss a number of examples; and compute the capacity for some of these examples.

\subsection{Related Work}\label{secI_subsecA}

 There is a connection between the role of states that are known strictly causally at an encoder and that of output feedback given to that encoder. In single-user channels, it is now well known that strictly causal feedback does not increase the capacity \cite{Sh56}. In multiuser channels or networks, however, the situation changes drastically, and output feedback can be beneficial --- but its role is still highly missunderstood. One has a similar picture with strictly causal states at the encoder. In single-user channels, independent and identically distributed states available only in a strictly causal manner at the encoder have no effect on the capacity. In multiuser channels or networks, however, like feedback, strictly causal states in general increase the capacity.

 The study of networks with strictly causal, or delayed, channel state information (CSI) has spurred much interest over the few recent years, due to its importance from both information-theoretic and communications aspects. Non-cooperative multiaccess channels with delayed state information are studied in \cite{LS13a} in the case in which the transmission is governed by a common state that is revealed with delay to both transmitters, and in \cite{LS13b,LSY13} in  the case in which the transmission is governed by independent states each revealed with delay to a different transmitter. The capacity region of a multiaccess channel with states known strictly causally at the encoder that sends only the common message and noncausally at the other encoder is established in \cite{ZPS13} (see also \cite{ZPS11a} and \cite{ZPS12a}). 
 
 A related line of research, initiated with the work of Maddah-Ali and Tse \cite{M-AT12}, investigates the usefulness of stale or outdated channel state information - typically outdated values of fading coefficients, in wireless networks. In such communication problems, the CSI is learned at the transmitters typically through output CSI feedback; and the utility of the outdated CSI at the transmitters is demonstrated typically by investigating gains in terms of the degrees of freedom or multiplexing \cite{ZT03} offered by the network. In this regard, the availability of outdated CSI at the transmitters is generally exploited through coding schemes that rely on some sorts of interferences alignment \cite{J10}. Examples include multiple-input multiple-output (MIMO) broadcast channels \cite{VV11a,AGK11,XAJ12}, MIMO interference channels \cite{VV11b,VM-AA13} and MIMO X channels with \cite{ZASV13a} and without \cite{GMK11,YZ12} security constraints.

 A growing body of work studies multi-user state-dependent models. The problem of joint communication and state estimation, initiated in \cite{SCCK05}, has been studied recently in \cite{CH-KM12} for the causal state case and in \cite{CSW12} in the presence of a helper node. Relay channels with states are studied in \cite{ZKLV10,ZKLV08a,ZSPV10a,ZSPV10b,ZV07b,ZV09b,AMA09,KE-GS13,LSY11}. Recent advances in the study of broadcast channels with states can be found in \cite{LW13,OS13} (see also the references therein); and other related contributions on multiaccess channels with noncausal states at the encoders can be found in \cite{K-FM11a,SCYA11a,ZVD07a,ZS13a,ZS14a}, among other works.   Finally, for related works on the connected area of multiuser information embedding the reader may refer to \cite{ZPD05b} and \cite{ZV09a} and the references therein.


\subsection{Outline and Notation}\label{secI_subsecB}

An outline of the remainder of this paper is as follows. Section \ref{secII} describes in more details the problem setup. In Section \ref{secIII} we study the setting in which the states are revealed (strictly causally) to both encoders; and in  Section \ref{secIV} we study the setting in which the states are revealed (strictly causally) to only one encoder. Section~\ref{secV} characterizes the capacity region of the cooperative multiaccess channel with states revealed causally to both encoders. Section~\ref{secVI} provides some concluding remarks. 

Throughout the paper we use the following notations. Upper case letters are used to denote random variables, e.g., $X$; lower case letters are used to denote realizations of random variables, e.g., $x$; and calligraphic letters designate alphabets, i.e., $\mc X$. The probability distribution of a random variable $X$ is denoted by $P_X(x)$. Sometimes, for convenience, we write it as $P_X$.  We use the notation $\mathbb{E}_{X}[\cdot]$ to denote the expectation of random variable $X$. A probability distribution of a random variable $Y$ given $X$ is denoted by $P_{Y|X}$. The set of probability distributions defined on an alphabet $\mc X$ is denoted by $\mc P(\mc X)$. The cardinality of a set $\mc X$ is denoted by $|\mc X|$. For convenience, the length $n$ vector $x^n$ will occasionally be denoted in boldface notation $\dv x$. For integers $i \leq j$, we define $[i:j]:=\{i,i+1,\hdots,j\}$. Throughout this paper, we use $h_2(\alpha)$ to denote the entropy of a Bernoulli\:$(\alpha)$ source, i.e., $h_2(\alpha) = - \alpha \log(\alpha) - (1-\alpha)\log(1-\alpha)$ and $p * q$ to denote the binary convolution, i.e., $p * q = p(1-q)+q(1-p)$. Finally, throughout the paper, logarithms are taken to base $2$, and the complement to unity of a scalar $u \in [0,1]$ is sometimes denoted by $\bar{u}$, i.e., $\bar{u}=1-u$. 


\section{Problem Setup}\label{secII}

We consider a stationary memoryless two-user state-dependent MAC $W_{Y|X_1,X_2,S}$  whose output $Y \in \mc Y$ is controlled by the channel inputs $X_1 \in \mc X_1$ and $X_2 \in \mc X_2$ from the encoders and the channel state $S \in \mc S$ which is drawn according to a memoryless probability law $Q_S$. The state is revealed -- depending on the scenario -- strictly causally or causally, to only one or both encoders. If the state is revealed causally to Encoder $k$, $k=1,2$, at time $i$ this encoder knows  the values of the state sequence up to and including time $i$, i.e., $S^{i}=(S_1,\hdots,S_{i-1},S_i)$. If the state is revealed only strictly causally to Encoder $k$, $k=1,2$, at time $i$ this encoder knows the values of the state sequence up to time $i-1$, i.e., $S^{i-1}=(S_1,\hdots,S_{i-1})$.

\begin{figure}[htpb]
\centering
\includegraphics[width=0.7\linewidth]{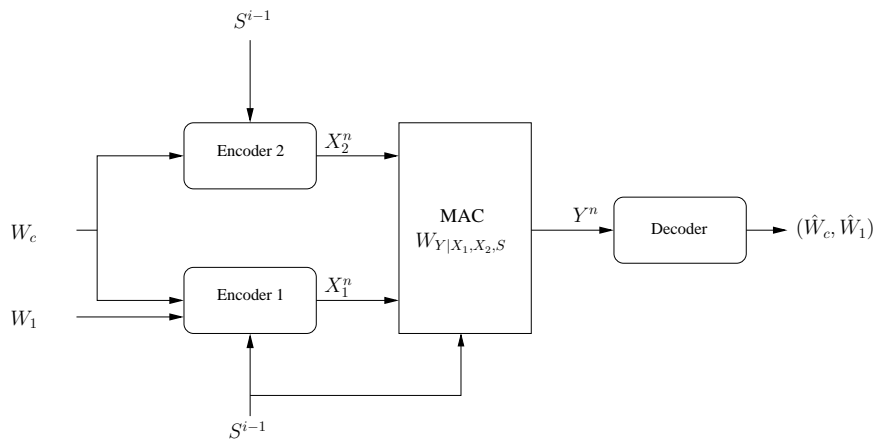}
\caption{State-dependent MAC with degraded messages sets and states known, strictly causally, to both the encoders.}
\label{ModelForMACwithAsymmetricCSI}
\end{figure}

Encoder 2 wants to send a common message $W_c$ and Encoder 1 wants to send an independent individual message $W_1$ along with the common message $W_c$. We assume that the common message $W_c$ and the individual message $W_1$ are independent random variables drawn uniformly from the sets $\mc W_c=\{1,\cdots,M_c\}$ and  $\mc W_1=\{1,\cdots,M_1\}$, respectively. The sequences $X_{1}^n$ and $X_{2}^n$ from the encoders are sent across a state-dependent multiple access channel modeled as a memoryless conditional probability distribution $W_{Y|X_1,X_2,S}$. The laws governing the state sequence and the output letters are given by
\begin{align}
W^n_{Y|X_1,X_2,S}(y^n| x^n_1, x^n_2, s^n) &= \prod_{i=1}^n W_{Y|X_1,X_2,S}(y_i|x_{1i},x_{2i},s_i)\\
Q^n_S(s^n) &= \prod_{i=1}^n Q_S(s_i).
\end{align}
The receiver guesses the pair $(\hat{W}_c,\hat{W}_1)$ from the channel output $Y^n$. 

In Figure~\ref{ModelForMACwithAsymmetricCSI}, the state may model some common information which is received, with delay, only by authorized (or connected) entities. Also, in a wireless context, while fading state variations are often measured at the receivers and then possibly fed back to the transmitters, certain interfering signals occurring at the vicinity of the transmitters may be measured or estimated more effectively directly by these, due to proximity, rather than at the end nodes.

\begin{definition}\label{basic-definitions-strictly-causal-case}
For positive integers $n$, $M_c$ and $M_1$, an $(M_c,M_1,n,\epsilon)$ code for the cooperative multiple access channel with states known strictly causally to both encoders consists of a sequence of mappings 
\begin{align}
\phi_{1,i}: \mc W_c{\times}\mc W_1{\times}\mc S^{i-1} \longrightarrow \mc X_1, , \quad i=1,\hdots,n
\label{encoding-function-encoder1-strictly-causal-states-setting}
\end{align}
at Encoder 1, a sequence of mappings 
\begin{align}
\phi_{2,i}: \mc W_c{\times}\mc S^{i-1} \longrightarrow \mc X_2, \quad i=1,\hdots,n
\label{encoding-function-encoder2-strictly-causal-states-setting}
\end{align}
at Encoder 2, and a decoder map
\begin{align}
\psi : \mc Y^n \longrightarrow \mc W_c{\times}\mc W_1
\label{decoding-function}
\end{align}
such that the average probability of error is bounded by $\epsilon$,
\begin{equation}
P_e^n = \mathbb{E}_{S}\big[\mathrm{Pr}\big(\psi(Y^n)\neq (W_c,W_1)|S^n=s^n\big)\big] \leq \epsilon.
\label{definition-probability-of-error}
\end{equation}
The rate of the common message and the rate of the individual message are defined as 
\begin{align*}
&R_c = \frac{1}{n}\log M_c \qquad \text{and} \qquad R_1 = \frac{1}{n}\log M_1,
\end{align*}
respectively.
A rate pair $(R_c,R_1)$ is said to be achievable if for every $\epsilon > 0$ there exists an $(2^{nR_c},2^{nR_1},n,\epsilon)$ code for the channel $W_{Y|X_1,X_2,S}$.  The capacity region $\mc C_{\text{s-c}}$ of the state-dependent MAC with strictly causal states is defined as the closure of the set of achievable rate pairs.
\end{definition}

\begin{definition}\label{basic-definitions-causal-case}
For positive integers $n$, $M_c$ and $M_1$, an $(M_c,M_1,n,\epsilon)$ code for the cooperative multiple access channel with states known causally to both encoders consists of a sequence of mappings      
\begin{align}
\phi_{1,i}: \mc W_c{\times}\mc W_1{\times}\mc S^{i} \longrightarrow \mc X_1, , \quad i=1,\hdots,n
\label{encoding-function-encoder1-causal-states-setting}
\end{align}
at Encoder 1, a sequence of mappings
\begin{align}
\phi_{2,i}: \mc W_c{\times}\mc S^{i} \longrightarrow \mc X_2, \quad i=1,\hdots,n
\label{encoding-function-encoder2-causal-states-setting}
\end{align}
at Encoder 2, and a decoder map \eqref{decoding-function} such that the probability of error is bounded as in \eqref{definition-probability-of-error}.

\noindent The definitions of a rate pair $(R_c,R_1)$ to be achievable as well as the capacity region, which we denote by $\mc C_{\text{c}}$ in this case, are similar to those in the strictly-causal states setting in Definition~\ref{basic-definitions-strictly-causal-case}.
\end{definition}

\noindent Similarly, in the case in which the states are revealed strictly causally to only one encoder, the definitions of a rate pair $(R_c,R_1)$ to be achievable as well as the capacity region can be obtained in a way that is similar to that in Definition~\ref{basic-definitions-strictly-causal-case}.

\section{Strictly Causal States at Both Encoders}\label{secIII}

In this section, it is assumed that the alphabets $\mc S, \mc X_1, \mc X_2$ are finite.

\subsection{Outer Bound on the Capacity Region}\label{secIII_subsecA}

Let $\tilde{\mc P}^{\text{out}}_{\text{s-c}}$ stand for the collection of all random variables $(S,U,V,X_1,X_2,Y)$ such that $U$, $V$, $X_1$ and $X_2$ take values in finite alphabets $\mc U$, $\mc V$, $\mc X_1$ and $\mc X_2$, respectively, and satisfy
\begin{subequations}
\begin{align}
P_{S,U,V,X_1,X_2,Y}(s,u,v,x_1,x_2,y) &= P_{S,U,V,X_1X_2}(s,u,v,x_1,x_2)W_{Y|X_1,X_2,S}(y|x_1,x_2,s)\\
P_{S,U,V,X_1,X_2}(s,u,v,x_1,x_2) &= Q_S(s)P_{X_2}(x_2)P_{X_1|X_2}(x_1|x_2)P_{V|S,X_1,X_2}(v|s,x_1,x_2)P_{U|S,V,X_1,X_2}(u|s,v,x_1,x_2).
\end{align}
\label{measure-temporary-outer-bound-strictly-causal-states-setting}
\end{subequations}
and
\begin{equation}
0 \leq I(V,X_2;Y)-I(V,X_2;S).
\label{nonnegativity-constraint-temporary-outer-bound}
\end{equation}

The relations in \eqref{measure-temporary-outer-bound-strictly-causal-states-setting} imply that $(U,V) \leftrightarrow (S,X_1,X_2) \leftrightarrow Y$ is a Markov chain, and $X_1$ and $X_2$ are independent of $S$.

Define $\tilde{\mc R}^{\text{out}}_{\text{s-c}}$ to be the set of all rate pairs $(R_c,R_1)$ such that
\begin{align}
R_1 \: &\leq \: I(U,X_1;Y|V,X_2)-I(U,X_1;S|V,X_2) \nonumber\\
R_c+ R_1 \: &\leq \: I(U,V,X_1,X_2;Y)-I(U,V,X_1,X_2;S)\nonumber\\
&\hspace{2cm} \text{for some}\:\: (S,U,V,X_1,X_2,Y) \in \tilde{\mc P}^{\text{out}}_{\text{s-c}}.
\label{temporary-outer-bound-strictly-causal-states-setting}
\end{align}

\noindent As stated in the following theorem, the set $\tilde{\mc R}^{\text{out}}_{\text{s-c}}$ is an outer bound on the capacity region of the state-dependent discrete memoryless MAC with strictly-causal states.

\vspace{0.3cm}

\begin{theorem}\label{theorem-temporary-outer-bound-strictly-causal-states-setting}
The capacity region of the multiple access channel with degraded messages sets and strictly causal states known only at the encoders satisfies 
\begin{equation}
\mc C_{\text{s-c}} \subseteq \tilde{\mc R}^{\text{out}}_{\text{s-c}}.
\end{equation}
\end{theorem}

\vspace{0.3cm}

\textbf{Proof:} The proof of Theorem~\ref{theorem-temporary-outer-bound-strictly-causal-states-setting} is given in Appendix~\ref{appendix-proof-theorem-temporary-outer-bound-strictly-causal-states-setting}.

We now recast the outer bound $\tilde{\mc R}^{\text{out}}$ into a form that will be shown to be more convenient (see Remark~\ref{remark1-outer-bounds-strictly-causal-states-setting} and Remark~\ref{remark2-outer-bounds-strictly-causal-states-setting} below). This is done by showing that the maximizing auxiliary random variable $U$ in $\tilde{\mc R}^{\text{out}}$ is a constant, i.e., $U=\emptyset$; and can be formalized as follows. Let $\mc P^{\text{out}}_{\text{s-c}}$ be the collection of all random variables $(S,V,X_1,X_2,Y)$ such that $V$, $X_1$ and $X_2$ take values in finite alphabets $\mc V$, $\mc X_1$ and $\mc X_2$, respectively, and satisfy
\begin{equation}
P_{S,V,X_1,X_2,Y} = Q_SP_{X_2}P_{X_1|X_2}P_{V|S,X_1,X_2}W_{Y|X_1,X_2,S}
\label{measure-outer-bound-strictly-causal-states-setting}
\end{equation}
and the constraint \eqref{nonnegativity-constraint-temporary-outer-bound}. Also, define ${\mc R}^{\text{out}}_{\text{s-c}}$ to be the set of all rate pairs $(R_c,R_1)$ such that
\begin{subequations}
\begin{align}
\label{outer-bound-strictly-causal-states-setting-individual-rate}
R_1 \: &\leq \: I(X_1;Y|V,X_2)\\
R_c+ R_1 \: &\leq \: I(V,X_1,X_2;Y)-I(V,X_1,X_2;S)\\
\label{outer-bound-strictly-causal-states-setting-sum-rate}
&\hspace{2cm} \text{for some}\:\: (S,V,X_1,X_2,Y) \in {\mc P}^{\text{out}}_{\text{s-c}}.\nonumber
\end{align}
\label{outer-bound-strictly-causal-states-setting}
\end{subequations}

It is easy to see that $\mc R^{\text{out}}_{\text{s-c}} \subseteq \tilde{\mc R}^{\text{out}}_{\text{s-c}}$, as $\mc R^{\text{out}}_{\text{s-c}}$ can be obtained from $\tilde{\mc R}^{\text{out}}_{\text{s-c}}$ by setting $U=\emptyset$. As shown in the proof of the theorem that will follow, $\tilde{\mc R}^{\text{out}}_{\text{s-c}} \subseteq \mc R^{\text{out}}_{\text{s-c}}$; and so $\mc R^{\text{out}}_{\text{s-c}} = \tilde{\mc R}^{\text{out}}_{\text{s-c}}$. Thus, by Theorem~\ref{theorem-temporary-outer-bound-strictly-causal-states-setting}, $\mc R^{\text{out}}$ is an outer bound on the capacity region of the state-dependent discrete memoryless MAC model with strictly-causal states.

\begin{theorem}\label{theorem-outer-bound-strictly-causal-states-setting}
The capacity region of the multiple access channel with degraded messages sets and strictly causal states known only at the encoders satisfies 
\begin{equation}
\mc C_{\text{s-c}} \subseteq \mc R^{\text{out}}_{\text{s-c}}.
\end{equation}
\end{theorem}

\vspace{0.3cm}

\textbf{Proof:} The proof of Theorem~\ref{theorem-outer-bound-strictly-causal-states-setting} is given in Appendix~\ref{appendix-proof-theorem-outer-bound-strictly-causal-states-setting}.

The outer bound can be expressed equivalently using $\tilde{\mc R}^{\text{out}}_{\text{s-c}}$ or $\mc R^{\text{out}}_{\text{s-c}}$, since the two sets coincide. However, the form $\mc R^{\text{out}}_{\text{s-c}}$ of the outer bound is more convenient and insightful. The following remarks aim at reflecting this.

\begin{remark}\label{remark1-outer-bounds-strictly-causal-states-setting}
As we already mentioned, some recent works have shown the utility of strictly causal states at the encoders in increasing the capacity region of multiaccess channels in certain settings. For example, this has been demonstrated for a MAC with independent inputs and states known strictly causally at the encoders \cite{LS13a,LS13b,LSY13}, and for a MAC with degraded messages sets with the states known strictly causally to the encoder that sends only the common-message and noncausally at the encoder that sends both messages \cite{ZPS13,ZPS11a,ZPS12a}. Also, in these settings, the increase in the capacity region is created by having the encoders cooperate in each block to convey a lossy version of the state of the previous block to the receiver. Furthermore, in the case of the MAC with independent inputs of \cite{LS13a}, it is shown that additional improvement can be obtained by having each encoder also sending a compressed version of the pair (input, state)  of the previous block, in addition to the cooperative transmission with the other encoder of the common compression of the state. (This is reflected in \cite{LS13a} through the improvement of the inner bound of Theorem 2 therein over that of Theorem 1). In our case, since one encoder knows the other encoder's message, it is not evident \`a-priori whether a similar additional improvement could be expected from having the encoder that transmits both messages also sending another compression of the state, in addition to that sent cooperatively. \qed 
\end{remark}

\begin{remark}\label{remark2-outer-bounds-strictly-causal-states-setting}
 A direct proof of the outer bound in its form $\mc R^{\text{out}}_{\text{s-c}}$ does not seem to be easy to obtain because of the necessity of introducing two auxiliary random variables in typical outer bounding approaches that are similar to that of Theorem~\ref{theorem-temporary-outer-bound-strictly-causal-states-setting}. In addition to that it is simpler comparatively, the form $\mc R^{\text{out}}_{\text{s-c}}$ of the outer bound is more convenient and insightful. It involves only one auxiliary random variable, $V$, (which, in a corresponding coding scheme, would represent intuitively the lossy version of the state that is to be sent by the two encoders cooperatively). Because the auxiliary random variable $U$ (which, in a corresponding coding scheme, would represent intuitively the additional compression of the state that is performed by the encoder that transmits both messages) can be set optimally to be a constant, the outer bound $\mc R^{\text{out}}_{\text{s-c}}$ suggests implicitly that there is no gain to be expected from additional compression at Encoder 1. That is, by opposition to the case of the non-cooperative MAC of \cite{LS13a}, for our model, for an efficient exploitation of the knowledge of the states strictly causally at the encoders it \textit{seems}\footnote{Note, however, that since the tightness of the outer bound of Theorem~\ref{theorem-outer-bound-strictly-causal-states-setting} is still to be shown in general, optimal state compressions for this model are still to be found.} enough to compress the state only cooperatively. We should mention that, although somewhat intuitive given known results on the role of feedback and strictly causal states at the encoder in point-to-point channels, a formal proof of the aforementioned fact for the model that we study does not follow directly from these existing results. \qed 
\end{remark}

We now state a proposition that provides an alternative outer bound on the capacity region of the multiaccess channel with degraded messages sets and states known only strictly causally at both encoders that we study. This proposition will turn out to be useful in Section~\ref{secIII_subsecD}.

\noindent Let $\breve{\mc R}^{\text{out}}_{\text{s-c}}$ be the set of all rate pairs $(R_c,R_1)$ satisfying
\begin{align}
R_1 &\leq I(X_1;Y|X_2,S)\nonumber\\
R_c + R_1 &\leq I(X_1,X_2;Y)
\label{alternative-outer-bound-strictly-causal-states-setting}
\end{align}
for some measure
\begin{equation}
P_{S,X_1,X_2,Y} = Q_SP_{X_1,X_2}W_{Y|S,X_1,X_2}.
\label{measure-alternative-outer-bound-strictly-causal-states-setting}
\end{equation}

\begin{proposition}\label{proposition-alternative-outer-bound-strictly-causal-states-setting}
The capacity region $\mc C_{\text{s-c}}$ of the multiple access channel with degraded messages sets and strictly causal states known only at the encoders satisfies
\begin{equation}
\mc C_{\text{s-c}} \subseteq \breve{\mc R}^{\text{out}}_{\text{s-c}}.
\end{equation}
\end{proposition}

\vspace{0.3cm}

\textbf{Proof:} The proof of Proposition~\ref{proposition-alternative-outer-bound-strictly-causal-states-setting} is given in Appendix~\ref{appendix-proof-proposition-alternative-outer-bound-strictly-causal-states-setting}.

The bound on the sum rate of Theorem~\ref{theorem-outer-bound-strictly-causal-states-setting} is at least as tight as that of Proposition~\ref{proposition-alternative-outer-bound-strictly-causal-states-setting}. This can be seen through the following inequalities.

\begin{align}
I(V,X_1,X_2;Y)-&I(V,X_1,X_2;S)\nonumber\\\
	\label{comparison-theorem2-theorem3-proof-step1}
	&= I(X_1,X_2;Y) + I(V;Y|X_1,X_2) - I(V;S|X_1,X_2)\\
	\label{comparison-theorem2-theorem3-proof-step2}
	&= I(X_1,X_2;Y) + I(V;Y|S,X_1,X_2) - I(V;S|X_1,X_2,Y)\\
	&= I(X_1,X_2;Y) - I(V;S|X_1,X_2,Y) + H(Y|S,X_1,X_2) - H(Y|V,S,X_1,X_2)\\
	\label{comparison-theorem2-theorem3-proof-step3}
	&= I(X_1,X_2;Y) - I(V;S|X_1,X_2,Y) + H(Y|S,X_1,X_2) - H(Y|S,X_1,X_2)\\
	&= I(X_1,X_2;Y)-I(V;S|X_1,X_2,Y)\\
	&\leq  I(X_1,X_2;Y)
\end{align}
where: \eqref{comparison-theorem2-theorem3-proof-step1} follows since $X_1$ and $X_2$ are independent of the state $S$; \eqref{comparison-theorem2-theorem3-proof-step2} follows since for all random variables $A$, $B$ and $C$, we have $I(A;B)-I(A;C)=I(A;B|C)-I(A;C|B)$; and \eqref{comparison-theorem2-theorem3-proof-step3} follows since $V \leftrightarrow (S,X_1,X_2) \leftrightarrow Y$ is a Markov chain.

For some channels, the sum-rate constraint of outer bound of Theorem~\ref{theorem-outer-bound-strictly-causal-states-setting} is \textit{strictly} tighter than that of the outer bound of Proposition~\ref{proposition-alternative-outer-bound-strictly-causal-states-setting}. The following example, illustrates this.

\begin{example}\label{example-comparison-of-outer-bounds} 
Consider the following discrete memoryless channel, considered initially in \cite{LS13a},
\begin{equation}
Y = X_S
\end{equation}
 where $\mc X_1=\mc X_2=\mc Y=\{0,1\}$, and the state $S$ is uniformly distributed over the set $\mc S=\{1,2\}$ and acts as a random switch that connects a randomly chosen transmitter to the output.

\noindent For this channel, the rate-pair $(R_c,R_1)=(1/2,1/2)$ is in the outer bound of Proposition~\ref{proposition-alternative-outer-bound-strictly-causal-states-setting}, but not in that of Theorem~\ref{theorem-outer-bound-strictly-causal-states-setting},  i.e., $(1/2,1/2) \in \breve{\mc R}^{\text{out}}_{\text{s-c}}$ and $(1/2,1/2) \notin \mc R^{\text{out}}_{\text{s-c}}$.
\end{example}

\textbf{Proof:} The analysis of Example~\ref{example-comparison-of-outer-bounds} appears in Appendix~\ref{appendix-analysis-example-comparison-of-outer-bounds}.

\subsection{Inner Bound on the Capacity Region}\label{secIII_subsecB}

Let $\mc P^{\text{in}}_{\text{s-c}}$ stand for the collection of all random variables $(S,V,X_1,X_2,Y)$ such that $V$, $X_1$ and $X_2$ take values in finite alphabets $\mc V$, $\mc X_1$ and $\mc X_2$, respectively, and satisfy 
\begin{subequations}
\begin{align}
P_{S,V,X_1,X_2,Y}(s,v,x_1,x_2,y) &= P_{S,V,X_1,X_2}(s,v,x_1,x_2)W_{Y|X_1,X_2,S}(y|x_1,x_2,s)\\
P_{S,V,X_1,X_2}(s,v,x_1,x_2) &= Q_S(s)P_{X_2}(x_2)P_{X_1|X_2}(x_1|x_2)P_{V|S,X_2}(v|s,x_2)
\end{align}
\label{measure-inner-bound-strictly-causal-states-setting}
\end{subequations}
and 
\begin{equation}
0 \leq I(V,X_2;Y)-I(V,X_2;S).
\label{nonnegativity-constraint-inner-bound}
\end{equation}
The relations in \eqref{measure-inner-bound-strictly-causal-states-setting} imply that $V \leftrightarrow (S,X_1,X_2) \leftrightarrow Y$ and $X_1 \leftrightarrow X_2 \leftrightarrow V$ are Markov chains; and $X_1$ and $X_2$ are independent of $S$.

Define $\mc R^{\text{in}}_{\text{s-c}}$ to be the set of all rate pairs $(R_c,R_1)$ such that
\begin{subequations}
\begin{align}
\label{inner-bound-strictly-causal-states-setting-individual-rate}
R_1 \: &\leq \: I(X_1;Y|V,X_2)\\
\label{inner-bound-strictly-causal-states-setting-sum-rate}
R_c+ R_1 \: &\leq \: I(V,X_1,X_2;Y)-I(V,X_1,X_2;S)\\
&\hspace{2cm} \text{for some}\:\: (S,V,X_1,X_2,Y) \in \mc P^{\text{in}}_{\text{s-c}}. \nonumber
\end{align}
\label{inner-bound-strictly-causal-states-setting}
\end{subequations}

\noindent As stated in the following theorem, the set $\mc R^{\text{in}}_{\text{s-c}}$ is an inner bound on the capacity region of the state-dependent discrete memoryless MAC with strictly-causal states.

\vspace{0.3cm}

\begin{theorem}\label{theorem-inner-bound-strictly-causal-states-setting}
The capacity region of the multiple access channel with degraded messages sets and strictly causal states known only at the encoders satisfies 
\begin{equation}
\mc R^{\text{in}}_{\text{s-c}} \subseteq \mc C_{\text{s-c}}.
\end{equation}
\end{theorem}

\vspace{0.3cm}

\textbf{Proof:} An outline proof of the coding scheme that we use for the proof of Theorem~\ref{theorem-inner-bound-strictly-causal-states-setting} will follow. The associated error analysis is provided in Appendix~\ref{appendix-proof-theorem-inner-bound-strictly-causal-states-setting}.

\vspace{0.3cm}

The following proposition states some properties of $\mc R^{\text{in}}_{\text{s-c}}$ and $\mc R^{\text{out}}_{\text{s-c}}$.

\begin{proposition}\label{proposition-bounds-auxiliary-random-variables-strictly-causal-states-setting}

{\color{white} (properties of inner and outer bounds)}

\begin{itemize}
\item[1.] The sets $\mc R^{\text{in}}_{\text{s-c}}$ and $\mc R^{\text{out}}_{\text{s-c}}$ are convex.
\item[2.] To exhaust $\mc R^{\text{in}}_{\text{s-c}}$ and $\mc R^{\text{out}}_{\text{s-c}}$, it is enough to restrict $\mc V$ to satisfy
\begin{equation}
|\mc V| \leq |\mc S||\mc X_1||\mc X_2|+2.
\end{equation}
\label{bounds-auxiliary-random-variables-inner-and-temporary-outer-bounds-strictly-causal-states-setting}
\end{itemize}
\end{proposition}

\vspace{0.3cm}

\textbf{Proof:} The proof of Proposition~\ref{proposition-bounds-auxiliary-random-variables-strictly-causal-states-setting} appears in Appendix~\ref{appendix-proposition-bounds-auxiliary-random-variables-strictly-causal-states-setting}.

\vspace{0.3cm}

\begin{remark}\label{remark-tightness-of-inner-bound-strictly-causal-states-setting}
The inner bound $\mc R^{\text{in}}_{\text{s-c}}$ differs from the outer bound $\mc R^{\text{out}}_{\text{s-c}}$ only through the Markov chain $X_1 \leftrightarrow X_2 \leftrightarrow V$. The outer bound requires arbitrary dependence of the auxiliary random variable $V$ on the inputs $X_1$ and $X_2$ by the encoders. For achievability results, while in block $i$ the dependence of $V$ on the input $X_2$ by the encoder that sends only the common message can be obtained by generating the covering codeword $\dv v$ on top of the input codeword $\dv x_2$ from the previous block $i-1$ and performing conditional compression of the state sequence from block $i-1$, i.e.,  conditionally on the input $\dv x_2$ by Encoder 2 in the previous block $i-1$, the dependence of $V$ on the input $X_1$ by the encoder that transmits both messages is not easy to obtain. Partly, this is because i) the codeword $\dv v$ can not be generated on top of $\dv x_1$ (since Encoder 2 does not know the individual message of Encoder 1), and ii) the input $\dv x_1$ by Encoder 1 has to be independent of the state sequence $\dv s$. \qed
\end{remark}

\begin{figure}[htpb]
\centering
\includegraphics[width=0.8\linewidth]{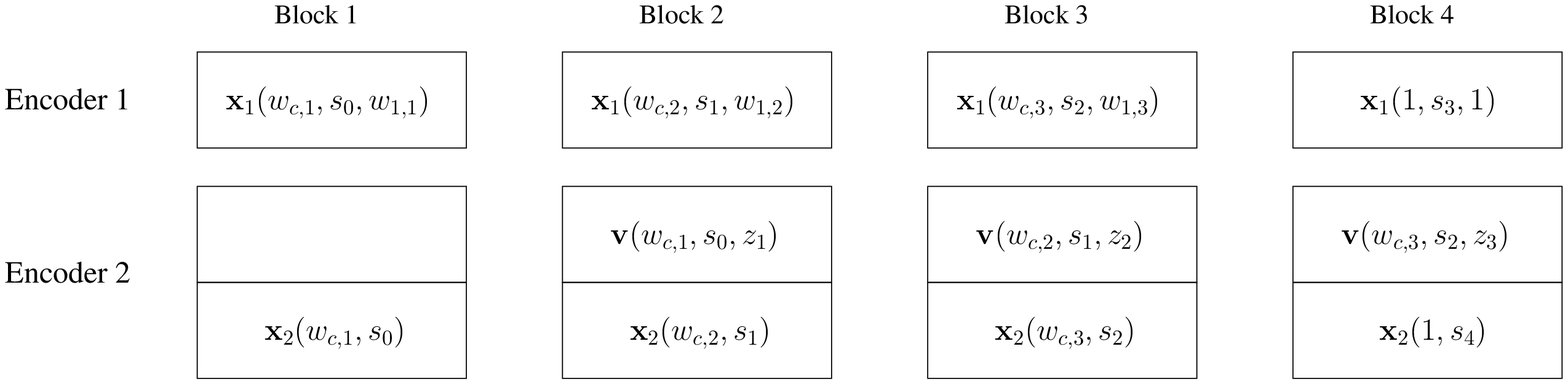}
\caption{Block Markov coding scheme employed for the inner bound of Theorem~\ref{theorem-inner-bound-strictly-causal-states-setting}, for $B=4$.}
\label{fig-example-block-markov-scheme}
\end{figure}

\begin{remark}\label{remark-main-idea-inner-bound-strictly-causal-states-setting}
The proof of Theorem~\ref{theorem-inner-bound-strictly-causal-states-setting} is based on a Block-Markov coding scheme in which the encoders collaborate to convey a lossy version of the state to the receiver, in addition to the information messages. The lossy version of the state is obtained through Wyner-Ziv compression. Also, in each block, Encoder 1 also transmits an individual information. However, in accordance with the aforementioned insights that we gain from the outer bound of Theorem~\ref{theorem-outer-bound-strictly-causal-states-setting}, the state is sent to the receiver \textit{only} cooperatively. That is, by opposition to the coding scheme of \cite[Theorem 2]{LS13a} for the MAC with independent inputs, Encoder 1 does not compress or convey the state to the receiver beyond what is done cooperatively with Encoder 2. More specifically, the encoding and transmission scheme is as follows.  Let $\dv s[i]$ denote the channel state in block $i$, and $s_{i}$ the index of the cell $\mc C_{s_i}$ containing the compression index $z_i$ of the state $\dv s[i]$, obtained through Wyner-Ziv compression. In block $i$, Encoder $2$, which has learned the state sequence $\dv s[i-1]$, knows $s_{i-2}$ and looks for a compression index $z_{i-1}$ such that $\dv v(w_{c,i-1},s_{i-2},z_{i-1})$ is strongly jointly typical with $\dv s[i-1]$ and $\dv x_2(w_{c,i-1},s_{i-2})$. It then transmits a codeword $\dv x_2(w_{c,i},s_{i-1})$ (drawn according to the appropriate marginal using \eqref{measure-inner-bound-strictly-causal-states-setting}), where the cell index $s_{i-1}$ is the index of the cell containing $z_{i-1}$, i.e., $z_{i-1} \in \mc C_{s_{i-1}}$. Encoder 1 finds $\dv x_2(w_{c,i},s_{i-1})$ similarly. It then transmits a vector $\dv x_1(w_{c,i},s_{i-1},w_{1i})$ (drawn according to the appropriate marginal using \eqref{measure-inner-bound-strictly-causal-states-setting}). For convenience, we list the codewords that are used for transmission in the first four blocks in Figure~\ref{fig-example-block-markov-scheme}. \qed
\end{remark}

The scheme of Theorem~\ref{theorem-inner-bound-strictly-causal-states-setting} utilizes Wyner-Ziv binning for the joint compression of the state by the two encoders. As it can be seen from the proof, the constraint 
\begin{equation}
0 \leq I(V,X_2;Y)-I(V,X_2;S)
\label{constraint-inner-bound}
\end{equation}
or, equivalently,
\begin{equation}
I(V;S|X_2)-I(V;Y|X_2) \leq I(X_2;Y),
\label{constraint-inner-bound-equivalent-form}
\end{equation}
is caused by having the receiver decode the compression index uniquely. One can devise an alternate coding scheme that achieves the region of Theorem~\ref{theorem-inner-bound-strictly-causal-states-setting} but without the constraint \eqref{nonnegativity-constraint-inner-bound}. More specifically, let $\tilde{\mc P}^{\text{in}}_{\text{s-c}}$ stand for the collection of all random variables $(S,V,X_1,X_2,Y)$ such that $V$, $X_1$ and $X_2$ take values in finite sets $\mc V$, $\mc X_1$ and $\mc X_2$, respectively, and satisfy \eqref{measure-inner-bound-strictly-causal-states-setting}. Also, define $\tilde{\mc R}^{\text{in}}_{\text{s-c}}$ to be the set of all rate pairs $(R_c,R_1)$ satisfying the inequalities in \eqref{inner-bound-strictly-causal-states-setting} for some $(S,V,X_1,X_2,Y) \in \tilde{\mc P}^{\text{in}}_{\text{s-c}}$. Because the constraint \eqref{nonnegativity-constraint-inner-bound} is relaxed, the set $\tilde{\mc R}^{\text{in}}_{\text{s-c}}$ satisfies 
\begin{equation}
\mc R^{\text{in}}_{\text{s-c}} \subseteq \tilde{\mc R}^{\text{in}}_{\text{s-c}} \subseteq \mc C_{\text{s-c}}.
\label{inner-bound-without-nonnegativity-constraint-strictly-causal-states-setting}
\end{equation}
The coding scheme that achieves the inner bound $\tilde{\mc R}^{\text{in}}_{\text{s-c}}$ is similar to that of Theorem~\ref{theorem-inner-bound-strictly-causal-states-setting}, but with the state compression performed \`a-la noisy network coding by Lim, Kim, El Gamal and Chung \cite{H-LKGC11} or the quantize-map-and-forward by Avestimeher, Diggavi and Tse \cite{ADT11}, i.e.,  with no binning. We omit it here for brevity.\footnote{The reader may refer to \cite{ZPS13} (see also \cite{ZPS11a} and \cite{ZPS12a}) where a setup with mixed -- strictly causal and noncausal states, is analyzed and the state compression is performed \`a-la noisy network coding.}

As the next example shows, the inner bound of Theorem~\ref{theorem-inner-bound-strictly-causal-states-setting} is strictly contained in the outer bound of Theorem~\ref{theorem-outer-bound-strictly-causal-states-setting}, i.e.,
\begin{equation}
\mc R^{\text{in}}_{\text{s-c}} \subsetneq \mc R^{\text{out}}_{\text{s-c}}.
\end{equation}

\begin{example}\label{example-comparison-of-inner-and-outer-bounds}
Consider a two-user cooperative MAC with binary inputs $\mc X_1=\mc X_2=\{0,1\}$ and output $Y=(Y_1,Y_2) \in \{0,1\}^2$ with
\begin{subequations}
\begin{align}
Y_1 &= X_1 + S_{X_1+X_2}\\
Y_2 &= X_2.
\end{align}
\label{output-example-comparison-of-inner-and-outer-bounds}
\end{subequations}
The transmission is controlled by a random state $S=(S_0,S_1) \in \{0,1\}^2$, where the state components $S_0$ and $S_1$ are i.i.d. $\text{Bernoulli}\:(p)$, where $p$ is the unique constant in the interval $[0,1/2]$ whose binary entropy is $1/2$, i.e.,
\begin{equation}
H(S_0) = H(S_1) = h_2(p) = \frac{1}{2}.
\end{equation}
In \eqref{output-example-comparison-of-inner-and-outer-bounds}, the addition is modulo two. Thus, if $X_1=X_2$ then $Y_1$ is the mod-2 sum of $X_1$ and $S_0$; otherwise, it is the mod-2 sum of $X_1$ and $S_1$. For this example the rate-pair $(R_c,R_1)=(1/2,1)$ is in the outer bound of Theorem~\ref{theorem-outer-bound-strictly-causal-states-setting}, but not in the inner bound of Theorem~\ref{theorem-inner-bound-strictly-causal-states-setting},  i.e., $(1/2,1) \in \mc R^{\text{out}}_{\text{s-c}}$ and $(1/2,1) \notin \mc R^{\text{in}}_{\text{s-c}}$.

\end{example}

\textbf{Proof:} The analysis of Example~\ref{example-comparison-of-inner-and-outer-bounds} appears in Appendix~\ref{appendix-analysis-example-comparison-of-inner-and-outer-bounds}. In what follows, we provide some intuition onto why the rate-pair $(R_c,R_1)=(1/2,1)$ is not in the inner bound $\mc R^{\text{in}}_{\text{s-c}}$. In order for the rate $R_1$ to be equal $1$, the receiver needs to learn $S_{X_1+X_2}$. In the coding scheme that yields the inner bound $\mc R^{\text{in}}_{\text{s-c}}$, the encoder that sends only the common message knows the values of the state $S=(S_0,S_1)$ as well as those of $X_2$ from the previous blocks, but not that of $X_1$; and, so, can not know the values of $S_{X_1+X_2}$ from the previous blocks.

\subsection{On the Utility of the Strictly Causal States}\label{secIII_subsecC}

The following example shows that revealing the states only strictly causally to both encoders increases the capacity region.

\begin{example}\label{example-uselfuness-of-knoweledge-of-states-strictly-causally}
Consider the memoryless binary MAC shown in Figure~\ref{BSCModelCounterExample-StrictlyCausalStates}. Here, all the random variables are binary $\{0,1\}$. The channel has two output components, i.e., $Y^n=(Y^n_1,Y^n_2)$. The component $Y^n_2$ is deterministic, $Y^n_2=X^n_2$, and the component $Y^n_1=X^n_1 + S^n + Z^n_1$, where the addition is modulo $2$. Encoder 2 has no message to transmit, and Encoder 1 transmits an individual message $W_1$. The encoders know the states only strictly causally. The state and noise vectors are independent and memoryless, with the state process $S_i$, $i \geq 1$, and the noise process $Z_{1,i}$, $i \geq 1$, assumed to be Bernoulli $(\frac{1}{2})$ and  Bernoulli $(p)$ processes, respectively. The vectors $X^n_1$ and $X^n_2$ are the channel inputs, subjected to the constraints
\begin{align}
\sum_{i=1}^{n} X_{1,i} &\leq nq_1 \quad \text{and} \quad \sum_{i=1}^{n} X_{2,i} \leq nq_2, \:\: q_2 \geq 1/2.
\label{BinaryChannel__InputsConstraints}
\end{align}

\begin{figure}[htpb]
\centering
\includegraphics[width=0.5\linewidth]{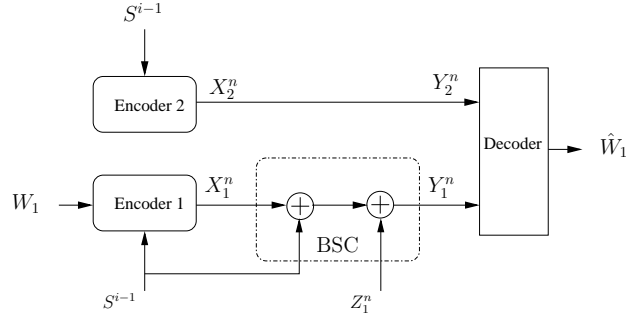}
\caption{Binary state-dependent MAC example with two output components, $Y^n=(Y^n_1,Y^n_2)$, with $Y^n_1=X^n_1 + S^n + Z^n_1$ and $Y^n_2=X^n_2$.}
\label{BSCModelCounterExample-StrictlyCausalStates}
\end{figure}

For this example, the strictly causal knowledge of the states at Encoder 2 increases the capacity, and in fact Encoder 1 can transmit at rates that are larger than the maximum rate that would be achievable had Encoder 2 been of no help.
\end{example}

\textit{Claim 1:} The capacity of the memoryless binary MAC with states known strictly causally at the encoders shown in Figure~\ref{BSCModelCounterExample-StrictlyCausalStates} is given by
\begin{align}
C_{\text{s-c}} &= \max_{p(x_1)} \:\: I(X_1;Y_1|S)
\label{Capacity__ModelCounterExample-StrictlyCausalStates}
\end{align}
where the maximization is over measures $p(x_1)$ satisfying the input constraint \eqref{BinaryChannel__InputsConstraints}.

\textbf{Proof:}  The proof of achievability is as follows. Set $R_c=0$, $V=S$ and $Y_2=X_2$, with $X_2$ independent of $(S,X_1)$ in the inner bound of Theorem~\ref{theorem-inner-bound-strictly-causal-states-setting}. 
Evaluating the first inequality, we obtain
\begin{align}
R_1 + \epsilon &\geq I(X_1;Y|V,X_2)\\
    &= I(X_1;Y_1,X_2|S,X_2)\\
    &= I(X_1;Y_1|S,X_2)\\
    &= I(X_1,X_2;Y_1|S)-I(X_2;Y_1|S)\\
    &= I(X_1;Y_1|S)+I(X_2;Y_1|X_1,S)-I(X_2;Y_1|S)\\
\label{Constraint1__MaximalIndividualRate__ModelCounterExample__Step1}
    &= I(X_1;Y_1|S)-I(X_2;Y_1|S)\\
    &= I(X_1;Y_1|S),
 \label{Constraint1__MaximalIndividualRate__ModelCounterExample}
\end{align}
where \eqref{Constraint1__MaximalIndividualRate__ModelCounterExample__Step1} follows since $X_2=Y_2$ and $Y_2 \leftrightarrow (X_1,S) \leftrightarrow Y_1$ is a Markov chain, and  the last equality follows by the Markov relation $X_2 \leftrightarrow  S \leftrightarrow Y_1$ for this example.

Evaluating the second inequality, we obtain
\begin{align}
R_1 + \epsilon &\geq I(V,X_1,X_2;Y)-I(V,X_1,X_2;S)\\
    &= I(X_1,S;Y_1,X_2)+H(X_2|X_1,S)-H(S)\\
    &= I(X_1,S;Y_1)+I(X_1,S;X_2|Y_1)+H(X_2|X_1,S)-H(S)\\
    &= I(X_1,S;Y_1)+H(X_2|Y_1)-H(X_2|X_1,S,Y_1)+H(X_2|X_1,S)-H(S)\\
\label{Constraint2__MaximalIndividualRate__ModelCounterExample__Step1}
    &= I(X_1;Y_1|S)+I(S;Y_1)+H(X_2|Y_1)-H(S)\\
    &= I(X_1;Y_1|S)+H(X_2|Y_1)-H(S|Y_1)\\
    &= I(X_1;Y_1|S)+H(Y_1|X_2)-H(Y_1|S)+H(X_2)-H(S)\\
    &= I(X_1;Y_1|S)+I(S;Y_1)+H(X_2)-H(S)
\label{Constraint2__MaximalIndividualRate__ModelCounterExample}
\end{align}
where \eqref{Constraint2__MaximalIndividualRate__ModelCounterExample__Step1} follows since $X_2$ is independent of $(X_1,S,Y_1)$.

Similarly, evaluating the constraint, we obtain
\begin{align}
I(V,X_2;Y)-I(V,X_2;S) &= I(S;Y_1|X_2)+H(X_2)-H(S).
\label{NonNegativityConstraint__ModelCounterExample}
\end{align}

Now, observe that with the choice $X_2 \sim \: \text{Bernoulli}\:(\frac{1}{2})$ independent of $(S,X_1)$, we have $H(X_2)=H(S)=1$ and, so, the RHS of \eqref{Constraint2__MaximalIndividualRate__ModelCounterExample} is larger than the RHS of \eqref{Constraint1__MaximalIndividualRate__ModelCounterExample}; and the RHS of \eqref{NonNegativityConstraint__ModelCounterExample} is nonnegative. This shows the achievability of the rate $R_1 = I(X_1;Y_1|S)$.

2) The converse follows straightforwardly by specializing Theorem 2 (or the cut-set upper bound) to this example,
\begin{align}
R  &\leq I(X_1;Y|X_2,S)\\
   &= I(X_1;Y_1|X_2,S)\\
   &= H(Y_1|X_2,S)-H(Y_1|X_1,X_2,S)\\
\label{ProofCutSetBoundCounterExample__Step1}
   &\leq H(Y_1|S)-H(Y_1|X_1,X_2,S)\\
   &\leq  H(Y_1|S)-H(Y_1|X_1,S)\\
\label{ProofCutSetBoundCounterExample__Step2}
   &= I(X_1;Y_1|S),
\end{align}
where \eqref{ProofCutSetBoundCounterExample__Step1} holds since conditioning reduces entropy, and \eqref{ProofCutSetBoundCounterExample__Step2} holds by the Markov relation $X_2 \leftrightarrow (X_1,S) \leftrightarrow Y_1$. \qed

\textit{Claim 2:} The capacity of the memoryless binary MAC with states known strictly causally at the encoders shown in Figure~\ref{BSCModelCounterExample-StrictlyCausalStates} satisfies
\begin{align}
C_{\text{s-c}} = \left\{
\begin{array}{lll}
h_2(p * q_1) - h_2(p) & \text{if} \quad 0 \leq q_1 \leq \frac{1}{2}\\
1 - h_2(p) & \text{if} \quad q_1 \geq \frac{1}{2}
\end{array}
\right\} \geq C_{\text{no-s}} = \max_{p(x_1)} I(X_1;Y_1).
\label{ExplicitCharacterization__Capacity__ModelCounterExample}
\end{align}

\textbf{Proof:} The explicit expression of $C_{\text{s-c}}$, i.e., $C_{\text{s-c}}=h_2(p * q_1) - h_2(p)$ if $ 0 \leq q_1 \leq 1/2$ and $C_{\text{s-c}}=1 - h_2(p)$ if $q_1 \geq 1/2$, follows straightforwardly from Claim 1 by simple algebra, where $h_2(\alpha)$ denotes the entropy of a Bernoulli\:$(\alpha)$ source and $p * q$ denotes the binary convolution, i.e., $p * q = p(1-q)+q(1-p)$, as defined in Section~\ref{secI_subsecB}. Let now $C_{\text{no-s}}$ denote the capacity of the same model had the states been known (strictly causally) only at Encoder 1. Since in this case the knowledge of the states only at Encoder 1 would not increase the capacity (see also Proposition~\ref{proposition-strictly-causal-states-at-only-strong-encoder} below), $C_{\text{no-s}}$ is also the capacity of the same model had the states been not known at all. Thus, $C_{\text{no-s}}$ is given by the RHS of \eqref{ExplicitCharacterization__Capacity__ModelCounterExample}. For this example, it is easy to see that $C_{\text{no-s}} = 0$. This holds since $h_2(q_1*1/2*p)-h_2(1/2*p)= 1-h_2(1/2*p)=0 \:\:\: \forall \:\:\: (p, q_1,q_2) \in [0,1]^2{\times}[1/2,1]$ -- recall that the state is Bernoulli $(\frac{1}{2})$ and is independent of the inputs $X_1$, $X_2$ and the noise $Z$. Thus, the inequality in \eqref{ExplicitCharacterization__Capacity__ModelCounterExample} holds irrespective to the values of the tuple $(p,q_1,q_2 \geq 1/2)$. \qed

Observe that the inequality in \eqref{ExplicitCharacterization__Capacity__ModelCounterExample} holds strictly if $p \neq 1/2$ and $q_1 \neq 0$; and, so, revealing the states strictly causally to Encoder 2 strictly increases the capacity in this case. 

\subsection{Capacity Results}\label{secIII_subsecD}

Example~\ref{example-uselfuness-of-knoweledge-of-states-strictly-causally} in Section~\ref{secIII_subsecC} shows that the knowledge of the states strictly causally at the encoders increases the capacity region of the cooperative MAC that we study. This fact has also been shown for other related models, such as a multiaccess channel with independent inputs and strictly causal or causal states at the encoders in \cite{LS13a, LS13b,LSY13}, and a multiaccess channel with degraded messages sets and states known noncausally to the encoder that sends both messages and only strictly causally at the encoder that sends only the common message in \cite{ZPS13,ZPS11a,ZPS12a}. Proposition~\ref{proposition-strictly-causal-states-at-only-strong-encoder} in Section~\ref{secIV} will show that, for the model with cooperative encoders that we study, the increase in the capacity holds precisely because the encoder that sends only the common message, i.e., Encoder 2, also knows the states. That is, if the states were known strictly causally to only Encoder 1, its availability would not increase the capacity of the corresponding model. Proposition~\ref{proposition-maximum-sum-rate-strictly-causal-states-setting} shows that, like for the model with independent inputs in \cite{LS13a}, the knowledge of the states strictly causally at the encoders does not increase the sum rate capacity, however.

\vspace{0.3cm}

\begin{proposition}\label{proposition-maximum-sum-rate-strictly-causal-states-setting}
The knowledge of the states only strictly causally at the encoders does not increase the sum capacity of the multiple access channel with degraded messages sets, i.e.,
\begin{equation}
\max_{(R_c,R_1) \: \in \: \mc C_{\text{s-c}}} R_c + R_1 = \max_{p(x_1,x_2)} I(X_1,X_2;Y).
\label{maximum-sum-rate-strictly-causal-states-setting}
\end{equation}
\end{proposition}

\vspace{0.3cm}

The converse proof of Proposition~\ref{proposition-maximum-sum-rate-strictly-causal-states-setting} follows immediately from Proposition~\ref{proposition-alternative-outer-bound-strictly-causal-states-setting}. The achievability proof of Proposition~\ref{proposition-maximum-sum-rate-strictly-causal-states-setting} follows simply by ignoring the state information at the encoders, since the RHS of \eqref{maximum-sum-rate-strictly-causal-states-setting} is the sum-rate capacity of the same MAC without states. 

\vspace{0.3cm}

 Proposition~\ref{proposition-maximum-sum-rate-strictly-causal-states-setting} shows that revealing the state that governs a MAC with degraded messages sets strictly causally to both encoders does not increase the sum-rate capacity. This is to be opposed to the case in which the encoders send only independent messages for which revealing the state strictly causally to both encoders can increase the sum-rate capacity \cite{LS13a}.

\vspace{0.3cm}

In what follows, we extend the capacity result derived for a memoryless Gaussian example in \cite[Example 2]{LS13a} to the case of cooperative encoders and then generalize it to a larger class of channels. Consider a class of discrete memoryless two-user cooperative MACs, denoted by $\mc D^{\text{sym}}_{\text{MAC}}$, in which the channel state $S$, assumed to be revealed strictly causally to both encoders, can be obtained as a deterministic function of the channel inputs $X_1$ and $X_2$ and the channel output $Y$, as
\begin{equation}
S =  f(X_1,X_2,Y).
\end{equation}

\begin{theorem}\label{theorem-capacity-region-special-case-strictly-causal-states-setting}
For any MAC in the class $\mc D^{\text{sym}}_{\text{MAC}}$ defined above, the capacity region $\mc C_{\text{s-c}}$ is given by the set of all rate pairs $(R_c,R_1)$ satisfying
\begin{align}
R_1 &\leq I(X_1;Y|X_2,S)\nonumber\\
R_c + R_1 &\leq I(X_1,X_2;Y)
\label{capacity-region-special-case-strictly-causal-states-setting}
\end{align}
for some measure 
\begin{equation}
P_{S,X_1,X_2,Y} = Q_SP_{X_1,X_2}W_{Y|S,X_1,X_2}.
\label{measure-capacity-region-special-case-strictly-causal-states-setting}
\end{equation}
\end{theorem}

\vspace{0.3cm}

\textbf{Proof:} The proof of the converse part of Theorem~\ref{theorem-capacity-region-special-case-strictly-causal-states-setting} follows by Proposition~\ref{proposition-alternative-outer-bound-strictly-causal-states-setting}. The proof of the direct part of Theorem~\ref{theorem-capacity-region-special-case-strictly-causal-states-setting} follows by setting $V=S$ in the region $\tilde{\mc R}^{\text{in}}_{\text{s-c}}$. (see \eqref{inner-bound-without-nonnegativity-constraint-strictly-causal-states-setting} and the discussion after Remark~\ref{remark-main-idea-inner-bound-strictly-causal-states-setting}).

\begin{remark} 
The class $\mc D^{\text{sym}}_{\text{MAC}}$ includes the following memoryless Gaussian example, which is similar to that in \cite[Example 2]{LS13a} but with the encoders being such that both of them send a common message and one of the two also sends an individual message,
\begin{equation}
Y=X_1+X_2+S
\label{memoryless-gaussian-example-remark5}
\end{equation}
where the inputs $X^n$ and $X^n_2$ are subjected to individual power constraints $(1/n)\sum_{i=1}^{n} \mathbb{E}[X^2_{k,i}] \leq P_k$, $k=1,2$, and the state $S^n$ is memoryless Gaussian, $S \sim \mc N(0,Q)$, and known strictly causally to both encoders. The capacity region of this model is given by the set of all rate pairs $(R_c,R_1)$ satisfying 
\begin{equation}
R_c+R_1 \leq \frac{1}{2}\log\big(1+\frac{(\sqrt{P_1}+\sqrt{P_2})^2}{Q}\big).
\label{capacity-region-gaussian-special-case-strictly-causal-states-setting}
\end{equation}
The region~\eqref{capacity-region-gaussian-special-case-strictly-causal-states-setting} can be obtained by first extending the result of Theorem~\ref{theorem-capacity-region-special-case-strictly-causal-states-setting} for the DM case to memoryless channels with discrete time and continuous alphabets using standard techniques \cite[Chapter 7]{G68}, and then maximizing each bound utilizing the \textit{Maximum Differential Entropy Lemma} \cite[Section 2.2]{GK11}. Note that, by doing so, the first condition on the individual rate in \eqref{capacity-region-special-case-strictly-causal-states-setting} appears to be redundant for this Gaussian model. \qed
\end{remark}

The class $\mc D^{\text{sym}}_{\text{MAC}}$ contains more channels along with the memoryless Gaussian model \eqref{memoryless-gaussian-example-remark5}.

\begin{example}\label{example-only-state-is-deterministic-symmetric-state-case}
Consider the Gaussian MAC with $Y=(Y_1,Y_2)$, and
\begin{subequations}
\begin{align}
Y_1 &= X_1 + X_2 + S\\
Y_2 &= X_2 + Z
\end{align}
\label{memoryless-gaussian-example-example4}
\end{subequations}
where the state process is memoryless Gaussian, with $S \sim \mc N(0,Q)$, and the noise process is memoryless Gaussian independent of all other processes, $Z ~\sim \mc N(0,N)$. Encoder 1 knows the state strictly causally, and transmits both common message $W_c \in [1,2^{nR_c}]$ and private message $W_1 \in [1,2^{nR_1}]$. Encoder 2 knows the state strictly causally, and transmits only the common message. We consider the input power constraints $\sum_{i=1}^{n} \mathbb{E}[X^2_{1,i}] \leq nP_1$ and $\sum_{i=1}^{n} \mathbb{E}[X^2_{2,i}] \leq nP_2$. The capacity region of this model can be computed using Theorem~\ref{theorem-capacity-region-special-case-strictly-causal-states-setting}. It is characterized as 
\begin{align}
\mc C^{\text{G}}_{\text{s-c}} = \left\{
\begin{array}{l}
(R_c,R_1) \in \mathbb{R}^2_{+}:\\
R_c+R_1 \leq \max_{0 \leq \rho_{12} \leq 1} \frac{1}{2}\log\big(1+\frac{P_2}{N}\big) \\
\hspace{1.5cm} + \frac{1}{2}\log\Big(1+\frac{(1-\rho^2_{12})P_1P_2+N((\sqrt{P_1}+\rho_{12}\sqrt{P_2})^2+(1-\rho^2_{12})P_2)}{Q(P_2+N)}\Big)
\end{array}
\right\}.
\label{capacity-region-example-only-state-is-deterministic-symmetric-state-case}
\end{align}
\end{example}


\vspace{0.2cm}

\textbf{Proof:} The analysis of Example~\ref{example-only-state-is-deterministic-symmetric-state-case} is given in Appendix~\ref{appendix-analysis-example-only-state-is-deterministic-symmetric-state-case}.
\vspace{0.3cm}

\section{Strictly Causal States at Only One Encoder}\label{secIV}

In this section we consider asymmetric state settings in which the state is revealed (strictly causally) to only one encoder.

\begin{proposition}\label{proposition-strictly-causal-states-at-only-strong-encoder}
The knowledge of the states strictly causally at only the encoder that sends both messages does not increase the capacity region of the cooperative MAC.
\end{proposition}

\vspace{0.3cm}

The proof of Proposition~\ref{proposition-strictly-causal-states-at-only-strong-encoder} appears in Appendix~\ref{appendix-proof-proposition-strictly-causal-states-at-only-strong-encoder}.

\vspace{0.3cm}

In the case in which the state is revealed strictly causally to only the encoder that sends only the common message, this increases the capacity region. In what follows, first we derive an inner bound on the capacity of this model. Next, we generalize the capacity result derived in \cite[Theorem 4]{LSY13} for discrete memoryless channels in which 1) the channel output is a deterministic function of the inputs and the state and 2) the state is a deterministic function of the channel output and inputs from the encoders, to a larger class of channels. For instance, in addition to that the model is different since the transmitters send a common message, the capacity result that will follow does not require that the channel output be a deterministic functions of the inputs and the state, which then is arbitrary.

Let $\mc P^{\text{in}}_{\text{asym,s-c}}$ stand for the collection of all random variables $(S,U,V,X_1,X_2,Y)$ such that $U$, $V$, $X_1$ and $X_2$ take values in finite alphabets $\mc U$, $\mc V$, $\mc X_1$ and $\mc X_2$, respectively, and satisfy
\begin{subequations}
\begin{align}
P_{S,U,V,X_1,X_2,Y}(s,u,v,x_1,x_2,y) &= P_{S,U,V,X_1,X_2}(s,u,v,x_1,x_2)W_{Y|X_1,X_2,S}(y|x_1,x_2,s)\\
P_{S,U,V,X_1,X_2}(s,u,v,x_1,x_2) &= Q_S(s)P_{U}(u)P_{X_2|U}(x_2|u)P_{X_1|U}(x_1|u)P_{V|S,U,X_2}(v|s,u,x_2).
\end{align}
\label{measure-inner-bound-asymmetric-strictly-causal-states-setting}
\end{subequations}
The relations in \eqref{measure-inner-bound-asymmetric-strictly-causal-states-setting} imply that $(U,V) \leftrightarrow (S,X_1,X_2) \leftrightarrow Y$, $X_1 \leftrightarrow U \leftrightarrow X_2$ and $X_1 \leftrightarrow (U,V,X_2) \leftrightarrow S$ are Markov chains; and $X_1$ and $X_2$ are independent of $S$.

Define $\mc R^{\text{in}}_{\text{asym,s-c}}$ to be the set of all rate pairs $(R_c,R_1)$ such that
\begin{align}
R_1 \: &\leq \: I(X_1;Y|U,V,X_2)\nonumber\\
R_1 \: &\leq \: I(V,X_1,X_2;Y|U)-I(V;S|U,X_2)\nonumber\\
R_c+ R_1 \: &\leq \: I(U,V,X_1,X_2;Y)-I(V;S|U,X_2)\nonumber\\
&\hspace{2cm} \text{for some}\:\: (S,U,V,X_1,X_2,Y) \in \mc P^{\text{in}}_{\text{asym,s-c}}.
\label{inner-bound-asymmetric-strictly-causal-states-setting}
\end{align}

\noindent As stated in the following theorem, the set $\mc R^{\text{in}}_{\text{asym,s-c}}$ is an inner bound on the capacity region of the state-dependent discrete memoryless MAC with strictly-causal states known only at the encoder that sends only the common message.

\vspace{0.3cm}

\begin{theorem}\label{theorem-inner-bound-asymmetric-strictly-causal-states-setting}
The capacity region of the cooperative multiple access channel with states revealed strictly causally to only the encoder that sends the common message satisfies
\begin{equation}
\mc R^{\text{in}}_{\text{asym,s-c}} \subseteq \mc C_{\text{asym,s-c}}.
\end{equation}
\end{theorem}

\vspace{0.3cm}

\textbf{Proof:} A description of the coding scheme that we use for the proof of Theorem~\ref{theorem-inner-bound-asymmetric-strictly-causal-states-setting}, as well a complete error analysis, are given in Appendix~\ref{appendix-proof-theorem-inner-bound-asymmetric-strictly-causal-states-setting}.

The following remark helps better understanding the coding scheme that we use for the proof of Theorem~\ref{theorem-inner-bound-asymmetric-strictly-causal-states-setting}.

\begin{remark}\label{remark-main-idea-inner-bound-cooperative-mac-with-asymmetric-state}
For the model of Theorem~\ref{theorem-inner-bound-asymmetric-strictly-causal-states-setting}, a good codebook at the encoder that sends only the common message should resolve a \textit{dilemma} among 1) exploiting the knowledge of the state that is available at this encoder and 2) sending information cooperatively with the other encoder (i.e., the common message). The coding scheme of Theorem~\ref{theorem-inner-bound-asymmetric-strictly-causal-states-setting} resolves this tension by splitting the common rate $R_c$ into two parts. More specifically, the common  message $W_c$ is divided into two parts, $W=(W_{c1},W_{c2})$. The part $W_{c1}$ is sent cooperatively by the two encoders, at rate $R_{c1}$; and the part $W_{c2}$ is sent only by the encoder that exploits the available state, at rate $R_{c2}$. The total rate for the common message is $R_c=R_{c1}+R_{c2}$. In Theorem~\ref{theorem-inner-bound-asymmetric-strictly-causal-states-setting}, the random variable $U$ stands for the information that is sent cooperatively by the two encoders, and the random variable $V$ stands for the compression of the state by the encoder that sends only the common message, in a manner that is similar to that of Theorem~\ref{theorem-inner-bound-strictly-causal-states-setting}. \qed 

\end{remark}

Consider the following class of discrete memoryless channels, which we denote as $\mc D_{\text{IH}}$. Encoder 1 does not know the state sequence at all, and transmits an individual message $W_1 \in [1,2^{nR_1}]$. Encoder 2 knows the state sequence strictly causally, and does not transmits any message. In this model, Encoder 2 plays the role of a helper that is informed of the channel state sequence only strictly causally. This network may model one in which there is an external node that interferes with the transmission from Encoder 1 to the destination, and that is overheard only by Encoder 2 which then assists the destination by providing some information about the interference. Furthermore, we assume that the state $S$ can be obtained as a deterministic function of the inputs $X_1$, $X_2$ and the channel output $Y$, as
\begin{equation}
S =  f(X_1,X_2,Y).
\end{equation}
For channels with a helper that knows the states strictly causally, the class of channels $\mc D_{\text{IH}}$ is larger than that considered in \cite{LSY13}, as the channel output needs not be a deterministic function of the channel inputs and the state. The following theorem characterizes the capacity region for the class of channels $\mc D_{\text{IH}}$.

The capacity of the class of channels $\mc D_{\text{IH}}$ can be characterized as follows.

\begin{theorem}\label{theorem-capacity-informed-helper}
For any channel in the class $\mc D_{\text{IH}}$ defined above, the capacity $C_{\text{s-c}}$ is given by
\begin{equation}
C_{\text{s-c}} = \min\: \big\{I(X_1;Y|S,X_2), \: I(X_1,X_2;Y)\big\}
\label{capacity-informed-helper}
\end{equation}
where the maximization is over measures of the form
\begin{equation}
P_{S,X_1,X_2,Y} = Q_SP_{X_1}P_{X_2}W_{Y|S,X_1,X_2}.
\label{measure-capacity-informed-helper}
\end{equation}
\end{theorem}

\vspace{0.3cm}

\textbf{Proof:} The proof of Theorem~\ref{theorem-capacity-informed-helper} is given in Appendix~\ref{appendix-proof-theorem-capacity-informed-helper}.

\begin{remark}\label{remark7}
The class $\mc D_{\text{IH}}$ includes the Gaussian model $Y=X_1+X_2+S$ where the state $S \sim \mc N(0,Q)$ comprises the channel noise, and the inputs are subjected to the input power constraints $(1/n)\sum_{i=1}^{n} \mathbb{E}[X^2_{k,i}] \leq P_k$, $k=1,2$. Encoder 1 does not know the state sequence and transmits message $W_1$. Encoder 2 knows the state sequence strictly causally, and does not transmit any message. The capacity of this model is given by 
\begin{equation}
C^{\text{G}}_{\text{s-c}} = \frac{1}{2}\log(1+\frac{P_1+P_2}{Q}).
\label{capacity-gaussian-informed-helper}
\end{equation}
The capacity~\eqref{capacity-gaussian-informed-helper} can be obtained from Theorem~\ref{theorem-capacity-informed-helper} by maximizing the two terms of the minimization utilizing the \textit{Maximum Differential Entropy Lemma} \cite[Section 2.2]{GK11}. Observe that the first term of the minimization in \eqref{capacity-informed-helper} is redundant in this case. Also, we note that the capacity \eqref{capacity-gaussian-informed-helper}  of this example can also be obtained as a special case of that of the Gaussian example considered in \cite[Remark 4]{LSY13}.\qed
\end{remark}

\begin{figure}[htpb]
\centering
\includegraphics[width=0.8\linewidth]{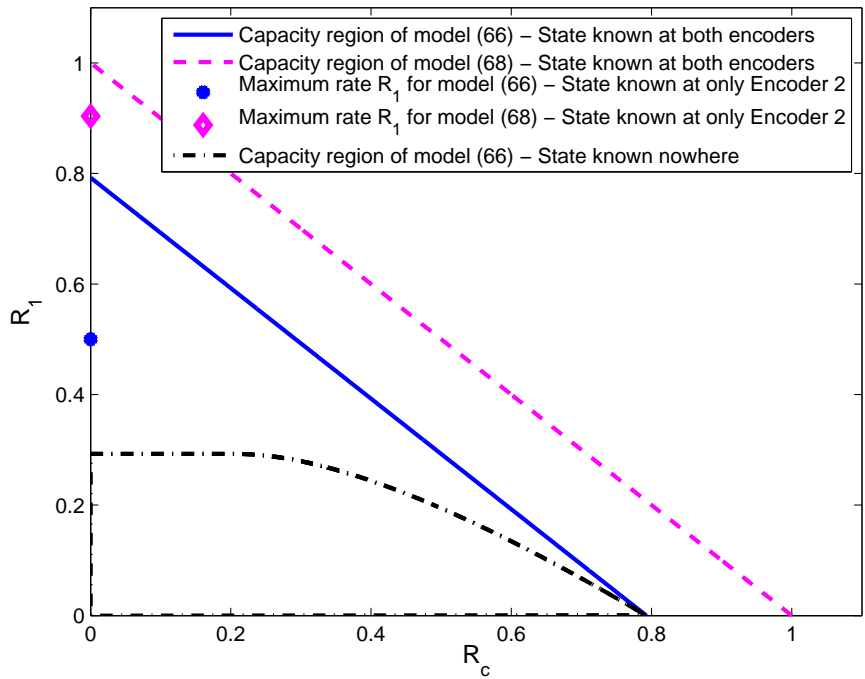}
\caption{Capacity of the models \eqref{memoryless-gaussian-example-remark5} and \eqref{memoryless-gaussian-example-example4}, with different degrees of knowledge of the state sequence at the encoders. Numerical values are: $P_1=P_2=N=1/2$ and $Q=1$.}
\label{fig-comparison}
\end{figure}

In the following example the channel output can \textit{not} be obtained as a deterministic function of the channel inputs and the channel state, and yet, its capacity can be characterized using Theorem~\ref{theorem-capacity-informed-helper}. 

\begin{example}\label{example-only-state-is-deterministic-asymmetric-state-case}
Consider the following Gaussian example with $Y=(Y_1,Y_2)$, and
\begin{subequations}
\begin{align}
Y_1 &= X_1 + X_2 + S\\
Y_2 &= X_2 + Z
\end{align}
\label{model-example5}
\end{subequations}
where the state process is memoryless Gaussian, with $S \sim \mc N(0,Q)$, and the noise process is memoryless Gaussian independent of all other processes, $Z ~\sim \mc N(0,N)$. Encoder 1 does not know the state sequence, and transmits message $W_1 \in [1,2^{nR_1}]$. Encoder 2 knows the state strictly causally, and does not transmit any message. The inputs are subjected to the input power constraints $\sum_{i=1}^{n} \mathbb{E}[X^2_{1,i}] \leq nP_1$ and $\sum_{i=1}^{n} \mathbb{E}[X^2_{2,i}] \leq nP_2$. The capacity of this model can be computed easily using Theorem~\ref{theorem-capacity-informed-helper}, as
\begin{equation}
C^{\text{G}}_{\text{s-c}} = \frac{1}{2}\log\big(1+\frac{P_1}{Q}+\frac{P_2}{Q}\frac{N}{P_2+N}\big) + \frac{1}{2}\log\big(1+\frac{P_2}{N}\big).
\end{equation}
Note that the knowledge of the states strictly causally at Encoder 2 makes it possible to send at positive rates by Encoder 1 even if the allowed average power $P_1$ is zero. The diamond on the y-axis of Figure~\ref{fig-comparison} shows the capacity of the model \eqref{model-example5} for the choice $P_1=P_2=N=1/2$ and $Q=1$. The figure also shows the capacity region \eqref{capacity-region-example-only-state-is-deterministic-symmetric-state-case} of the same model had the state sequence been known (strictly causally) to both encoders. The gap on the y-axis is precisely the gain in capacity enabled by also revealing the state to the encoder that sends both messages. A similar improvement can be observed for the Gaussian model $Y=X_1+X_2+S$ of Remark~\ref{remark7}. The dot-dashed curve depicts the capacity region of this model had the state sequence been not known at all, neither to encoders nor to the decoder \cite{BLW08,W83} -- which is the same capacity region has the state sequence been known (strictly causally) only to the encoder that transmits both messages (see Proposition~\ref{proposition-strictly-causal-states-at-only-strong-encoder}). Note that for both models, of Remark~\ref{remark7} and \eqref{model-example5}, if the state sequence is known non-causally to the encoder that sends only the common message, a standard dirty paper coding scheme \cite{C83} at this encoder cancels completely the effect of the state. The reader may refer to  \cite{ZKLV09a,ZKLV08a,ZKLV10} where a related model is referred to as the \textit{deaf helper problem}. A related Gaussian Z-channel with mismatched side information, revealed non-causally to one encoder, and interference is studied in \cite{DLKS13}. Other related multiaccess models with states revealed non-causally to one encoder can be found in \cite{SBSV07a,KL07, KL07a}.
\end{example}

\begin{example}\label{example2-only-state-is-deterministic-asymmetric-state-case}
Consider the following binary example in which the state models fading. The channel output has two components, i.e., $Y=(Y_1,Y_2)$, with
\begin{subequations}
\begin{align}
Y_1 &= S{\cdot}X_1\\
Y_2 &= X_2 + Z
\end{align}
\label{output-example2-only-state-is-deterministic-asymmetric-state-case}
\end{subequations}
where $\mc X_1=\mc X_2=\mc S=\mc Z= \{+1,-1\}$, and the noise $Z$ is independent of $(S,X_1,X_2)$ with $\text{Pr}\{Z=1\}=p$ and $\text{Pr}\{Z=-1\}=1-p$, $0 \leq p \leq 1$, and the state $S$, known strictly causally to only Encoder 2, is such that $\text{Pr}\{S=1\}=\text{Pr}\{S=-1\}=1/2$. Using Theorem~\ref{theorem-capacity-informed-helper}, it is easy to compute the capacity of this example, as
\begin{equation}
C^{\text{B}}_{\text{s-c}} = \max_{0 \leq q_1,q_2 \leq 1}\:\min\:\big\{h_2(q_1),g(p,q_2)-h_2(p)\big\}
\end{equation}
where
\begin{equation}
g(p,q_2) = -pq_2\log(pq_2)-(1-p)(1-q_2)\log((1-p)(1-q_2))-p*q_2\log(p*q_2).
\label{entropy-second-output-component-example2-only-state-is-deterministic-asymmetric-state-case}
\end{equation}
Observe that $C^{\text{B}}_{\text{s-c}} \geq 1-\frac{1}{2} h_2(p) \geq 0.5$.
\end{example}

\vspace{0.2cm}

\textbf{Proof:} Using \eqref{output-example2-only-state-is-deterministic-asymmetric-state-case}, we have $S=Y_1/X_1$, and, so, $S$ is a deterministic function of $(X_1,X_2,Y)$. Thus, the capacity of this channel can be computed using Theorem~\ref{theorem-capacity-informed-helper}. Let $0 \leq q_1 \leq 1$ such that $\text{Pr}\{X_1=1\}=q_1$ and $\text{Pr}\{X_1=-1\}=1-q_1$. Also, let $0 \leq q_2 \leq 1$ such that $\text{Pr}\{X_2=1\}=q_2$ and $\text{Pr}\{X_2=-1\}=1-q_2$. Then, considering the first term on the RHS of \eqref{capacity-informed-helper}, we get 
\begin{align}
I(X_1;Y|S,X_2) &= H(Y|S,X_2)-H(Y|S,X_1,X_2)\\
               &= H(SX_1,X_2+Z|S,X_2)-H(Z|S,X_1,X_2)\\
\label{analysis-example2-only-state-is-deterministic-asymmetric-state-case-step1}
	       &= H(X_1,Z|S,X_2)-H(Z)\\
\label{analysis-example2-only-state-is-deterministic-asymmetric-state-case-step2}
	       &= H(X_1,Z)-H(Z)\\
\label{analysis-example2-only-state-is-deterministic-asymmetric-state-case-step3}
	       &= H(X_1)\\
	       &= h_2(q_1)
\end{align}
where \eqref{analysis-example2-only-state-is-deterministic-asymmetric-state-case-step1} holds since $Z$ is independent of $(S,X_1,X_2)$, \eqref{analysis-example2-only-state-is-deterministic-asymmetric-state-case-step2} holds since $(X_1,Z)$ is independent of $(S,X_2)$, and \eqref{analysis-example2-only-state-is-deterministic-asymmetric-state-case-step3} holds since $X_1$ and $Z$ are independent. 

Similarly, considering the second term on  the RHS of \eqref{capacity-informed-helper}, we get
\begin{align}
I(X_1,X_2;Y) &= H(Y) - H(Y|X_1,X_2)\\
             &= H(Y) - (SX_1,Z|X_1,X_2)\\
\label{analysis-example2-only-state-is-deterministic-asymmetric-state-case-step4}
	     &= H(Y) - H(Z) - H(S)\\
\label{analysis-example2-only-state-is-deterministic-asymmetric-state-case-step5}
             &= H(SX_1)+H(X_2+Z) - H(Z) - H(S)\\
\label{analysis-example2-only-state-is-deterministic-asymmetric-state-case-step6}
             &= H(X_2+Z)-H(Z)\\
\label{analysis-example2-only-state-is-deterministic-asymmetric-state-case-step7}
             &= g(p,q_2) - h_2(p)
\end{align}
where \eqref{analysis-example2-only-state-is-deterministic-asymmetric-state-case-step4} holds since $S$ and $Z$ are independent of $(X_1,X_2)$ and independent of each other, \eqref{analysis-example2-only-state-is-deterministic-asymmetric-state-case-step5} holds since $Y_1=SX_1$ and $Y_2=X_2+Z$ are independent, \eqref{analysis-example2-only-state-is-deterministic-asymmetric-state-case-step6} follows because 
\begin{equation}
\text{Pr}\{SX_1=1\}=\text{Pr}\{SX_1=-1\}=\frac{1}{2}
\end{equation}
and, so, $H(SX_1)=1=H(S)$, and \eqref{analysis-example2-only-state-is-deterministic-asymmetric-state-case-step7} follows because 
\begin{equation}
\text{Pr}\{X_2+Z=0\}=p*q_2, \quad \text{Pr}\{X_2+Z=2\}=pq_2,\quad \text{Pr}\{X_2+Z=-2\}= (1-p)(1-q_2)
\end{equation}
and, so, $H(X_2+Z)=g(p,q_2)$ as given by \eqref{entropy-second-output-component-example2-only-state-is-deterministic-asymmetric-state-case}. \qed

\begin{remark}\label{}
The result of Theorem~\ref{theorem-capacity-informed-helper} can be extended to the case in which the encoders send separate messages and each observes (strictly causally) an independent state. In this case, denoting by $S_1$ the state that is observed by Encoder 1 and by $S_2$  the state that is observed by Encoder 2, it can be shown that, if both $S_1$ and $S_2$ can be obtained as deterministic functions of the inputs $X_1$ and $X_2$ and the channel output $Y$, then the capacity region is given by the convex hull of the set of all rates satisfying
\begin{subequations}
\begin{align}
R_1 &\leq I(X_1;Y|X_2,S_2)\\
R_2 &\leq I(X_2;Y|X_1,S_1)\\
R_1+R_2 &\leq I(X_1,X_2;Y)
\end{align}
\label{capacity-region-specific-case-mac-with-independent-states}
\end{subequations}
for some measure of the form $Q_{S_1,S_2,X_1,X_2}=Q_{S_1}Q_{S_2}P_{X_1}P_{X_2}$. This result can also be obtained by noticing that, if both $S_1$ and $S_2$ are deterministic functions of $(X_1,X_2,Y)$, then the inner bound of \cite[Theorem 2]{LSY13} reduces to \eqref{capacity-region-specific-case-mac-with-independent-states}, which is also an outer bound as stated in \cite[Proposition 3]{LS13b}.
\end{remark}

\section{Causal States}\label{secV}

Let $\mc P_{\text{c}}$ stand for the collection of all random variables $(S,U,V,X_1,X_2,Y)$ such that $U$, $V$, $X_1$ and $X_2$ take values in finite alphabets $\mc U$, $\mc V$, $\mc X_1$ and $\mc X_2$, respectively, and
\begin{subequations}
\begin{align}
P_{S,U,V,X_1,X_2,Y}(s,u,v,x_1,x_2,y) &= P_{S,U,V,X_1X_2}(s,u,v,x_1,x_2)W_{Y|X_1,X_2,S}(y|x_1,x_2,s)\\
P_{S,U,V,X_1,X_2}(s,u,v,x_1,x_2) &= Q_S(s)P_{V}(v)P_{U|V}(u|v)P_{X_2|V,S}(x_2|v,s)P_{X_1|S,V,U}(x_1|s,v,u).
\end{align}
\label{measure-capacity-region-causal-states-setting}
\end{subequations}
The relations in \eqref{measure-capacity-region-causal-states-setting} imply that $(U,V) \leftrightarrow (S,X_1,X_2) \leftrightarrow Y$ is a Markov chain; and that $(V,U)$ is independent of $S$.

Define $\mc C_{\text{c}}$ to be the set of all rate pairs $(R_c,R_1)$ such that
\begin{align}
R_1 \: &\leq \: I(U;Y|V) \nonumber\\
R_c+ R_1 \: &\leq \: I(U,V;Y)\nonumber\\
&\hspace{2cm} \text{for some}\:\: (S,U,V,X_1,X_2,Y) \in \mc P_{\text{c}}.
\label{capacity-region-causal-states-setting}
\end{align}

\noindent As stated in the following theorem, the set $\mc C_{\text{c}}$ is the capacity region of the state-dependent discrete memoryless MAC model with causal states.

\vspace{0.3cm}

\begin{theorem}\label{theorem-capacity-region-causal-states-setting}
The capacity region of the multiple access channel with degraded messages sets and states known causally at both encoders is given by $\mc C_{\text{c}}$.
\end{theorem}

\vspace{0.3cm}

\textbf{Proof:} The proof of Theorem~\ref{theorem-capacity-region-causal-states-setting} is given in Appendix~\ref{appendix-proof-theorem-capacity-region-causal-states-setting}.

\vspace{0.3cm}

\begin{remark}
For the proof of Theorem~\ref{theorem-capacity-region-causal-states-setting} , the converse part can be shown in a way very that is essentially very similar to \cite{SK05}. The coding scheme that we use to prove the achievability part is based on Shannon strategies \cite{Sh58}. By opposition to the case of MAC with independent inputs in \cite{SK05} or that with one common message and two individual messages \cite{K-FM10}, in our case one of the two encoders knows the other encoder's message, and this permits to create the desired correlation among the auxiliary codewords that is required by the outer bound. Also, we should mention that the fact that Shannon strategies are optimal for the MAC with degraded messages sets that we study is in opposition with the case of the MAC with independent inputs, for which it has been shown in \cite[Section III]{LS13a} that Shannon strategies are suboptimal in general. \qed
\end{remark}

\section{Concluding Remarks}\label{secVI} 

In this paper we study the transmission over a state-controlled two-user cooperative multiaccess channel with the states known -- depending on the scenario, strictly causally or causally to only one or both transmitters. While, like the MAC with non-degraded messages sets of \cite{LS13a} (and also the related models of \cite{LS13b,LSY13} and \cite{ZPS13}),  it can be expected that conveying a description of the state by the encoders to the decoder can be beneficial in general, it is not clear how the state compression should be performed \textit{optimally}, especially at the encoder that sends both messages in the model in which the state is revealed strictly causally to both transmitters. The role of this encoder is seemingly similar to that of each of the two encoders in the model of \cite{LS13a}. However, because in our case the other encoder only sends a common message, the outer bound of Theorem~\ref{theorem-outer-bound-strictly-causal-states-setting} \textit{suggests} that, by opposition to the setting of\cite{LS13a}, in each block the private information of the encoder that sends both messages needs not carry an individual description of the state. Intuitively, this holds because, in our model in order to help the other encoder transmit at a larger rate, the encoder that transmits both messages better exploits any fraction of its individual message's rate by directly transmitting the common message, rather than compressing the state any longer so that the decoder obtains an estimate of the state that is better than what is possible using only the cooperative compression. Although a formal proof of this, as well as exact characterizations of the capacity regions of some of the models studied in this paper, are still to be found, this work enlightens different aspects relative to the utility of delayed CSI at transmitters in a cooperative multiaccess channel.   


\section*{Acknowledgement}
Insightful discussions with P. Piantanida are gratefully acknowledged. This work has been supported by the European Commission in the framework of the FP7 Network of Excellence in Wireless Communications (NEWCOM\#).


\appendix
Throughout this section we denote the set of strongly jointly $\epsilon$-typical sequences \cite[Chapter 14.2]{CT91} with respect to the distribution $P_{X,Y}$ as $\mc T_{\epsilon}^n(P_{X,Y})$.

\renewcommand{\theequation}{A-\arabic{equation}}
\setcounter{equation}{0}  
\subsection{Proof of Theorem~\ref{theorem-temporary-outer-bound-strictly-causal-states-setting}}\label{appendix-proof-theorem-temporary-outer-bound-strictly-causal-states-setting}

We prove that for any $(M_c,M_1,n,\epsilon)$ code consisting of a sequence of mappings $\phi_{1,i}:\mc W_c{\times}\mc W_1{\times}\mc S^{i-1} \longrightarrow \mc X_1$ at Encoder 1, a sequence of mappings $\phi_{2,i}: \mc W_c{\times}\mc S^{i-1} \longrightarrow \mc X_2$, $i=1,\hdots,n$, at Encoder 2, and a mapping $\psi : \mc Y^n \longrightarrow \mc W_c{\times}\mc W_1$ at the decoder with average error probability $P_e^n \rightarrow 0$ as $n \rightarrow 0$ and rates $R_c=n^{-1}\log_2M_c$ and $R_1=n^{-1}\log_2M_1$, there exists random variables $(V,U,X_1,X_2) \in {\mc V}{\times}{\mc U}{\times}{\mc X_1}{\times}{\mc X_2}$ such that the joint distribution $P_{S,V,U,X_1,X_2}$ is of the form 
\begin{align} 
P_{S,V,U,X_1,X_2}=Q_SP_{X_2}P_{V|S,X_2}P_{X_1|V,X_2}P_{U|V,S,X_1,X_2},
\end{align} 
the marginal distribution of $S$ is $Q_S(s)$, i.e.,
 \begin{align}
	\sum_{v,u,x_1,x_2}P_{S,V,U,X_1,X_2}(s,v,u,x_1,x_2)=Q_S(s)
\end{align}
and the rate pair $(R_c,R_1)$ must satisfy \eqref{measure-temporary-outer-bound-strictly-causal-states-setting}.

\noindent Fix $n$ and consider a given code of block length $n$. The joint probability mass function on $\mc W_c{\times}\mc W_1{\times}\mc S^n{\times}\mc X^n_1{\times}\mc X^n_2{\times}\mc Y^n$ is given by
\begin{align}
p(w_c,w_1,s^n,x^n_1,x^n_2,y^n)=p(w_c,w_1)\prod_{i=1}^np(s_i)p(x_{1i}|w_c,w_1,s^{i-1})p(x_{2i}|w_c,s^{i-1})p(y_i|x_{1i},x_{2i},s_i),
\end{align}
where, $p(x_{1i}|w_c,w_1,s^{i-1})$ is equal $1$ if $x_{1i}=f_1(w_c,w_1,s^{i-1})$ and $0$ otherwise; and $p(x_{2i}|w_c,s^{i-1})$ is equal $1$ if $x_{2i}=f_2(w_c,s^{i-1})$ and $0$ otherwise.

The decoder map $\psi$ recovers $(W_c,W_1)$ from $Y^n$ with the vanishing average error probability $P_e$.  By Fano's inequality, we have
\begin{align}
H(W_c,W_1|Y^n) \leq n\epsilon_n,
\end{align}
where $\epsilon_n \rightarrow 0$ as $P_e^n \rightarrow 0$.

Define the random variables
\begin{align}
\bar{V}_i &= (W_c,S^{i-1},Y^n_{i+1})\nonumber\\
\bar{U}_i &= (W_1,\bar{V}_i).
\label{Definition__RabdomVariables__OuterBound}
\end{align}
Observe that the random variables so defined satisfy
\begin{align}
(S_i,\bar{U}_i,\bar{V}_i, X_{1,i},X_{2,i},Y_i) \in \tilde{\mc P}^{\text{out}}_{\text{s-c}}, \quad \forall i \in\{1,\hdots,n\}.
\label{MeasureRandomVariables__OuterBound}
\end{align}

i) We can bound the sum rate as follows.

{\allowdisplaybreaks
\begin{align}
 n(R_c+R_1)  &\leq H(W_c,W_1) \\
     &= I(W_c,W_1;Y^n)+H(W_c,W_1|Y^n)\nonumber\\
     &\stackrel{(a)}{\leq}I(W_c,W_1;Y^n)+n\epsilon_n\nonumber\\
     &\stackrel{(b)}{=} I(W_c,W_1;Y^n)-I(W_c,W_1;S^n)+n\epsilon_n\\
     &= \sum_{i=1}^{n} I(W_c,W_1;Y_i|Y^n_{i+1})-I(W_c,W_1;S_i|S^{i-1})+n\epsilon_n\\
     &= \sum_{i=1}^{n} I(W_c,W_1,S^{i-1};Y_i|Y^n_{i+1})-I(S^{i-1};Y_i|W_c,W_1,Y^n_{i+1})-I(W_c,W_1;S_i|S^{i-1})+n\epsilon_n\\
     &= \sum_{i=1}^{n} I(W_c,W_1,S^{i-1};Y_i|Y^n_{i+1})-I(W_c,W_1;S_i|S^{i-1})-\sum_{i=1}^{n} I(S^{i-1};Y_i|W_c,W_1,Y^n_{i+1})+n\epsilon_n\\
     &\stackrel{(c)}{=} \sum_{i=1}^{n} I(W_c,W_1,S^{i-1};Y_i|Y^n_{i+1})-I(W_c,W_1;S_i|S^{i-1})-\sum_{i=1}^{n} I(Y^n_{i+1};S_i|W_c,W_1,S^{i-1})+n\epsilon_n\\
     &=\sum_{i=1}^{n} I(W_c,W_1,S^{i-1};Y_i|Y^n_{i+1})-H(S_i|S^{i-1})+H(S_i|W_c,W_1,S^{i-1},Y^n_{i+1})+n\epsilon_n\\
     &\stackrel{(d)}{=}\sum_{i=1}^{n} I(W_c,W_1,S^{i-1};Y_i|Y^n_{i+1})-H(S_i)+H(S_i|W_c,W_1,S^{i-1},Y^n_{i+1})+n\epsilon_n\\
     &\leq \sum_{i=1}^{n} I(W_c,W_1,S^{i-1},Y^n_{i+1};Y_i)-I(W_c,W_1,S^{i-1},Y^n_{i+1};S_i)+n\epsilon_n\\
     &\stackrel{(e)}{=} \sum_{i=1}^{n} I(\bar{U}_i,\bar{V}_i;Y_i)-I(\bar{U}_i,\bar{V}_i;S_i)+n\epsilon_n\\
     &\stackrel{(f)}{=} \sum_{i=1}^{n} I(\bar{U}_i,\bar{V}_i,X_{1i},X_{2i};Y_i)-I(\bar{U}_i,\bar{V}_i,X_{1i},X_{2i};S_i)+n\epsilon_n\\
     &\stackrel{(g)}{=} \sum_{i=1}^{n} I(\bar{U}_i,\bar{V}_i,X_{1i},X_{2i};Y_i)-I(\bar{U}_i,\bar{V}_i;S_i|X_{1i},X_{2i})+n\epsilon_n
\label{ProofOuterBoundDiscreteMemorylessChannel__FirstTerm}
\end{align}
where $(a)$ follows by Fano's inequality; $(b)$ follows from the fact that messages $W_c$ and $W_1$ are independent of the state sequence $S^n$; $(c)$ follows from  Csiszar and Korner's Sum Identity \cite{CK78}
\begin{align}
\sum_{i=1}^{n} I(Y^n_{i+1};S_i|W_c,W_1,S^{i-1})=\sum_{i=1}^{n} I(S^{i-1};Y_i|W_c,W_1,Y^n_{i+1})
\label{csiszar-korner-sum-identity}
\end{align}
$(d)$ follows from the fact that state $S^n$ is i.i.d.; $(e)$ follows by the definition of the random variables $\bar{U}_i$ and $\bar{V}_i$ in \eqref{Definition__RabdomVariables__OuterBound}; $(f)$ follows from the fact that $X_{1i}$ is a deterministic function of $(W_c,W_1,S^{i-1})$, and $X_{2i}$ is a deterministic function of $(W_c,S^{i-1})$, and $(g)$ follows from the fact that $X_{1i}$ and $X_{2i}$ are independent of $S_i$.

\vspace{0.3cm}

ii) Also, we can bound the individual rate as follows.

{\allowdisplaybreaks
\begin{align}
 nR_1  &\leq H(W_1|W_c) \\
     &= I(W_1;Y^n|W_c)+H(W_1|Y^n,W_c)\nonumber\\
     &\stackrel{(a)}{\leq} I(W_1;Y^n|W_c)+n\epsilon_n\nonumber\\
     &\stackrel{(b)}{=} I(W_1;Y^n|W_c)-I(W_1;S^n|W_c)+n\epsilon_n\\
     &= \sum_{i=1}^{n} I(W_1;Y_i|W_c,Y^n_{i+1})-I(W_1;S_i|W_c,S^{i-1})+n\epsilon_n\\
     &= \sum_{i=1}^{n} I(W_1,S^{i-1};Y_i|W_c,Y^n_{i+1})-I(S^{i-1};Y_i|W_c,W_1,Y^n_{i+1})-I(W_1;S_i|W_c,S^{i-1})+n\epsilon_n\\
     &\stackrel{(c)}{=} \sum_{i=1}^{n} I(W_1,S^{i-1};Y_i|W_c,Y^n_{i+1})-I(S_i;W_1,Y^n_{i+1}|W_c,S^{i-1})+n\epsilon_n\\
     &= \sum_{i=1}^{n} I(W_1;Y_i|W_c,S^{i-1},Y^n_{i+1})+I(S^{i-1};Y_i|W_c,Y^n_{i+1})-I(S_i;Y^n_{i+1}|W_c,S^{i-1})-I(S_i;W_1|W_c,S^{i-1},Y^n_{i+1})+n\epsilon_n\\
     &\stackrel{(d)}{=} \sum_{i=1}^{n} I(W_1;Y_i|W_c,S^{i-1},Y^n_{i+1})-I(W_1;S_i|W_c,S^{i-1},Y^n_{i+1})+n\epsilon_n\\
     &\stackrel{(e)}{=} \sum_{i=1}^{n} I(W_1;Y_i|W_c,S^{i-1},Y^n_{i+1},X_{2,i})-I(S_i;W_1|W_c,S^{i-1},Y^n_{i+1},X_{2,i})+n\epsilon_n\\
     &\stackrel{(f)}{=} \sum_{i=1}^{n} I(\bar{U}_i;Y_i|\bar{V}_i,X_{2,i})-I(\bar{U}_i;S_i|\bar{V}_i,X_{2,i})+n\epsilon_n\nonumber\\
     &\stackrel{(g)}{=} \sum_{i=1}^{n} I(\bar{U}_i,X_{1i};Y_i|\bar{V}_i,X_{2,i})-I(\bar{U}_i,X_{1i};S_i|\bar{V}_i,X_{2,i})+n\epsilon_n
\label{ProofOuterBoundDiscreteMemorylessChannel__SecondTerm}
\end{align}
where $(a)$ follows by Fano's inequality; $(b)$ follows from the fact that messages $W_c$ and $W_1$ are independent of the state sequence $S^n$; $(c)$ and $(d)$ follow from  Csiszar and Korner's Sum Identity \eqref{csiszar-korner-sum-identity}; $(e)$ follows since $X_{2i}$ is a deterministic function of $(W_c,S^{i-1})$; $(f)$ follows by the definition of the random variables $\bar{U}_i$ and $\bar{V}_i$ in \eqref{Definition__RabdomVariables__OuterBound}; and $(g)$ follows since $X_{1i}$ is a deterministic function of $(W_c,W_1,S^{i-1})$. 

From the above, we get that
\begin{align}
R_1 &\leq \frac{1}{n} \sum_{i=1}^{n} I(\bar{U}_i;Y_i|\bar{V}_i,X_{2,i})-I(\bar{U}_i;S_i|\bar{V}_i,X_{2,i})+\epsilon_n\nonumber\\
R_c+R_1 &\leq \frac{1}{n} \sum_{i=1}^{n} I(\bar{U}_i,\bar{V}_i,X_{2,i};Y_i)-I(\bar{U}_i,\bar{V}_i,X_{2,i};S_i)+\epsilon_n.
\label{MultiLetter__UpperBound__DiscreteChannel}
\end{align}
Also, observe that the auxiliary random variable $\bar{V}_i$ satisfies
\begin{align}
\sum_{i=1}^{n} I(\bar{V}_i,X_{2,i};Y_i)-I(\bar{V}_i,X_{2,i};S_i) &\geq 0.
\label{ConverseProof__ConstraintOuterBound__CompressionIndexDecoding}
\end{align}
This can be seen by noticing that
\begin{align}
I(W_1;Y^n|W_c) &= \sum_{i=1}^{n} I(\bar{U}_i;Y_i|\bar{V}_i,X_{2,i})-I(\bar{U}_i;S_i|\bar{V}_i,X_{2,i})\\
I(W_c,W_1;Y^n) &\leq \sum_{i=1}^{n} I(\bar{U}_i,\bar{V}_i,X_{2i};Y_i)-I(\bar{U}_i,\bar{V}_i,X_{2i};S_i).
\end{align}
and then noticing that, since $I(W_1;Y^n|W_c) \leq I(W_c,W_1;Y^n)$, the constraint \eqref{ConverseProof__ConstraintOuterBound__CompressionIndexDecoding} should hold.

The statement of the converse follows now by applying to \eqref{MultiLetter__UpperBound__DiscreteChannel} and \eqref{ConverseProof__ConstraintOuterBound__CompressionIndexDecoding} the standard time-sharing argument and taking the limits of large $n$. This is shown briefly here. We introduce a random variable $T$ which is independent of $S$, and uniformly  distributed over $\{1,\cdots,n\}$. Set $S=S_T$, $\bar{U}=\bar{U}_T$, $\bar{V}=\bar{V}_T$, $X_1=X_{1,T}$, $X_2=X_{2,T}$, and $Y=Y_T$. Then, considering the first bound in \eqref{MultiLetter__UpperBound__DiscreteChannel}, we obtain
\begin{align}
\frac{1}{n} &\sum_{i=1}^{n} I(\bar{U}_i;Y_i|\bar{V}_i,X_{2,i})-I(\bar{U}_i;S_i|\bar{V}_i,X_{2,i})\nonumber\\
&= I(\bar{U};Y|\bar{V},X_2,T)-I(\bar{U};S|\bar{V},X_2,T)\nonumber\\
&= I(\bar{U},T;Y|\bar{V},X_2,T)-I(\bar{U},T;S|\bar{V},X_2,T).
\label{FirstTermUpperBound__WithTimeSharingVariable__DiscreteChannel}
\end{align}

\noindent Similarly, considering the second bound in \eqref{MultiLetter__UpperBound__DiscreteChannel}, we obtain
\begin{align}
\frac{1}{n} &\sum_{i=1}^{n} I(\bar{U}_i,\bar{V}_i,X_{2,i};Y_i)-I(\bar{U}_i,\bar{V}_i,X_{2,i};S_i)\nonumber\\
            &= I(\bar{U},\bar{V},X_2;Y|T)-I(\bar{U},\bar{V},X_2;S|T)\nonumber\\
	    &= I(T,\bar{U},\bar{V},X_2;Y)-I(T;Y)-I(T,\bar{U},\bar{V},X_2;S)+I(T;S)\nonumber\\
	    &\leq I(T,\bar{U},\bar{V},X_2;Y)-I(T,\bar{U},\bar{V},X_2;S).
\label{SecondTermUpperBound__WithTimeSharingVariable__DiscreteChannel}
\end{align}
Also, considering the constraint \eqref{ConverseProof__ConstraintOuterBound__CompressionIndexDecoding}, we obtain
\begin{align}
0 &\leq \frac{1}{n} \sum_{i=1}^{n} I(\bar{V}_i,X_{2,i};Y_i)-I(\bar{V}_i,X_{2,i};S_i)\nonumber\\
            &= I(\bar{V},X_2;Y|T)-I(\bar{V},X_2;S|T)\nonumber\\
	    &= I(T,\bar{V},X_2;Y)-I(T;Y)-I(T,\bar{V},X_2;S)+I(T;S)\nonumber\\
	    &\leq I(T,\bar{V},X_2;Y)-I(T,\bar{V},X_2;S).
\end{align}

The distribution on $(T,S,\bar{U},\bar{V},X_1,X_2,Y)$ from the given code is of the form
\begin{align}
P_{T,S,\bar{U},\bar{V},X_1,X_2,Y} &= Q_SP_TP_{X_2|T}P_{X_1|X_2,T}P_{\bar{V}|X_1,X_2,S,T}P_{\bar{U}|\bar{V},S,X_1,X_2,T}W_{Y|X_1,X_2,S}.
\label{MeasureOuterBound__with__TimeSharingRandomVariable}
\end{align}

\noindent Let us now define $U=(\bar{U},T)$ and $V=(\bar{V},T)$. Using \eqref{MultiLetter__UpperBound__DiscreteChannel}, \eqref{FirstTermUpperBound__WithTimeSharingVariable__DiscreteChannel} and \eqref{SecondTermUpperBound__WithTimeSharingVariable__DiscreteChannel}, we then get
\begin{align}
R_1 &\leq I(U;Y|V,X_2)-I(U;S|V,X_2)+\epsilon_n\nonumber\\
R_c+R_1 &\leq I(U,V,X_2;Y)-I(U,V,X_2;S)+\epsilon_n,
\end{align}
where the distribution on $(S,U,V,X_1,X_2,Y)$, obtained by marginalizing \eqref{MeasureOuterBound__with__TimeSharingRandomVariable} over the time sharing random variable $T$, satisfies $(S,U,V,X_1,X_2,Y) \in \tilde{\mc P}^{\text{out}}_{\text{s-c}}$.

So far we have shown that, for a given sequence of $(\epsilon_n,n,R_c,R_1)-$codes with $\epsilon_n$ going to zero as $n$ goes to infinity, there exist random variables $(S,U,V,X_1,X_2,Y) \in \tilde{\mc P}^{\text{out}}_{\text{s-c}}$ such that the rate pair $(R_c,R_1)$ essentially satisfies the inequalities in \eqref{temporary-outer-bound-strictly-causal-states-setting}, i.e., $(R_c,R_1) \in \tilde{\mc R}^{\text{out}}_{\text{s-c}}$.

\renewcommand{\theequation}{B-\arabic{equation}}
\setcounter{equation}{0}  
\subsection{Proof of Theorem~\ref{theorem-outer-bound-strictly-causal-states-setting}}\label{appendix-proof-theorem-outer-bound-strictly-causal-states-setting}

Recall the set $\tilde{\mc R}^{\text{out}}_{\text{s-c}}$ which is an outer bound on the capacity region $\mc C_{\text{s-c}}$ as stated in  Theorem~\ref{theorem-temporary-outer-bound-strictly-causal-states-setting}. Let a rate-pair $(R_c,R_1) \in \tilde{\mc R}^{\text{out}}_{\text{s-c}}$. Then we have 
\begin{subequations}
\begin{align}
\label{ineq1-temporary-outer-bound-strictly-causal-states-setting}
R_1 \: &\leq \: I(U,X_1;Y|V,X_2)-I(U,X_1;S|V,X_2)\\
R_c+ R_1 \: &\leq \: I(U,V,X_1,X_2;Y)-I(U,V,X_1,X_2;S)
\label{ineq2-temporary-outer-bound-strictly-causal-states-setting}
\end{align}
\label{ineqs-temporary-outer-bound-strictly-causal-states-setting}
\end{subequations}
Consider the first inequality \eqref{ineq1-temporary-outer-bound-strictly-causal-states-setting}. We have

\begin{align}
R_1 &\leq  I(U,X_1;Y|V,X_2)-I(U,X_1;S|V,X_2)\nonumber\\
    &= I(X_1;Y|V,X_2)-I(X_1;S|V,X_2)+I(U;Y|V,X_1,X_2)-I(U;S|V,X_1,X_2)\nonumber\\
    &\leq I(X_1;Y|V,X_2)+ I(U;Y|V,X_1,X_2)-I(U;S|V,X_1,X_2)\nonumber\\
    &\leq I(X_1;Y|V,X_2)+ I(U;Y,S|V,X_1,X_2)-I(U;S|V,X_1,X_2)\nonumber\\
    &= I(X_1;Y|V,X_2)+ I(U;Y|S,V,X_1,X_2)\nonumber\\
    &= I(X_1;Y|V,X_2)+ H(Y|S,V,X_1,X_2)-H(Y|S,U,V,X_1,X_2)\nonumber\\
    &\stackrel{(b)}{=} I(X_1;Y|V,X_2)+ H(Y|S,X_1,X_2)-H(Y|S,X_1,X_2)\nonumber\\
    &= I(X_1;Y|V,X_2)
\label{ineq1-outer-bound-strictly-causal-states-setting}
\end{align}
where $(a)$ follows since $(U,V) \leftrightarrow (S,X_1,X_2) \leftrightarrow Y$ is a Markov chain

Similarly, considering the second inequality \eqref{ineq2-temporary-outer-bound-strictly-causal-states-setting}, we have
\begin{align}
R_c+ R_1 &\leq I(U,V,X_1,X_2;Y)-I(U,V,X_1,X_2;S)\nonumber\\
         &= I(V,X_1,X_2;Y)-I(V,X_1,X_2;S) + [I(U;Y|V,X_1,X_2)-I(U;S|V,X_1,X_2)]\nonumber\\
	 &\stackrel{(b)}{\leq} I(V,X_1,X_2;Y)-I(V,X_1,X_2;S)
\label{ineq2-outer-bound-strictly-causal-states-setting}
\end{align}
where $(b)$ follows by following straightforwardly the lines of \eqref{ineq1-outer-bound-strictly-causal-states-setting}.

Finally, using \eqref{ineq1-outer-bound-strictly-causal-states-setting} and \eqref{ineq2-outer-bound-strictly-causal-states-setting} we obtain the desired simpler outer bound form \eqref{outer-bound-strictly-causal-states-setting}. Summarizing, the above shows that the region $\mc R^{\text{out}}_{\text{s-c}}$ is an outer bound on the capacity region of the multiaccess channel with degraded messages sets and states known strictly causally at only the encoders. This completes the proof of Theorem~\ref{theorem-outer-bound-strictly-causal-states-setting}.

\renewcommand{\theequation}{C-\arabic{equation}}
\setcounter{equation}{0}  
\subsection{Proof of Proposition~\ref{proposition-alternative-outer-bound-strictly-causal-states-setting}}\label{appendix-proof-proposition-alternative-outer-bound-strictly-causal-states-setting}

We prove that for any $(M_c,M_1,n,\epsilon)$ code consisting of sequences of mappings $\phi_{1,i}:\mc W_c{\times}\mc W_1{\times}\mc S^{i-1} \longrightarrow \mc X_1$ at Encoder 1, and $\phi_{2,i}: \mc W_c{\times}\mc S^{i-1} \longrightarrow \mc X_2$ at Encoder 2, $i=1,\hdots,n$, and a mapping $\psi : \mc Y^n \longrightarrow \mc W_c{\times}\mc W_1$ at the decoder with average error probability $P_e^n \rightarrow 0$ as $n \rightarrow 0$ and rates $R_c=n^{-1}\log_2M_c$ and $R_1=n^{-1}\log_2M_1$, the rate pair $(R_c,R_1)$ must satisfy \eqref{measure-alternative-outer-bound-strictly-causal-states-setting}.

\noindent Fix $n$ and consider a given code of block length $n$. The joint probability mass function on $\mc W_c{\times}\mc W_1{\times}\mc S^n{\times}\mc X^n_1{\times}\mc X^n_2{\times}\mc Y^n$ is given by
\begin{align}
p(w_c,w_1,s^n,x^n_1,x^n_2,y^n)=p(w_c,w_1)\prod_{i=1}^np(s_i)p(x_{1i}|w_c,w_1,s^{i-1})p(x_{2i}|w_c,s^{i-1})p(y_i|x_{1i},x_{2i},s_i),
\end{align}
where, $p(x_{1i}|w_c,w_1,s^{i-1})$ is equal $1$ if $x_{1i}=f_1(w_c,w_1,s^{i-1})$ and $0$ otherwise; and $p(x_{2i}|w_c,s^{i-1})$ is equal $1$ if $x_{2i}=f_2(w_c,s^{i-1})$ and $0$ otherwise.

The decoder map $\psi$ recovers $(W_c,W_1)$ from $Y^n$ with the vanishing average error probability $P_e$.  By Fano's inequality, we have
\begin{align}
H(W_c,W_1|Y^n) \leq n\epsilon_n,
\end{align}
where $\epsilon_n \rightarrow 0$ as $P_e^n \rightarrow 0$.

\noindent The proof of the bound on $R_1$ follows trivially by revealing the state $S^n$ to the decoder.

\noindent The proof of the bound on the sum rate $(Rc+R_1)$ follows as follows.

\begin{align}
 n(R_c+R_1)  &\leq H(W_c,W_1) \nonumber\\
      &= I(W_c,W_1;Y^n)+H(W_c,W_1|Y^n)\nonumber\\
      &\leq I(W_c,W_1;Y^n)+n\epsilon_n\nonumber\\
      &= \sum_{i=1}^{n} I(W_c,W_1;Y_i|Y^{i-1})+n\epsilon_n\nonumber\\
      &= \sum_{i=1}^{n} H(Y_i|Y^{i-1})- H(Y_i|W_c,W_1,Y^{i-1})+n\epsilon_n\nonumber\\
      &\stackrel{(a)}{\leq} \sum_{i=1}^{n} H(Y_i)- H(Y_i|W_c,W_1,Y^{i-1})+n\epsilon_n\nonumber\\
      &\stackrel{(b)}{\leq} \sum_{i=1}^{n} H(Y_i)- H(Y_i|W_c,W_1,Y^{i-1},S^{i-1})+n\epsilon_n\nonumber\\
      &\stackrel{(c)}{=} \sum_{i=1}^{n} H(Y_i)- H(Y_i|W_c,W_1,Y^{i-1},S^{i-1},X_{1i},X_{2i})+n\epsilon_n\nonumber\\
      &\stackrel{(d)}{=} \sum_{i=1}^{n} H(Y_i)- H(Y_i|X_{1i},X_{2i})+n\epsilon_n\nonumber\\
      &= \sum_{i=1}^{n} I(X_{1i},X_{2i};Y_i)+n\epsilon_n
\end{align}
where $(a)$ and $(b)$ follow from the fact that conditioning reduces the entropy; $(c)$ follows from the fact that $X_{1i}$ is a deterministic function of $(W_c,W_1,S^{i-1})$, and $X_{2i}$ is a deterministic function of $(W_c,S^{i-1})$, and $(d)$ follows from the fact that  $(W_c,W_1,Y^{i-1},S^{i-1}) \leftrightarrow (X_{1i},X_{2i},S_i) \leftrightarrow Y_i$  and $(W_c,W_1,Y^{i-1},S^{i-1},X_{1i},X_{2i})$ is independent of $S_i$.

\noindent The rest of the proof of Proposition~\ref{proposition-alternative-outer-bound-strictly-causal-states-setting} follows by standard single-letterization.

\renewcommand{\theequation}{D-\arabic{equation}}
\setcounter{equation}{0}  
\subsection{Analysis of Example~\ref{example-comparison-of-outer-bounds}}\label{appendix-analysis-example-comparison-of-outer-bounds}

Recall Example 1. For this example, it is easy to see that $\breve{\mc R}^{\text{out}}_{\text{s-c}} = \{(R_1,R_2)\: : \: 0 \leq R_1 \leq 1/2, \: 0 \leq R_2 \leq 1/2\}$. Thus, $(1/2,1/2) \in \breve{\mc R}^{\text{out}}_{\text{s-c}}$. We now show that $(1/2,1/2) \notin \mc R^{\text{out}}_{\text{s-c}}$.

Assume that the rate-pair $(R_c,R_1)=(1/2,1/2) \in \mc R^{\text{out}}_{\text{s-c}}$ for some measure of the form \eqref{measure-outer-bound-strictly-causal-states-setting} and that satisfies the constraint \eqref{nonnegativity-constraint-temporary-outer-bound}. 
Since $R_c+R_1=1$, the constraint on the sum rate
\begin{equation}
I(V,X_1,X_2;Y)-I(V,X_1,X_2;S) = H(Y)-H(Y|V,X_1,X_2)-I(V,X_1,X_2;S)
\label{contraint-on-sum-rate-example1}
\end{equation}
leads to
\begin{equation}
1 \leq H(Y)-H(Y|V,X_1,X_2)-I(V,X_1,X_2;S) \leq H(Y) \leq 1
\label{analysis-of-example1-step1}
\end{equation}
where the last inequality holds since $|\mc Y|=2$. Thus,
\begin{subequations}
\begin{align}
\label{analysis-of-example1-step2-subeq1}
& H(Y) = 1\\
\label{analysis-of-example1-step2-subeq2}
& S \:\:\:\text{is independent of}\:\:\: (V,X_1,X_2)\\
& H(Y|V,X_1,X_2)=0.
\end{align}
\label{analysis-of-example1-step2}
\end{subequations}

\noindent Observing that the constraint on the sum rate \eqref{contraint-on-sum-rate-example1} can also be written equivalently as
\begin{equation}
I(V,X_1,X_2;Y)-I(V,X_1,X_2;S) = H(Y)-H(Y|X_1,X_2)-I(V;S|X_1,X_2,Y)
\label{contraint-on-sum-rate-example1-equivalent-form}
\end{equation}
we obtain that
\begin{equation}
H(Y|X_1,X_2)=0.
\label{analysis-of-example1-step3}
\end{equation}
\noindent Using \eqref{analysis-of-example1-step2-subeq2} and \eqref{analysis-of-example1-step3}, and the fact that $Y=X_S$, it follows that 
\begin{equation}
\text{Pr}\{X_1=X_2\} =1.
\label{analysis-of-example1-step4}
\end{equation}
Now, using \eqref{analysis-of-example1-step4}, the constraint on the individual rate $R_1$ leads to
\begin{align}
0 \leq R_1 &\leq I(X_1;Y|V,X_2)\\
           &= H(X_1|V,X_2) - H(X_1|V,X_2,Y)\\
	   &\leq H(X_1|V,X_2)\\
\label{analysis-of-example1-step5}
	   &\leq H(X_1|X_2)\\
	   &= 0
\label{analysis-of-example1-step6}
\end{align}
where \eqref{analysis-of-example1-step5} follows from the fact that conditioning reduces entropy, and \eqref{analysis-of-example1-step6} follows by \eqref{analysis-of-example1-step4}.

\noindent The above shows that $R_1=0$. This contradicts the fact that the rate-pair $(R_c,R_1)=(1/2,1/2) \in \mc R^{\text{out}}_{\text{s-c}}$. We conclude that the rate-pair $(1/2,1/2) \notin \mc R^{\text{out}}_{\text{s-c}}$.
\renewcommand{\theequation}{E-\arabic{equation}}
\setcounter{equation}{0}  
\subsection{Proof of Theorem~\ref{theorem-inner-bound-strictly-causal-states-setting}}\label{appendix-proof-theorem-inner-bound-strictly-causal-states-setting}


The transmission takes place in $B$ blocks. The common message $W_c$ is divided into $B-1$ blocks $w_{c,1},\hdots,w_{c,B-1}$ of $nR_c$ bits each, and the individual message $W_1$ is divided into $B-1$ blocks $w_{1,1},\hdots,w_{1,B-1}$ of $nR_1$ bits each. For convenience, we let $w_{c,B}=w_{1,B}=1$ (default values). We thus have $B_{W_c}=n(B-1){R_c}$, $B_{W_1}=n(B-1){R_1}$, $N=nB$, $R_{W_c}=B_{W_c}/N=R_c{\cdot}(B-1)/B$ and $R_{W_1}=B_{W_1}/N=R_1{\cdot}(B-1)/B$, where $B_{W_c}$ is the number of common message bits, $B_{W_1}$ is the number of individual message bits, $N$ is the number of channel uses and $R_{W_c}$ and $R_{W_1}$ are the overall rates of the common and individual messages, respectively. For fixed $n$, the average rate pair $(R_{W_c}, R_{W_1})$ over $B$ blocks can be made as close to $(R_c,R_1)$ as desired by making $B$ large.

\noindent \textbf{Codebook Generation:} Fix a measure $P_{S,V,X_1,X_2,Y} \in \mc P^{\text{in}}_{\text{s-c}}$. Fix $\epsilon > 0$ and denote $M_c = 2^{n[R_c-\eta_c\epsilon]}$, $M_1 = 2^{n[R_1-\eta_1\epsilon]}$, $K = 2^{n[T+\mu_c\epsilon]}$ and $\hat{K} = 2^{n[\hat{T}+\hat{\mu}_c\epsilon]}$.

\begin{itemize}
\item[1)] We generate $M_cK$ independent and identically distributed (i.i.d.) codewords $\dv x_2(w_c,s)$ indexed by $w_c=1,\hdots,M_c$, $s=1,\hdots,K$, each with i.i.d. components drawn according to $P_{X_2}$.
\item[2)] For each codeword $\dv x_2(w_c,s)$,  we generate $\hat{K}$ independent and identically distributed (i.i.d.) codewords $\dv v(w_c,s,z)$ indexed by $z=1,\hdots,\hat{K}$, each with i.i.d. components drawn according to $P_{V|X_2}$.
\item[3)] For each codeword $\dv x_2(w_c,s)$, we generate $M_1$ independent and identically distributed (i.i.d.) codewords $\dv x_1(w_c,s,w_1)$ indexed by $w_1=1,\hdots,M_1$, each with i.i.d. components drawn according to $P_{X_1|X_2}$.
\item[4)] Randomly partition the set $\{1,\hdots,\hat{K}\}$ into $K$ cells $\mc C_s$, $s \in [1,K]$.
\end{itemize}

\textbf{Encoding:} Suppose that a common message $W_c=w_c$ and an individual message $W_1=w_1$ are to be transmitted. As we mentioned previously, message $w_c$ is divided into $B-1$ blocks $w_{c,1},\hdots,w_{c,B-1}$ and message $w_1$ is divided into $B-1$ blocks $w_{1,1},\hdots,w_{1,B-1}$, with $(w_{c,i},w_{1,i})$ the pair messages sent in block $i$. We denote by $\dv s[i]$ the channel state in block $i$, $i=1,\hdots,B$. For convenience, we let $\dv s[0]=\emptyset$ and $z_0=1$ (a default value), and $s_0$ the index of the cell containing $z_0$, i.e., $z_0 \in \mc C_{s_0}$. The encoding at the beginning of the block $i$, $i=1,\hdots,B-1$, is as follows.

\noindent Encoder $2$, which has learned the state sequence $\dv s[i-1]$, knows $s_{i-2}$ and looks for a compression index $z_{i-1} \in [1,\hat{K}]$ such that $\dv v(w_{c,i-1},s_{i-2},z_{i-1})$ is strongly jointly typical with $\dv s[i-1]$ and $\dv x_2(w_{c,i-1},s_{i-2})$. If there is no such index or the observed state $\dv s[i-1]$ is not typical, $z_{i-1}$ is set to $1$ and an error is declared. If there is more than one such index $z_{i-1}$, choose the smallest. One can show that the probability of error of this event is arbitrarily small provided that $n$ is large and
\begin{align}
\hat{T} &> I(V;S|X_2).
\label{constraint-common-compression}
\end{align}

\noindent Encoder 2 then transmits the vector $\dv x_2(w_{c,i},s_{i-1})$, where the cell index $s_{i-1}$ is chosen such that $z_{i-1} \in \mc C_{s_{i-1}}$.

\noindent Encoder 1 finds $\dv x_2(w_{c,i},s_{i-1})$ similarly. It then transmits the vector $\dv x_1(w_{c,i},s_{i-1},w_{1i})$.

\noindent (Note that, other than the information messages, Encoder 1 sends only the cooperative compression index $s_{i-1}$; no other individual compression index is sent by this encoder).

\textbf{Decoding:} Let $\dv y[i]$ denote the information received at the receiver at block $i$, $i=1,\hdots,B$. The receiver collects these information until the last block of transmission is completed. The decoder then performs Willem's backward decoding \cite{W82}, by first decoding the pair $(w_{c,B-1},w_{1,B-1})$ from $\dv y[B-1]$.

\textit{1) Decoding in Block $B-1$:}

The decoding of the pair $(w_{c,B-1},w_{1,B-1})$ is performed in five steps, as follows.

\noindent \underline{\textit{Step (a):}} The decoder knows $w_{c,B}=1$ and looks for the unique cell index $\hat{s}_{B-1}$ such that the vector $\dv x_2(w_{c,B},\hat{s}_{B-1})$ is jointly typical with $\dv y[B]$.  The decoding operation in this step incurs small probability of error as long as $n$ is sufficiently large and
\begin{align}
T &< I(X_2;Y).
\label{DecodingOfCellIndex}
\end{align}

\noindent \underline{\textit{Step (b):}} The decoder now knows the cell index $\hat{s}_{B-1}$ (but not the exact compression index $z_{B-1}$). It then decodes message $w_{c,B-1}$ by looking for the unique $\hat{w}_{c,B-1}$ such that $\dv x_2(\hat{w}_{c,B-1},s_{B-2})$, $\dv v(\hat{w}_{c,B-1},s_{B-2},z_{B-1})$, $\dv x_1(\hat{w}_{c,B-1},s_{B-2},w_{1,B-1})$ and $\dv y[B-1]$ are jointly typical for some $s_{B-2} \in [1,K]$, $z_{B-1} \in \mc C_{\hat{s}_{B-1}}$ and $w_{1,B-1} \in [1,M_1]$. One can show that the decoder obtains the correct $w_{c,B-1}$ as long as $n$ and $B$ are large and
\begin{align}
R_c + R_1 +T+(\hat{T}-T) &\leq I(V,X_1,X_2;Y).
\label{Constraint__On__SumRate}
\end{align}

\noindent \underline{\textit{Step (c):}} The decoder now knows message $\hat{w}_{c,B-1}$ and, by proceeding as in the step a), finds the correct cell index $\hat{s}_{B-2}$ as long as $n$ is sufficiently large and \eqref{DecodingOfCellIndex} is true. 

\noindent \underline{\textit{Step (d):}} The decoder calculates a set $\mc L(\dv y[B-1])$ of $z_{B-1}$ such that $z_{B-1} \in \mc L(\dv y[B-1])$ if $\dv v(\hat{w}_{c,B-1},\hat{s}_{B-2},z_{B-1})$, $\dv x_2(\hat{w}_{c,B-1},\hat{s}_{B-2})$, $\dv y[B-1]$ are jointly typical. It then declares that $z_{B-1}$ was sent in block $B-1$ if
\begin{align}
\hat{z}_{B-1} \in \mc C_{\hat{s}_{B-1}} \cap \mc L(\dv y[B-1]).
\end{align}
One can show that, for large $n$, $\hat{z}_{B-1}=z_{B-1}$ with arbitrarily high probability provided that $n$ is sufficiently large and 
\begin{equation}
\hat{T} < I(V;Y|X_2) + T.
\label{Constraint__Of__NonNegativity}
\end{equation}

\noindent \underline{\textit{Step (e):}} Finally, the decoder, which now knows message $\hat{w}_{c,B-1}$, the cell indices $(\hat{s}_{B-2},\hat{s}_{B-1})$ and the exact compression index $z_{B-1} \in \mc C_{\hat{s}_{B-1}}$, estimates message $w_{1,B-1}$ using $\dv y[B-1]$. It declares that $\hat{w}_{1,B-1}$ was sent if there exists a unique $\hat{w}_{1,B-1}$ such that  $\dv x_2(\hat{w}_{c,B-1},\hat{s}_{B-2})$,  $\dv v(\hat{w}_{c,B-1},\hat{s}_{B-2},\hat{z}_{B-1})$, $\dv x_1(\hat{w}_{c,B-1},\hat{s}_{B-2},\hat{w}_{1,B-1})$ and $\dv y[B-1]$ are jointly typical. The decoding in this step incurs small probability of error as long as $n$ is sufficiently large and
\begin{align}
R_1 &\leq I(X_1;Y|V,X_2).
\label{Constraint1__On__IndividualRate}
\end{align}

\textit{2) Decoding in Block $b$, $b=B-1,B-2,\hdots,2$:}

Next, for $b$ ranging from $B-1$ to $2$, the decoding of the pair $(w_{c,b-1},w_{1,b-1})$ is performed similarly, in four steps, by using the information $\dv y[b]$ received in block $b$ and the information $\dv y[b-1]$ received in block $b-1$. More specifically, this is done as follows.

\noindent \underline{\textit{Step (a):}} The decoder knows $w_{c,b}$ and looks for the unique cell index $\hat{s}_{b-1}$ such that the vector $\dv x_2(w_{c,b},\hat{s}_{b-1})$ is jointly typical with $\dv y[b]$. The decoding error in this step is small for sufficiently large $n$ if \eqref{DecodingOfCellIndex} is true.

\noindent \underline{\textit{Step (b):}} The decoder now knows the cell index $\hat{s}_{b-1}$ (but not the exact compression indices $z_{b-1}$). It then decodes message $w_{c,b-1}$ by looking for the unique $\hat{w}_{c,b-1}$ such that $\dv x_2(\hat{w}_{c,b-1},s_{b-2})$, $\dv v(\hat{w}_{c,b-1},s_{b-2},z_{b-1})$, $\dv x_1(\hat{w}_{c,b-1},s_{b-2},w_{1,b-1})$ and $\dv y[b-1]$ are jointly typical for some $s_{b-2} \in [1,K]$, $z_{b-1} \in \mc C_{\hat{s}_{b-1}}$ and $w_{1,b-1} \in [1,M_1]$. One can show that the decoder obtains the correct $w_{c,b-1}$ as long as $n$ and $B$ are large and \eqref{Constraint__On__SumRate} is true.

\noindent \underline{\textit{Step (c):}} The decoder knows message $\hat{w}_{c,b-1}$ and, by proceeding as in the step a), finds the correct cell index $\hat{s}_{b-2}$ as long as $n$ is sufficiently large and \eqref{DecodingOfCellIndex} is true.

\noindent \underline{\textit{Step (d):}} The decoder calculates a set $\mc L(\dv y[b-1])$ of $z_{b-1}$ such that $z_{b-1} \in \mc L(\dv y[b-1])$ if $\dv v(\hat{w}_{c,b-1},\hat{s}_{b-2},z_{b-1})$, $\dv x_2(\hat{w}_{c,b-1},\hat{s}_{b-2})$, $\dv y[b-1]$ are jointly typical. It then declares that $z_{b-1}$ was sent in block $b-1$ if
\begin{align}
\hat{z}_{b-1} \in \mc C_{\hat{s}_{b-1}} \cap \mc L(\dv y[b-1]).
\end{align}
One can show that, for large $n$, $\hat{z}_{b-1}=z_{b-1}$ with arbitrarily high probability provided that $n$ is sufficiently large and \eqref{Constraint__Of__NonNegativity} is true. 

\noindent \underline{\textit{Step (e):}} Finally, the decoder, which now knows message $\hat{w}_{c,b-1}$, the cell indices $(\hat{s}_{b-2},\hat{s}_{b-1})$ as well as the exact compression index $\hat{z}_{b-1} \in \mc C_{\hat{s}_{b-1}}$, estimates message $w_{1,b-1}$ using $\dv y[b-1]$. It declares that $\hat{w}_{1,b-1}$ was sent if there exists a unique $\hat{w}_{1,b-1}$ such that  $\dv x_2(\hat{w}_{c,b-1},\hat{s}_{b-2})$, $\dv v(\hat{w}_{c,b-1},\hat{s}_{b-2},\hat{z}_{b-1})$, $\dv x_1(\hat{w}_{c,b-1},\hat{s}_{b-2},\hat{w}_{1,b-1})$ and $\dv y[b-1]$ are jointly typical. One can show that the decoder obtains the correct $w_{c,b-1}$ as long as $n$ and $B$ are large and \eqref{Constraint1__On__IndividualRate} is true.

\textbf{Fourier-Motzkin Elimination:} From the above, we get that the error probability is small provided that $n$ is large and
\begin{subequations}
\begin{align}
\hat{T} &> I(V;S|X_2)\\
T &< I(X_2;Y)\\
\hat{T} &< I(V;Y|X_2)+T\\
R_c+R_1+\hat{T} &\leq I(V,X_1,X_2;Y)\\
R_1 &\leq I(X_1;Y|V,X_2)\\
\end{align}
\label{fme-theorem-inner-bound-step1}
\end{subequations}

We now apply Fourier-Motzkin Elimination (FME) to successively project out  $T$ and $\hat{T}$ from \eqref{fme-theorem-inner-bound-step1}. Projecting out $T$ from \eqref{fme-theorem-inner-bound-step1}, we get
\begin{subequations}
\begin{align}
\hat{T} &> I(V;S|X_2)\\
\hat{T} &< I(V,X_2;Y)\\
R_c+R_1+\hat{T} &< I(V,X_1,X_2;Y)\\
R_1 &< I(X_1;Y|V,X_2)
\end{align}
\label{fme-theorem-inner-bound-step2}
\end{subequations}
Next, projecting out $\hat{T}$ from \eqref{fme-theorem-inner-bound-step1}, we get
\begin{subequations}
\begin{align}
0 &\leq I(V,X_2;Y)-I(V;S|X_2)\\
R_1 &\leq I(X_1;Y|V,X_2)\\
R_c+R_1 &< I(V,X_1,X_2;Y)-I(V;S|X_2)
\end{align}
\label{fme-theorem-inner-bound-step3}
\end{subequations}
Finally, recalling that the measure $P_{S,V,X_1,X_2,Y} \in \mc P^{\text{in}}_{\text{s-c}}$ satisfies that $X_1$ and $X_2$ are independent of $S$ and also implies that $X_1 \leftrightarrow (V,X_2) \leftrightarrow S$ is a Markov chain, it can be seen easily that the inequalities in \eqref{fme-theorem-inner-bound-step3} can be rewritten equivalently as
\begin{subequations}
\begin{align}
0 &\leq I(V,X_2;Y)-I(V,X_2;S)\\
R_1 &\leq I(X_1;Y|V,X_2)\\
R_c+R_1 &\leq I(V,X_1,X_2;Y)-I(V,X_1,X_2;S).
\end{align}
\label{fme-theorem-inner-bound-step4}
\end{subequations}

\noindent This completes the proof of Theorem~\ref{theorem-inner-bound-strictly-causal-states-setting}.


\renewcommand{\theequation}{F-\arabic{equation}}
\setcounter{equation}{0}  
\subsection{Proof of Proposition~\ref{proposition-bounds-auxiliary-random-variables-strictly-causal-states-setting}}\label{appendix-proposition-bounds-auxiliary-random-variables-strictly-causal-states-setting}

In what follows we show that the outer bound $\mc R^{\text{out}}_{\text{s-c}}$ is convex, and that it is enough to restrict $\mc V$ to satisfy \eqref{bounds-auxiliary-random-variables-inner-and-temporary-outer-bounds-strictly-causal-states-setting}. The proof for the inner bound $\mc R^{\text{in}}_{\text{s-c}}$ follows similarly.

\textit{Part 1-- Convexity:} Consider the region $\mc R^{\text{out}}_{\text{s-c}}$. To prove the convexity of the region $\mc R^{\text{out}}_{\text{s-c}}$, we use a standard argument. We introduce a time-sharing random variable $T$ and define the joint distribution
\begin{align}
P_{T,S,V,X_1,X_2,Y}(t,s,v,x_1,x_2,y) &= P_{T,S,V,X_1,X_2}(t,s,v,x_1,x_2)W_{Y|X_1,X_2,S}(y|x_1,x_2,s)\\
\sum_{v,x_1,x_2} P_{T,S,V,X_1,X_2}(t,s,v,x_1,x_2) &= P_T(t)Q_S(s). 
\end{align}

Let now $(R^{T}_c,R^{T}_1)$ be the common and individual rates resulting from time sharing. Then,
\begin{align}
R^{T}_1 \: &\leq \: I(X_1;Y|V,X_2,T)\\
           &= \: I(X_1;Y|\tilde{V},X_2)\\
R^{T}_c+ R^{T}_1 \: &\leq \: I(V,X_1,X_2;Y|T)-I(V,X_1,X_2;S|T)\\
	   &= \: I(V,X_1,X_2;Y|T)-I(V,X_1,X_2,T;S)\\
	   &\leq \: I(V,X_1,X_2,T;Y)-I(V,X_1,X_2,T;S)\\
	   &= \: I(\tilde{V},X_1,X_2;Y)-I(\tilde{V},X_1,X_2;S)
\end{align}
where $\tilde{V} := (V,T)$. Also, we have
\begin{align}
I(\tilde{V},X_2;Y)-I(\tilde{V},X_2;S) &= I(V,X_2,T;Y)-I(V,X_2,T;S)\\
                                      &= I(V,X_2;Y|T)-I(V,X_2;S|T)+I(T;Y)\\
				      &\geq  I(V,X_2;Y|T)-I(V,X_2;S|T)
\end{align}
where the second equality follows since $T$ and $S$ are independent.

\noindent The above shows that the time sharing random variable $T$ is incorporated into the auxiliary random variable $V$. This shows that time sharing cannot yield rate pairs that are not included in $\mc R^{\text{out}}_{\text{s-c}}$ and, hence, $\mc R^{\text{out}}_{\text{s-c}}$ is convex. 

\textit{Part 2-- Bound on $|\mc V|$:} To prove that the region $\mc R^{\text{out}}_{\text{s-c}}$ is not altered if one restricts the random variable $V$ to have its alphabet restricted as indicated in \eqref{bounds-auxiliary-random-variables-inner-and-temporary-outer-bounds-strictly-causal-states-setting}, we invoke the support lemma \cite[p. 310]{CK81}. Fix a distribution $\mu \in \mc P^{\text{out}}_{\text{s-c}}$ of $(S,V,X_1,X_2,Y)$ and, without loss of generality, let us denote the product set $\mc S\times\mc X_1\times\mc X_2=\{1,\hdots,m\}$, $m=|\mc S{\times}\mc X_1{\times}\mc X_2|$.

\noindent To prove the bound \eqref{bounds-auxiliary-random-variables-inner-and-temporary-outer-bounds-strictly-causal-states-setting} on $|\mc V|$, note that we have
\begin{align}
I_{\mu}(X_1;Y|V,X_2) &= I_{\mu}(X_1,X_2;Y|V)-I_{\mu}(X_2;Y|V)\nonumber\\
&= H_{\mu}(X_2,Y|V)-H_{\mu}(X_1,X_2,Y|V)+H_{\mu}(X_1,X_2|V)-H_{\mu}(X_2|V)
\end{align}
and
\begin{align}
& I_{\mu}(V,X_1,X_2;Y)-I_{\mu}(V,X_1,X_2;S) \nonumber\\
&\hspace{1cm}= I_{\mu}(X_1,X_2;Y|V)-I_{\mu}(X_1,X_2;S|V)+I_{\mu}(V;Y)-I_{\mu}(V;S)\nonumber \\
&\hspace{1cm}= H_{\mu}(X_1,X_2,S|V)-H_{\mu}(X_1,X_2,Y|V)+H_{\mu}(Y)-H_{\mu}(S).
\end{align}
Similarly, we have
\begin{align}
I_{\mu}(V,X_2;Y)-I_{\mu}(V,X_2;S) &= H_{\mu}(X_2,S|V)-H_{\mu}(X_2,Y|V)+H_{\mu}(Y)-H_{\mu}(S).
\end{align}

\noindent Hence, it suffices to show that the following functionals of $\mu(S,V,X_1,X_2,Y)$
\begin{subequations}
\begin{align}
r_{i}(\mu) &= \mu(s,x,x'), \quad i=1,\hdots,m-1\\
r_m(\mu) &= \int_{v}d_{\mu}(v)[H_{\mu}(X_2,Y|v)-H_{\mu}(X_1,X_2,Y|v)+H_{\mu}(X_1,X_2|v)-H_{\mu}(X_2|v)]\\
r_{m+1}(\mu) &= \int_{v}d_{\mu}(v)[H_{\mu}(X_1,X_2,S|v)-H_{\mu}(X_1,X_2,Y|v)+H_{\mu}(Y)-H_{\mu}(S)]\\
r_{m+2}(\mu) &= \int_{v}d_{\mu}(v)[H_{\mu}(X_2,S|v)-H_{\mu}(X_2,Y|v)+H_{\mu}(Y)-H_{\mu}(S)]
\end{align}
\label{functionals-bounds-auxiliary-random-variables-inner-and-temporary-outer-bounds-strictly-causal-states-setting}
\end{subequations}
can be preserved with another measure $\mu' \in \mc P^{\text{out}}_{\text{s-c}}$. Observing that  there is a total of $\Big(|\mc S||\mc X_1||\mc X_2|+2\Big)$ functionals in \eqref{functionals-bounds-auxiliary-random-variables-inner-and-temporary-outer-bounds-strictly-causal-states-setting}, this is ensured by a standard application of the support lemma; and this shows that the alphabet of the auxiliary random variable $V$ can be restricted as indicated in \eqref{bounds-auxiliary-random-variables-inner-and-temporary-outer-bounds-strictly-causal-states-setting} without altering the region $\mc R^{\text{out}}_{\text{s-c}}$.

\renewcommand{\theequation}{G-\arabic{equation}}
\setcounter{equation}{0}  
\subsection{Analysis of Example~\ref{example-comparison-of-inner-and-outer-bounds}}\label{appendix-analysis-example-comparison-of-inner-and-outer-bounds}

First observe that for a given measure $p_{(S,X_1,X_2)}(s,x_1,x_2)$ of the form $p_{(S,X_1,X_2)}(s,x_1,x_2)=p_S(s)p_{(X_1,X_2)}(x_1,x_2)$, i.e., with arbitrary joint $p_{(X_1,X_2)}(x_1,x_2)$, we have
\begin{subequations}
\begin{align}
H(S_{X_1+X_2}|X_1,X_2) &= \mathbb{E}_{(X_1,X_2)} \big[H(S_{X_1+X_2}|X_1=x_1,X_2=x_2)\big]\\
&= p_{X_1,X_2}(0,0)H(S_0|X_1=0,X_2=0) + p_{X_1,X_2}(1,1)H(S_0|X_1=1,X_2=1) \nonumber\\ 
& + p_{X_1,X_2}(1,0)H(S_1|X_1=1,X_2=0) + p_{X_1,X_2}(0,1)H(S_1|X_1=0,X_2=1)\\
\label{preliminary-term1-proof-example-comparison-of-inner-and-outer-bounds-step1}
&= H(S_0)[p_{X_1,X_2}(0,0) + p_{X_1,X_2}(1,1)] + H(S_1)[p_{X_1,X_2}(1,0) + p_{X_1,X_2}(0,1)]\\
&= \text{Pr}\{X_1 = X_2\}H(S_0) + \text{Pr}\{X_1 \neq X_2\}H(S_1) \\
\label{preliminary-term1-proof-example-comparison-of-inner-and-outer-bounds-step2}
&= H(S_0)\\
\label{preliminary-term1-proof-example-comparison-of-inner-and-outer-bounds-step3}
&= \frac{1}{2}
\end{align} 
\label{preliminary-term1-proof-example-comparison-of-inner-and-outer-bounds}
\end{subequations}
where \eqref{preliminary-term1-proof-example-comparison-of-inner-and-outer-bounds-step1} holds since $S_0$ and $S_1$ are independent of the events $\{X_1=i,X_2=j\}$, $(i,j) \in \{1,2\}^2$, \eqref{preliminary-term1-proof-example-comparison-of-inner-and-outer-bounds-step2} holds since $S=(S_0,S_1)$ is such that $H(S_0)=H(S_1)$, and \eqref{preliminary-term1-proof-example-comparison-of-inner-and-outer-bounds-step3} holds since $H(S_0)=1/2$.

\noindent Similarly, we have
\begin{subequations}
\begin{align}
& H(S|X_1,X_2,S_{X_1+X_2}) = \mathbb{E}_{(X_1,X_2)} \big[H(S_0,S_1|X_1=x_1,X_2=x_2,S_{x_1+x_2})\big]\\
&= p_{X_1,X_2}(0,0)H(S_1|X_1=0,X_2=0,S_0) + p_{X_1,X_2}(1,1)H(S_1|X_1=1,X_2=1,S_0) \nonumber\\
&+  p_{X_1,X_2}(1,0)H(S_0|X_1=1,X_2=0,S_1) + p_{X_1,X_2}(0,1)H(S_0|X_1=0,X_2=1,S_1)\\
&= p_{X_1,X_2}(0,0)H(S_1) + p_{X_1,X_2}(1,1)H(S_1) + p_{X_1,X_2}(1,0)H(S_0) + p_{X_1,X_2}(0,1)H(S_0)\\
&= \text{Pr}\{X_1 \neq X_2\}H(S_0) + \text{Pr}\{X_1 = X_2\}H(S_1)\\
&= H(S_0)\\
&= \frac{1}{2}.
\end{align}
\label{preliminary-term2-proof-example-comparison-of-inner-and-outer-bounds}
\end{subequations}

We first prove that $(R_c,R_1)=(1/2,1) \in  \mc R^{\text{out}}_{\text{s-c}}$. This can be seen by setting in \eqref{outer-bound-strictly-causal-states-setting} $V=S_{X_1+X_2}$ and the inputs $X_1$ and $X_2$ to be i.i.d with $X_1 \sim \: \text{Bernoulli}\:(1/2)$ and $X_2 \sim \: \text{Bernoulli}\:(1/2)$. More specifically, it is easy to show that with this choice we have
\begin{align}
\label{intermediary-term1-proof-example-comparison-of-inner-and-outer-bounds}
H(X_2,X_1+S_{X_1+X_2}) &= 2 \\
\label{intermediary-term2-proof-example-comparison-of-inner-and-outer-bounds}
H(X_1|S_{X_1+X_2},X_2) &= 1
\end{align}
and, so, we also have
\begin{subequations}
\begin{align}
I(X_1,X_2;Y) &= H(Y) - H(Y|X_1,X_2)\\
             &= H(X_2,X_1+S_{X_1+X_2}) - H(S_{X_1+X_2}|X_1,X_2)\\
\label{intermediary-term3-proof-example-comparison-of-inner-and-outer-bounds-step1}
             &= 2 - \frac{1}{2}\\
\label{intermediary-term3-proof-example-comparison-of-inner-and-outer-bounds-step2}
             &= \frac{3}{2}
\end{align}
\label{intermediary-term3-proof-example-comparison-of-inner-and-outer-bounds}
\end{subequations}
where \eqref{intermediary-term3-proof-example-comparison-of-inner-and-outer-bounds-step1} follows  by using \eqref{preliminary-term1-proof-example-comparison-of-inner-and-outer-bounds} and \eqref{intermediary-term1-proof-example-comparison-of-inner-and-outer-bounds}.

Thus, we have
\begin{subequations}
\begin{align}
I(X_1;Y|V,X_2) &= I(X_1;Y_1,X_2|S_{X_1+X_2},X_2) \\
\label{term1-proof-example-comparison-of-inner-and-outer-bounds-step1}
               &= I(X_1;X_1|S_{X_1+X_2},X_2) \\
               &= H(X_1|S_{X_1+X_2},X_2)\\
\label{term1-proof-example-comparison-of-inner-and-outer-bounds-step2}
	       &= 1
\end{align}
\label{term1-proof-example-comparison-of-inner-and-outer-bounds}
\end{subequations}
where \eqref{term1-proof-example-comparison-of-inner-and-outer-bounds-step1} follows by setting $V=S_{X_1+X_2}$, and \eqref{term1-proof-example-comparison-of-inner-and-outer-bounds-step2} follows by \eqref{intermediary-term2-proof-example-comparison-of-inner-and-outer-bounds}.

Similarly, we have
\begin{subequations}
\begin{align}
I(V,X_1,X_2;Y)-I(V,X_1,X_2;S) &= I(X_1,X_2;Y) - I(V;S|X_1,X_2,Y)\\
\label{term2-proof-example-comparison-of-inner-and-outer-bounds-step1}
                              &= I(X_1,X_2;Y) - I(S_{X_1+X_2};S|X_1,X_2,Y_1)\\
\label{term2-proof-example-comparison-of-inner-and-outer-bounds-step2}
&=  I(X_1,X_2;Y) - I(S_{X_1+X_2};S|X_1,X_2,S_{X_1+X_2},Y_1)\\ 
\label{term2-proof-example-comparison-of-inner-and-outer-bounds-step3}
&=  I(X_1,X_2;Y)\\
&= \frac{3}{2} 
\end{align}
\label{term2-proof-example-comparison-of-inner-and-outer-bounds}
\end{subequations}
where \eqref{term2-proof-example-comparison-of-inner-and-outer-bounds-step1} follows by setting $V=S_{X_1+X_2}$, \eqref{term2-proof-example-comparison-of-inner-and-outer-bounds-step2} follows since $V=S_{X_1+X_2}$ is a deterministic function of $(X_1,Y_1)$, and \eqref{term2-proof-example-comparison-of-inner-and-outer-bounds-step3} follows by \eqref{intermediary-term3-proof-example-comparison-of-inner-and-outer-bounds}.

\noindent It remains to show that the constraint \eqref{nonnegativity-constraint-temporary-outer-bound} is satisfied with the choice $V=S_{X_1+X_2}$ and the inputs $X_1$ and $X_2$ to be i.i.d. with $X_1 \sim \: \text{Bernoulli}\:(1/2)$ and $X_2 \sim \: \text{Bernoulli}\:(1/2)$. This can be seen as follows

\begin{subequations}
\begin{align}
I(V,X_2;Y) - I(V,X_2;S) &= I(S_{X_1+X_2},X_2;Y) - I(S_{X_1+X_2},X_2;S)\\
                        &= H(X_2,X_1+S_{X_1+X_2}) - H(X_2,X_1+S_{X_1+X_2}|X_2,S_{X_1+X_2}) - H(S) + H(S|X_2,S_{X_1+X_2}) \\
\label{proof-example-nonnegativity-constraint-outer-bound-step1}
			&= H(X_2,X_1+S_{X_1+X_2}) - H(X_1|X_2,S_{X_1+X_2}) - H(S) + H(S|X_2,S_{X_1+X_2}) \\
\label{proof-example-nonnegativity-constraint-outer-bound-step2}
			&= H(X_2) + H(X_1+S_{X_1+X_2}) - H(X_1) - H(S) + H(S|X_2,S_{X_1+X_2}) \\
\label{proof-example-nonnegativity-constraint-outer-bound-step3}
			&= H(X_1+S_{X_1+X_2}) + H(S|X_2,S_{X_1+X_2}) - 1 \\ 
\label{proof-example-nonnegativity-constraint-outer-bound-step4}
			&= 1 + \frac{1}{2} - 1 \\ 
			&= \frac{1}{2}\\
			&\geq 0
\end{align}
\end{subequations}
where \eqref{proof-example-nonnegativity-constraint-outer-bound-step2} follows since $X_1$ and $X_2$ are independent of each other and independent of $(S,S_{X_1+X_2})$; \eqref{proof-example-nonnegativity-constraint-outer-bound-step3} follows by substituting $H(X_1)=1$, $H(X_2)=1$ and $H(S)=1$; \eqref{proof-example-nonnegativity-constraint-outer-bound-step4} follows by straightforward algebra to obtain and then substitute using $H(X_1+S_{X_1+X_2})=1$ and $H(S|X_2,S_{X_1+X_2})=1/2$. 

\noindent The above shows that the outer bound $\mc R^{\text{out}}_{\text{s-c}}$ contains the rate pair $(R_c,R_1)=(1/2,1)$.

We now turn to proving that $(R_c,R_1)=(1/2,1) \notin  \mc R^{\text{in}}_{\text{s-c}}$. Fix a distribution $P_{S,V,X_1,X_2,Y}$ of the form \eqref{measure-inner-bound-strictly-causal-states-setting}; and, assume that $(R_c,R_2) \in \mc R^{\text{in}}_{\text{s-c}}$ with $R_1=1$. We will show that $R_c$ must be zero. 

First, note that, by $R_1=1$ and  \eqref{inner-bound-strictly-causal-states-setting-individual-rate}, $X_1$ is not deterministic, i.e., $p_{X_1}(x_1) > 0$ for all $x_1 \in \{0,1\}$. Also, it can be seen easily that if $X_2$ is deterministic then one immediately gets $R_c+R_1 \leq H(X_1) \leq 1$ from \eqref{inner-bound-strictly-causal-states-setting-individual-rate}, where the last inequality follows since $\mc X_1$ is binary. Therefore, in the rest of this proof we assume that both $X_1$ and $X_2$ are not deterministic. 

First consider \eqref{inner-bound-strictly-causal-states-setting-sum-rate}. We get
\begin{subequations}
\begin{align}
R_c+R_1 &\leq I(V,X_1,X_2;Y) - I(V,X_1,X_2;S) \\
\label{term3-proof-example-comparison-of-inner-and-outer-bounds-step1}
&= I(X_1,X_2;Y) - I(V;S|X_1,X_2,Y) \\
\label{term3-proof-example-comparison-of-inner-and-outer-bounds-step2}
&= H(Y_1,Y_2)-H(X_1+S_{X_1+X_2}|X_1,X_2) - I(V;S|X_1,X_2,Y_1) \\
&= H(Y_1,Y_2)-H(S_{X_1+X_2}|X_1,X_2) - H(S|X_1,X_2,Y_1) + H(S|X_1,X_2,Y_1,V) \\
\label{term3-proof-example-comparison-of-inner-and-outer-bounds-step3}
&= H(Y_1,Y_2)-H(S_{X_1+X_2}|X_1,X_2) - H(S|X_1,X_2,Y_1,S_{X_1+X_2}) + H(S|X_1,X_2,Y_1,V,S_{X_1+X_2}) \\
\label{term3-proof-example-comparison-of-inner-and-outer-bounds-step4}
&= H(Y_1,Y_2)-H(S_{X_1+X_2}|X_1,X_2) - H(S|X_1,X_2,S_{X_1+X_2}) + H(S|X_1,X_2,V,S_{X_1+X_2}) \\
\label{term3-proof-example-comparison-of-inner-and-outer-bounds-step5}
&\leq 2 - H(S_{X_1+X_2}|X_1,X_2) - H(S|X_1,X_2,S_{X_1+X_2}) + H(S|X_1,X_2,V,S_{X_1+X_2})\\
\label{term3-proof-example-comparison-of-inner-and-outer-bounds-step6}
&\leq 2 - \frac{1}{2} - \frac{1}{2} + H(S|X_1,X_2,V,S_{X_1+X_2})\\
\label{term3-proof-example-comparison-of-inner-and-outer-bounds-step7}
&= 1+ H(S|X_1,X_2,V,S_{X_1+X_2})
\end{align}
\label{term3-proof-example-comparison-of-inner-and-outer-bounds}
\end{subequations}
where \eqref{term3-proof-example-comparison-of-inner-and-outer-bounds-step3} follows since $S_{X_1+X_2}=X_1+Y_1$ is a deterministic function of $X_1$ and $Y_1$, \eqref{term3-proof-example-comparison-of-inner-and-outer-bounds-step4} follows since $Y_1$ is a deterministic function of $X_1$ and $S_{X_1+X_2}$, \eqref{term3-proof-example-comparison-of-inner-and-outer-bounds-step5} follows since the alphabet $\mc Y_1{\times}\mc Y_2$ has four elements, and \eqref{term3-proof-example-comparison-of-inner-and-outer-bounds-step6} follows since $H(S_{X_1+X_2}|X_1,X_2)=1/2$ by \eqref{preliminary-term1-proof-example-comparison-of-inner-and-outer-bounds} and $H(S|X_1,X_2,S_{X_1+X_2})=1/2$ by \eqref{preliminary-term2-proof-example-comparison-of-inner-and-outer-bounds}.

In what follows we will show that the term $H(S|X_1,X_2,V,S_{X_1+X_2})$ on the RHS of \eqref{term3-proof-example-comparison-of-inner-and-outer-bounds-step7} is zero, which together with $R_1=1$ will then imply that $R_c=0$. 

\noindent To this end, consider now \eqref{inner-bound-strictly-causal-states-setting-individual-rate}. Since $R_1=1$ and $\mc X_1$ is binary, we have
\begin{equation}
1 \leq H(X_1|V,X_2) - H(X_1|V,X_2,Y) \leq H(X_1) \leq 1.
\label{term4-proof-example-comparison-of-inner-and-outer-bounds}
\end{equation}
Thus, $H(X_1|V,X_2)=H(X_1)$; and, so $X_1$ is independent of $(V,X_2)$. Since $X_1$ is also independent of $S$, we then have that 
\begin{equation}
 X_1\:\: \text{is independent of}\:\: (S,V,X_2).
\label{term5-proof-example-comparison-of-inner-and-outer-bounds}
\end{equation}

\noindent From \eqref{inner-bound-strictly-causal-states-setting-individual-rate}, we also obtain
\begin{subequations}
\begin{align}
 1 = R_1 &\leq H(Y|V,X_2) - H(Y|V,X_1,X_2)\\
&= H(Y_1|V,X_2) - H(Y_1|V,X_1,X_2) \\
&\leq H(Y_1) - H(Y_1|V,X_1,X_2) \\
&\leq H(Y_1)\\
&\leq 1
\end{align}
\label{term6-proof-example-comparison-of-inner-and-outer-bounds}
\end{subequations}
where the last inequality follows since $\mc Y_1$ is binary. This implies that $H(Y_1)=1$ and $H(Y_1|V,X_1,X_2)=0$. Thus,
\begin{subequations}
\begin{align}
0 &= H(Y_1|V,X_1,X_2)\\
&= H(X_1+S_{X_1+X_2}|V,X_1,X_2)\\
&= H(S_{X_1+X_2}|V,X_1,X_2).
\end{align}
\label{term7-proof-example-comparison-of-inner-and-outer-bounds}
\end{subequations}

\noindent The joint distribution of $X_1$ and $X_2$ satisfies
\begin{equation}
p_{X_1,X_2}(x_1,x_2)=p_{X_1}(x_1)p_{X_2}(x_2) > 0, \:\: \forall (x_1,x_2) \in \{0,1\}^2
\label{term8-proof-example-comparison-of-inner-and-outer-bounds}
\end{equation}
where the equality follows by \eqref{term5-proof-example-comparison-of-inner-and-outer-bounds} from the independence of $X_1$ and $X_2$, and the strict positivity follows since both $X_1$ and $X_2$ are assumed to be non deterministic. Next, from \eqref{term7-proof-example-comparison-of-inner-and-outer-bounds}, we get
\begin{subequations}
\begin{align}
0 &= H(S_{X_1+X_2}|V,X_1,X_2)\\
&= \mathbb{E}_{(X_1,X_2)} \big[H(S_{x_1+x_2}|V,X_1=x_1,X_2=x_2)\big]\\
&= p_{X_1,X_2}(0,0)H(S_0|V,X_1=0,X_2=0) + p_{X_1,X_2}(1,1)H(S_0|V,X_1=1,X_2=1) \nonumber\\
&\:\: + p_{X_1,X_2}(1,0)H(S_1|V,X_1=1,X_2=0) + p_{X_1,X_2}(0,1)H(S_1|V,X_1=0,X_2=1) \\
\label{term3-proof-example-comparison-of-inner-and-outer-bounds-step-step1}
&= p_{X_1,X_2}(0,0)H(S_0|V,X_2=0) + p_{X_1,X_2}(1,1)H(S_0|V,X_2=1) \nonumber\\
&\:\: + p_{X_1,X_2}(1,0)H(S_1|V,X_2=0) + p_{X_1,X_2}(0,1)H(S_1|V,X_2=1) 
\end{align}
\label{term9-proof-example-comparison-of-inner-and-outer-bounds}
\end{subequations}
where the last equality follows since, by \eqref{measure-inner-bound-strictly-causal-states-setting}, $X_1 \leftrightarrow (V,X_2) \leftrightarrow S$ is a Markov chain.

\noindent From \eqref{term9-proof-example-comparison-of-inner-and-outer-bounds}, and the fact that, by \eqref{term8-proof-example-comparison-of-inner-and-outer-bounds}, $p_{X_1,X_2}(x_1,x_2) > 0$ for all $(x_1,x_2) \in \{0,1\}^2$, we get that all the conditional entropy terms on the RHS of \eqref{term9-proof-example-comparison-of-inner-and-outer-bounds} are zero,
\begin{subequations}
\begin{align}
H(S_0|V,X_2=0)=0, &\qquad H(S_0|V,X_2=1)=0 \\
H(S_1|V,X_2=0)=0, &\qquad H(S_1|V,X_2=1)=0.
\end{align}
\label{term10-proof-example-comparison-of-inner-and-outer-bounds}
\end{subequations}

Consider now the term $H(S|X_1,X_2,V,S_{X_1+X_2})$ on the RHS of \eqref{term3-proof-example-comparison-of-inner-and-outer-bounds-step7}. We have
\begin{subequations}
\begin{align}
0 \leq H(S|X_1,X_2,V,S_{X_1+X_2}) &= \mathbb{E}_{(X_1,X_2)} \big[H(S|X_1=x_1,X_2=x_2,V,S_{x_1+x_2})\big]\\
&= p_{X_1,X_2}(0,0)H(S|X_1=0,X_2=0,V,S_0) + p_{X_1,X_2}(1,1)H(S|X_1=1,X_2=1,V,S_0) \nonumber\\
\label{term11-proof-example-comparison-of-inner-and-outer-bounds-step1}
&\:\: + p_{X_1,X_2}(0,1)H(S|X_1=0,X_2=1,V,S_1) + p_{X_1,X_2}(1,0)H(S|X_1=1,X_2=0,V,S_1) \\
&= p_{X_1,X_2}(0,0)H(S_1|X_1=0,X_2=0,V,S_0) + p_{X_1,X_2}(1,1)H(S_1|X_1=1,X_2=1,V,S_0) \nonumber\\
\label{term11-proof-example-comparison-of-inner-and-outer-bounds-step2}
&\:\: + p_{X_1,X_2}(0,1)H(S_0|X_1=0,X_2=1,V,S_1) + p_{X_1,X_2}(1,0)H(S_0|X_1=1,X_2=0,V,S_1) \\
&\leq p_{X_1,X_2}(0,0)H(S_1|X_1=0,X_2=0,V) + p_{X_1,X_2}(1,1)H(S_1|X_1=1,X_2=1,V) \nonumber\\
\label{term11-proof-example-comparison-of-inner-and-outer-bounds-step3}
&\:\: + p_{X_1,X_2}(0,1)H(S_0|X_1=0,X_2=1,V) + p_{X_1,X_2}(1,0)H(S_0|X_1=1,X_2=0,V) \\
&= p_{X_1,X_2}(0,0)H(S_1|X_2=0,V) + p_{X_1,X_2}(1,1)H(S_1|X_2=1,V) \nonumber\\
\label{term11-proof-example-comparison-of-inner-and-outer-bounds-step4}
&\:\: + p_{X_1,X_2}(0,1)H(S_0|X_2=1,V) + p_{X_1,X_2}(1,0)H(S_0|X_2=0,V) \\
\label{term11-proof-example-comparison-of-inner-and-outer-bounds-step5}
&= 0
\end{align}
\label{term11-proof-example-comparison-of-inner-and-outer-bounds}
\end{subequations}
where \eqref{term11-proof-example-comparison-of-inner-and-outer-bounds-step2} follows by substituting $S=(S_0,S_1)$, \eqref{term11-proof-example-comparison-of-inner-and-outer-bounds-step3} follows since conditioning reduces entropy, \eqref{term11-proof-example-comparison-of-inner-and-outer-bounds-step4} follows since by \eqref{measure-inner-bound-strictly-causal-states-setting}, $X_1 \leftrightarrow (V,X_2) \leftrightarrow S$ is a Markov chain, and \eqref{term11-proof-example-comparison-of-inner-and-outer-bounds-step5} follows by substituting using \eqref{term10-proof-example-comparison-of-inner-and-outer-bounds}.

Finally, combining \eqref{term3-proof-example-comparison-of-inner-and-outer-bounds-step7} and \eqref{term11-proof-example-comparison-of-inner-and-outer-bounds-step5}, we get that
\begin{equation}
0 \leq R_c+R_1 \leq 1
\label{term12-proof-example-comparison-of-inner-and-outer-bounds}
\end{equation}
which, together with $R_1=1$, implies that $R_c=0$.

The above shows that the inner bound $\mc R^{\text{in}}_{\text{s-c}}$ does not contain any rate pair of the form $(R_c,R_1=1)$ with $R_c > 0$; and, so, in particular  $(R_c,R_1)=(1/2,1) \notin \mc R^{\text{in}}_{\text{s-c}}$.

\renewcommand{\theequation}{H-\arabic{equation}}
\setcounter{equation}{0}  
\subsection{Proof of Proposition~\ref{proposition-strictly-causal-states-at-only-strong-encoder}}\label{appendix-proof-proposition-strictly-causal-states-at-only-strong-encoder}

We show that the capacity region of the state-dependent MAC with strictly causal states known only at the encoder that sends both messages is given by the set of all rate pairs $(R_c,R_1)$ such that
\begin{subequations}
\begin{align}
\label{ineq1-capacity-region-mac-without-states}
R_1 \: &\leq \: I(X_1;Y|X_2)\\
R_c+ R_1 \: &\leq \: I(X_1,X_2;Y)
\label{ineq2-capacity-region-mac-without-states}
\end{align}
\label{capacity-region-mac-without-states}
\end{subequations}
for some measure of the form
\begin{equation}
P_{S,X_1,X_2,Y} = Q_SP_{X_2}P_{X_1|X_2}W_{Y|X_1,X_2,S}.
\label{measure-capacity-region-mac-without-states}
\end{equation}

i) The region described by \eqref{capacity-region-mac-without-states} is the capacity region of the same MAC model without states; and, so, it is also achievable in the presence of (strictly causal) states, as these states can always be ignored by the transmitters.

ii) The proof of the converse is as follows.

\noindent The bound \eqref{ineq2-capacity-region-mac-without-states} on the sum rate $(R_c+R_1)$ follows by using the result of Proposition~\ref{proposition-maximum-sum-rate-strictly-causal-states-setting}. 

\noindent The bound \eqref{ineq1-capacity-region-mac-without-states} on the individual rate $R_1$ follows as follows.

{\allowdisplaybreaks
\begin{align}
 nR_1 &\leq H(W_1) \nonumber\\
      &= H(W_1|W_c) \nonumber\\
      &= I(W_1;Y^n|W_c)+H(W_1|W_c,Y^n)\nonumber\\
      &\leq I(W_1;Y^n|W_c)+n\epsilon_n\nonumber\\
      &= \sum_{i=1}^{n} I(W_1;Y_i|W_c,Y^{i-1})+n\epsilon_n\nonumber\\
      &= \sum_{i=1}^{n} H(Y_i|W_c,Y^{i-1})- H(Y_i|W_c,W_1,Y^{i-1})+n\epsilon_n\nonumber\\
      &\stackrel{(a)}{=} \sum_{i=1}^{n} H(Y_i|W_c,Y^{i-1},X_{2i})- H(Y_i|W_c,W_1,Y^{i-1},X_{2i})+n\epsilon_n\nonumber\\
      &\stackrel{(b)}{\leq} \sum_{i=1}^{n} H(Y_i|X_{2i})- H(Y_i|W_c,W_1,Y^{i-1},X_{1i},X_{2i})+n\epsilon_n\nonumber\\
      &\stackrel{(c)}{\leq} \sum_{i=1}^{n} H(Y_i|X_{2i})- H(Y_i|X_{1i},X_{2i})+n\epsilon_n\nonumber\\
      &= \sum_{i=1}^{n} I(X_{1i};Y_i|X_{2i})+n\epsilon_n
\end{align}
where $(a)$ follows by the fact that $X_{1i}$ is a deterministic function of $W_c$ for the model in which the states are known, strictly causally, at only the encoder that sends both messages; $(b)$ follows since conditioning reduces the entropy; and $(c)$  follows from the fact that  $(W_c,W_1,Y^{i-1},S^{i-1}) \leftrightarrow (X_{1i},X_{2i},S_i) \leftrightarrow Y_i$ and $(W_c,W_1,Y^{i-1},S^{i-1},X_{1i},X_{2i})$ is independent of $S_i$.

\noindent The rest of the proof of Proposition~\ref{proposition-strictly-causal-states-at-only-strong-encoder} follows by standard single-letterization.

\renewcommand{\theequation}{I-\arabic{equation}}
\setcounter{equation}{0}  
\subsection{Analysis of Example~\ref{example-only-state-is-deterministic-symmetric-state-case}}\label{appendix-analysis-example-only-state-is-deterministic-symmetric-state-case}

In this section, we use the result of Theorem~\ref{theorem-capacity-region-special-case-strictly-causal-states-setting} to show that the capacity region of the model of Example~\ref{example-only-state-is-deterministic-symmetric-state-case} is given by \eqref{capacity-region-example-only-state-is-deterministic-symmetric-state-case}.

1) Fix a joint distribution of $(X_1,X_2,S,Y)$ of the form \eqref{measure-capacity-region-special-case-strictly-causal-states-setting} and satisfying
\begin{equation}
\mathbb{E}[X^2_1]=\tilde{P}_1 \leq P_1, \quad \mathbb{E}[X^2_2]=\tilde{P}_2 \leq P_2, \quad \mathbb{E}[X_1X_2]=\sigma_{12}
\label{FixedSecondMomentsOuterBoundFullDuplexGaussianCase}
\end{equation}

We shall also use the correlation coefficient $\rho_{12}$ defined as
\begin{equation}
\rho_{12}=\frac{\sigma_{12}}{\sqrt{\tilde{P}_1\tilde{P}_2}}.
\end{equation}

We first compute the RHS of the bound on the sum rate. 
\begin{align}
I(X_1,X_2;Y_1,Y_2) &= h(Y_1,Y_2)-h(Y_1,Y_2|X_1,X_2) \\
&= h(Y_1,Y_2)-h(S,Z|X_1,X_2) \\
&\stackrel{(a)}{=} h(Y_1,Y_2) - h(S) - h(Z)\\
&= h(Y_1|X_2+Z) + h(X_2+Z) - h(S) - h(Z)\\
& \stackrel{(b)}{\leq} \frac{1}{2}\log\frac{\big|\mathbb{E}[Y_1Y^T_1]-\mathbb{E}[Y_1\mathbb{E}[Y_1|X_2+Z]]\big|}{Q} + \frac{1}{2}\log\big(1+\frac{\tilde{P}_2}{N}\big)\\
& \stackrel{(c)}{=} \frac{1}{2}\log\Big(1+\frac{\tilde{P}_1+\tilde{P}_2+2\sigma_{12}}{Q}-\frac{(\sigma_{12}+\tilde{P}_2)^2}{Q(\tilde{P}_2+N)}\Big) +  \frac{1}{2}\log\big(1+\frac{\tilde{P}_2}{N}\big)\\
&= \frac{1}{2}\log\Big(1+\frac{(1-\rho^2_{12})\tilde{P}_1\tilde{P}_2+N((\sqrt{\tilde{P}_1}+\rho_{12}\sqrt{\tilde{P}_2})^2+(1-\rho^2_{12})\tilde{P}_2)}{Q(\tilde{P}_2+N)}\Big) +  \frac{1}{2}\log\big(1+\frac{\tilde{P}_2}{N}\big)
\label{capacity-region-example-symmetric-state-case-step1}
\end{align}
where $(a)$ follows since the state $S$ and the noise $Z$ are independent of each other and independent of the channel inputs $X_1$ and $X_2$; $(b)$ follows by the \textit{Maximum Differential Entropy Lemma} \cite[Section 2.2]{GK11}; and $(c)$ follows by straightforward algebra, noticing that the minimum mean square estimator (MMSE) of $Y_1$ given $Y_2=X_2+Z$ is
\begin{equation}
\mathbb{E}[Y_1|X_2+Z] = \frac{\sigma_{12}+\tilde{P}_2}{\tilde{P}_2+N}(X_2+Z).
\end{equation}

\noindent It is easy to see that the RHS of the bound on the individual rate is redundant. 

\noindent For convenience, let us define the function $\Theta(\tilde{P}_1,\tilde{P}_2,\rho_{12})$ as the RHS of \eqref{capacity-region-example-symmetric-state-case-step1}. The above shows that the capacity region of the model of Example~\ref{example-only-state-is-deterministic-symmetric-state-case} is outer-bounded by the set of pairs $(R_c,R_1)$ satisfying
\begin{equation}
0 \leq R_c + R_1 \leq \max\: \Theta(\tilde{P}_1,\tilde{P}_2,\rho_{12})
\label{capacity-region-example-symmetric-state-case-step2}
\end{equation}
where the maximization is over $0 \leq \tilde{P}_1$, $0 \leq \tilde{P}_2$ and $-1 \leq \rho_{12} \leq 1$. 

\noindent Investigating $\Theta(\tilde{P}_1,\tilde{P}_2,\rho_{12})$, it can be see that it suffices to consider $\rho_{12} \in [0,1]$ for the maximization in \eqref{capacity-region-example-symmetric-state-case-step2}; and that $\Theta(\tilde{P}_1,\tilde{P}_2,\rho_{12})$ is maximized at $\tilde{P}_1=P_1$ and $\tilde{P}_2=P_2$.

2) As for the proof of achievability, choose in the inner bound of Theorem~\ref{theorem-capacity-region-special-case-strictly-causal-states-setting}, random variables $S$, $X_1$, $X_2$ and $Y$ that are jointly Gaussian with second moments $\mathbb{E}[X^2_1]=\tilde{P}_1$ and $\mathbb{E}[X^2_2]=\tilde{P}_2$, and with $\mathbb{E}[X_1X_2]=\rho_{12}\sqrt{P_1P_2}$. The rest of the proof of the direct part follows by straightforward algebra that is very similar to that for the converse proof above and that we omit for brevity.

This completes the proof of Theorem~\ref{theorem-capacity-region-special-case-strictly-causal-states-setting}.

\renewcommand{\theequation}{J-\arabic{equation}}
\setcounter{equation}{0}  
\subsection{Proof of Theorem~\ref{theorem-inner-bound-asymmetric-strictly-causal-states-setting}}\label{appendix-proof-theorem-inner-bound-asymmetric-strictly-causal-states-setting}

The transmission takes place in $B$ blocks. The common message $W_c$ and the individual message $W_1$ are sent over \textit{all} blocks. We thus have $B_{W_c}=nB{R_c}$, $B_{W_1}=nB{R_1}$, $N=nB$, $R_{W_c}=B_{W_c}/N=R_c$ and $R_{W_1}=B_{W_1}/N=R_1$, where $B_{W_c}$ is the number of common message bits, $B_{W_1}$ is the number of individual message bits, $N$ is the number of channel uses and $R_{W_c}$ and $R_{W_1}$ are the overall rates of the common and individual messages, respectively.

\noindent \textbf{Codebook Generation:} Fix a measure $P_{S,U,V,X_1,X_2,Y} \in {\mc P}^{\text{in}}_{\text{asym,s-c}}$. Fix $\epsilon > 0$, $\eta_c > 0$, $\eta_1 > 0$, $\hat{\eta} > 0$, $\delta > 1$ and denote $M_c = 2^{nB[R_c-\eta_c\epsilon]}$, $M_1 = 2^{nB[R_1-\eta_1\epsilon]}$, and $\hat{M} = 2^{n[\hat{R}+\hat{\eta}\epsilon]}$. Also, let $\eta_{c1} > 0$, $\eta_{c2} > 0$ and $M_{c1}=2^{nB[R_{c1}-\eta_{c1}\epsilon]}$ and  $M_{c2}=2^{nB[R_{c2}-\eta_{c2}\epsilon]}$ such that $R_c=R_{c1}+R_{c2}$.

\noindent We randomly and independently generate a codebook for each block.

\begin{itemize}
\item[1)] For each block $i$, $i=1,\hdots,B$, we generate $M_{c1}$ independent and identically distributed (i.i.d.) codewords $\dv u_i(w_{c1})$ indexed by $w_{c1}=1,\hdots,M_{c1}$, each with i.i.d. components drawn according to $P_{U}$.
\item[2)] For each block $i$, for each codeword $\dv u_i(w_{c1})$, we generate $M_{c2}\hat{M}$ independent and identically distributed (i.i.d.) codewords $\dv x_{2,i}(w_{c1},w_{c2},t'_i)$ indexed by $w_{c2}=1,\hdots,M_{c2}$, $t'_i=1,\hdots,\hat{M}$, each with i.i.d. components drawn according to $P_{X_2|U}$.
\item[3)] For each block $i$, for each pair of codewords $(\dv u_i(w_{c1}), \dv x_{2,i}(w_{c1},w_{c2},t'_i))$,  we generate $\hat{M}$ i.i.d. codewords $\dv v_i(w_{c1},w_{c2},t'_i,t_i)$ indexed by $t_i=1,\hdots,\hat{M}$, each with i.i.d. components drawn according to $P_{V|U,X_2}$.
\item[4)] For each block $i$, for each codeword $\dv u_i(w_{c1})$, we generate $M_1$ independent and identically distributed (i.i.d.) codewords $\dv x_{1,i}(w_{c1},w_1)$ indexed by $w_1=1,\hdots,M_1$, each with i.i.d. components drawn according to $P_{X_1|U}$.
\end{itemize}

\textbf{Encoding:} Suppose that a common message $W_c=w_c=(w_{c1},w_{c2})$ and an individual message $W_1=w_1$ are to be transmitted. As we mentioned previously, $w_c$ and $w_1$ will be sent over \textit{all} blocks. We denote by $\dv s[i]$ the state affecting the channel in block $i$, $i=1,\hdots,B$. For convenience, we let $\dv s[0]=\emptyset$ and $t_{-1}=t_0=1$ (a default value). The encoding at the beginning of block $i$, $i=1,\hdots,B$, is as follows.

\noindent Encoder $2$, which has learned the state sequence $\dv s[i-1]$, knows $t_{i-2}$ and looks for a compression index $t_{i-1} \in [1:\hat{M}]$ such that $\dv v_{i-1}(w_{c1},w_{c2},t_{i-2},t_{i-1})$ is strongly jointly typical with $\dv s[i-1]$, $\dv u_{i-1}(w_{c1})$ and $\dv x_{2,i-1}(w_{c1},w_{c2},t_{i-2})$. If there is no such index or the observed state $\dv s[i-1]$ is not typical, $t_{i-1}$ is set to $1$ and an error is declared. If there is more than one such index $t_{i-1}$, choose the smallest. It can be shown that the error in this step has vanishing probability as long as $n$ and $B$ are large and
\begin{equation} 
\hat{R} > I(V;S|U,X_2).
\end{equation}

\noindent Encoder 2 then transmits the vector $\dv x_{2,i}(w_{c1},w_{c2},t_{i-1})$. Encoder 1 transmits the vector $\dv x_{1,i}(w_{c1},w_1)$.

\textbf{Decoding:} At the end of the transmission, the decoder has collected all the blocks of channel outputs $\dv y[1],\hdots,\dv y[B]$.

\noindent \underline{\textit{Step (a):}} The decoder estimates message $w_c=(w_{c1},w_{c2})$ using \text{all} blocks $i=1,\hdots,B$, i.e., simultaneous decoding. It declares that $\hat{w}_c=(\hat{w}_{c1},\hat{w}_{c2})$ is sent if there exist $t^B=(t_1,\hdots,t_B) \in [1:\hat{M}]^{B}$ and $w_1 \in [1:M_1]$ such that $\dv u_i(\hat{w}_{c1})$, $\dv x_{2,i}(\hat{w}_{c1},\hat{w}_{c2},t_{i-1})$, $\dv v_i(\hat{w}_{c1},\hat{w}_{c2},t_{i-1},t_i)$, $\dv x_{1,i}(\hat{w}_{c1},w_1)$ and $\dv y[i]$ are jointly typical for all $i=1,\hdots,B$. One can show that the decoder obtains the correct $w_c=(w_{c1},w_{c2})$ as long as $n$ and $B$ are large and
\begin{align}
R_{c2} + R_1 &\leq I(V,X_1,X_2;Y|U)-\hat{R}\\
R_{c1} + R_{c2} + R_1 &\leq I(U,V,X_1,X_2;Y)-\hat{R}.
\end{align}

\noindent \underline{\textit{Step (b):}} Next, the decoder  estimates message $w_1$ using again \text{all} blocks $i=1,\hdots,B$, i.e., simultaneous decoding. It declares that $\hat{w}_1$ is sent if there exist $t^B=(t_1,\hdots,t_B) \in [1:\hat{M}]^{B}$ such that $\dv u_i(\hat{w}_{c1})$, $\dv x_{2,i}(\hat{w}_{c1},\hat{w}_{c2},t_{i-1})$, $\dv v_i(\hat{w}_{c1},\hat{w}_{c2},t_{i-1},t_i)$, $\dv x_{1,i}(\hat{w}_{c1},w_1)$ and $\dv y[i]$ are jointly typical for all $i=1,\hdots,B$. One can show that the decoder obtains the correct $w_1$ as long as $n$ and $B$ are large and
\begin{align}
R_1 &\leq I(X_1;Y|U,V,X_2)\\
R_1 &\leq I(V,X_1,X_2;Y|U)-\hat{R}.
\end{align}

\textbf{Probability of error analysis:} We examine the probability of error associated with each of the encoding and decoding procedures. The events $E_1$, $E_2$ and $E_3$ correspond to encoding errors, and the events $E_4$, $E_5$, $E_6$ and $E_7$ correspond to decoding errors. To bound the probability of error, we assume without loss of generality that the messages equal to unity, i.e., $w_{c1}=w_{c2}=w_1=1$; and, except for the anlysis of the event $E_1$, we also assume that the compression indices are all equal unity, i.e., $t_1=t_2=\hdots=t_B=1$.

\begin{itemize}
\item Let $E_1=\cup_{i=1}^{B}E_{1i}$ where $E_{1i}$ is the event that, for the encoding in block $i$, there is no covering codeword $\dv v_{i-1}(1,1,t_{i-2},t_{i-1})$ strongly jointly typical with $\dv s[i-1]$ given $\dv u_{i-1}(1)$ and $\dv x_{2,i-1}(1,1,t_{i-2})$, i.e.,
\begin{align}
E_1 &= \bigcup_{i=1}^{B} \Big\{\nexists \:\:t_{i-1} \in [1:\hat{M}] \:\: \text{s.t.:}\:\: \Big(\dv v_{i-1}(1,1,t_{i-2},t_{i-1}), \dv s[i-1], \dv u_{i-1}(1), \dv x_{2,i-1}(1,1,t_{i-2})\Big) \in \mc T_{\epsilon}^n(P_{V,S,U,X_2})\Big\}.
\end{align}
\noindent For $i \in [1:B]$, the probability that $(\dv s[i-1], \dv u_{i-1}(1), \dv x_{2,i-1}(1,1,t_{i-2}))$ is not jointly typical goes to zero as $n \rightarrow \infty$, by the asymptotic equipartition property (AEP) \cite[p. 384]{CT91}. Then, for $(\dv s[i-1], \dv u_{i-1}(1), \dv x_{2,i-1}(1,1,t_{i-2}))$ jointly typical, the covering lemma \cite[Lecture Note 3]{GK11} ensures that the probability that there is no $t_{i-1} \in [1:\hat{M}]$ such that $(\dv v_{i-1}(1,1,t_{i-2},t_{i-1}),\dv s[i-1])$ is strongly jointly typical given $\dv u_{i-1}(1)$ and $\dv x_{2,i-1}(1,1,t_{i-2})$ is exponentially small for large $n$ provided that the number of covering codewords $\dv v_{i-1}$ is greater than $2^{nI(V;S|U,X_2)}$, i.e.,
\begin{equation}
\hat{R} > I(V;S|U,X_2).
\label{Condition__on__CoveringRate}
\end{equation}
Thus, if  \eqref{Condition__on__CoveringRate} holds, $\text{Pr}(E_{1i}) \rightarrow 0 \quad \text{as}\quad  n \rightarrow \infty$ and, so, by the union of bound over the $B$ blocks, $\text{Pr}(E_{1}) \rightarrow 0 \quad \text{as}\quad  n \rightarrow \infty$.
\item For the decoding of the common message $w_c=(1,1)$ at the receiver, let $E_2=\cup_{i=1}^{B}E_{2i}$ where $E_{4i}$ is the event that $\Big(\dv u_{i-1}(1), \dv x_{2,i}(1,1,1)$, $\dv v_i(1,1,1,1)$, $x_{1,i}(1,1)$, $\dv y[i]\Big)$ is not jointly typical, i.e.,
\begin{align}
E_2 &= \bigcup_{i=1}^{B} \Big\{\Big(\dv u_{i-1}(1), \dv x_{2,i}(1,1,1), \dv v_i(1,1,1,1), x_{1,i}(1,1), \dv y[i]\Big) \notin \mc T_{\epsilon}^n(P_{U,X_2,V,X_1,Y})\Big\}.
\end{align}

Conditioned on $E^c_{1i}$, the vectors $\dv s[i]$, $\dv u_{i-1}(1)$, $\dv x_{2,i}(1,1,1)$ and $\dv v_i(1,1,1,1)$ are jointly typical, and with $x_{1,i}(1,1)$. Then, conditioned on $E^c_{1i}$, the vectors $\dv s[i]$, $\dv u_{i-1}(1)$, $\dv x_{2,i}(1,1,1)$ and $\dv v_i(1,1,1,1)$, $x_{1,i}(1,1)$ and $\dv y[i]$ are jointly typical by the Markov lemma \cite[p. 436]{CT91}, i.e., $\text{Pr}(E_{2i}|E^c_{1i}) \rightarrow 0 \quad \text{as}\quad  n \rightarrow \infty$. Thus, by the union bound over the $B$ blocks, $\text{Pr}(E_2|E^c_1) \rightarrow 0 \quad \text{as}\quad  n \rightarrow \infty$.
\item For the decoding of the common message $w_c=(1,1)$ at the receiver, let $E_3$ be the event that $\dv u_i(w_{c1})$, $\dv x_{2,i}(w_{c1},w_{c2},t_{i-1})$, $\dv v_i(w_{c1},w_{c2},t_{i-1},t_i)$, $x_{1,i}(w_{c1},w_1)$ and $\dv y[i]$ are jointly typical for all $i=1,\hdots,B$ and some $w_{c1} \in [1:M_{c1}]$, $w_{c2} \in [1:M_{c2}]$, $w_1 \in [1:M_1]$ and $t^B=(t_1,\hdots,t_B) \in [1:\hat{M}]^{B}$ such that $w_{c1} \neq 1$, i.e.,
\begin{align}
E_3 = \bigg\{ &\: \exists \: w_{c1} \in [1:M_{c1}],\: w_{c2} \in [1:M_{c2}],\: w_1 \in [1:M_1],\: t^B=(t_1,\hdots,t_B) \in [1:\hat{M}]^{B} \:\text{s.t.:} \: w_{c1} \neq 1,\nonumber\\
& \bigcap_{i=1}^{B} \Big\{\Big(\dv u_i(w_{c1}), \dv x_{2,i}(w_{c1},w_{c2},t_{i-1}), \dv v_i(w_{c1},w_{c2},t_{i-1},t_i), \dv x_{1,i}(w_{c1},w_1), \dv y[i]\Big) \in \mc T_{\epsilon}^n(P_{U,X_2,V,X_1,Y})\Big\}\bigg\}.
\end{align}

To bound the probability of the event $E_3$, define the following event for given $w_{c1} \in [1:M_{c1}]$, $w_{c2} \in [1:M_{c2}]$, $w_1 \in [1:M_1]$ and $(t_{i-1},t_i) \in [1:\hat{M}]^2$ such that $w_{c1} \neq 1$,
\begin{align*}
E_{3i}(w_{c1},w_{c2},t_{i-1},t_i,w_1) = \Big\{ &\:\Big(\dv u_i(w_{c1}), \dv x_{2,i}(w_{c1},w_{c2},t_{i-1}), \dv v_i(w_{c1},w_{c2},t_{i-1},t_i), \dv x_{1,i}(w_{c1},w_1), \dv y[i]\Big) \in \mc T_{\epsilon}^n(P_{U,X_2,V,X_1,Y})\Big\}.
\end{align*}
Note that for $w_{c1} \neq 1$ the vectors $\dv u_i(w_{c1})$, $\dv x_{2,i}(w_{c1},w_{c2},t_{i-1})$, $\dv v_i(w_{c1},w_{c2},t_{i-1},t_i)$ and $\dv x_{1,i}(w_{c1},w_1)$ are generated independently of $\dv y[i]$. Hence, by the joint typicality lemma \cite[Lecture Note 2]{GK11}, we get
\begin{align}
\text{Pr}\Big(E_{3i}(w_{c1},w_{c2},t_{i-1},t_i,w_1)|E_1^c,E_2^c\Big) & \leq 2^{-n[I(U,V,X_1,X_2;Y)-\epsilon]}.
\label{BoundingProbabilityIntermediaryEventE3_Step1}
\end{align}

Then, conditioned on the events $E_{1}^c$ and $E_{2}^c$, the probability of the event $E_3$ can be bounded as
\begin{align}
\text{Pr}(E_3|E_1^c,E_2^c) &= \text{Pr}\Big( \bigcup_{w_{c1} \neq 1} \bigcup_{w_{c2} \in [1,M_{c2}]} \bigcup_{w_1 \: \in \: [1:M_1]} \bigcup_{t^B \: \in \: [1:\hat{M}]^{B}} \bigcap_{i=1}^{B} E_{3i}(w_{c1},w_{c2},t_{i-1},t_i,w_1)|E_1^c,E_2^c\Big)\nonumber\\
&\stackrel{(a)}{\leq} \sum_{w_{c1} \neq 1} \sum_{w_{c2} \in [1,M_{c2}]} \sum_{w_1 \in [1:M_1]} \sum_{t^B \: \in \: [1:\hat{M}]^{B}}  \text{Pr}\Big(\bigcap_{i=1}^{B} E_{3i}(w_{c1},w_{c2},t_{i-1},t_i,w_1)|E_1^c,E_2^c\Big)\nonumber\\
&\stackrel{(b)}{=} \sum_{w_{c1} \neq 1} \sum_{w_{c2} \in [1,M_{c2}]} \sum_{w_1 \in [1:M_1]} \sum_{t^B \: \in \: [1:\hat{M}]^{B}}  \prod_{i=1}^{B} \text{Pr}\Big(E_{3i}(w_{c1},w_{c2},t_{i-1},t_i,w_1)|E_1^c,E_2^c\Big)\nonumber\\
&\leq \sum_{w_{c1} \neq 1} \sum_{w_{c2} \in [1,M_{c2}]} \sum_{w_1 \in [1:M_1]} \sum_{t^B \: \in \: [1:\hat{M}]^{B}}  \prod_{i=2}^{B} \text{Pr}\Big(E_{3i}(w_{c1},w_{c2},t_{i-1},t_i,w_1)|E_1^c,E_2^c\Big)\nonumber\\
&\stackrel{(c)}{\leq} \sum_{w_{c1} \neq 1} \sum_{w_{c2} \in [1,M_{c2}]}  \sum_{w_1 \in [1:M_1]} \sum_{t^B \: \in \: [1:\hat{M}]^{B}} \prod_{i=2}^{B} 2^{-n\big[I(U,V,X_1,X_2;Y)-\epsilon\big]}\nonumber\\
&= \sum_{w_{c1} \neq 1} \sum_{w_{c2} \in [1,M_{c2}]} \sum_{w_1 \in [1:M_1]} \sum_{t_B \: \in \: [1:\hat{M}]} 2^{n(B-1)\big[\hat{R}+\hat{\eta}\epsilon\big]}2^{-n(B-1)\big[I(U,V,X_1,X_2;Y)-\epsilon\big]}\nonumber\\
&\leq M_{c1}M_{c2}M_1\hat{M}2^{-n(B-1)\big[I(U,V,X_1,X_2;Y)-\hat{R}-(\hat{\eta}+1)\epsilon\big]}\nonumber\\
&= 2^{-nB\big[\frac{B-1}{B}\big(I(U,V,X_1,X_2;Y)-\hat{R}\big)-(R_{c1}+R_{c2}+R_1)-\frac{\hat{R}}{B}+\big(\eta_{c1}+\eta_{c2}+\eta_1-\hat{\eta}-\frac{B-1}{B}\big)\epsilon\big]}
\label{BoundingProbabilityIntermediaryEventE3_Step2}
\end{align}
where: $(a)$ follows by the union bound; $(b)$ follows since the codebook is generated independently for each block $i \in [1:B]$ and the channel is memoryless; and $(c)$ follows by \eqref{BoundingProbabilityIntermediaryEventE3_Step1}.

The right hand side (RHS) of \eqref{BoundingProbabilityIntermediaryEventE3_Step2} tends to zero as $n \rightarrow \infty$ if
\begin{align}
  R_{c1}+R_{c2}+R_1 \leq \frac{B-1}{B}\big(I(U,V,X_1,X_2;Y)-\hat{R}\big)-\frac{\hat{R}}{B}.
  \label{BoundingProbabilityIntermediaryEventE3_Step3}
\end{align}
Taking $B \rightarrow \infty$, we get $\text{Pr}(E_3|E^c_1,E^c_2) \rightarrow 0$ as long as
\begin{equation}
R_c+R_1 \leq I(U,V,X_1,X_2;Y)-\hat{R}.
\label{condition1-on-sum-rate}
\end{equation}

\item For the decoding of the common message $w_c=(1,1)$ at the receiver, let $E_4$ be the event that $\dv u_i(1)$, $\dv x_{2,i}(1,w_{c2},t_{i-1})$, $\dv v_i(1,w_{c2},t_{i-1},t_i)$, $x_{1,i}(1,w_1)$ and $\dv y[i]$ are jointly typical for all $i=1,\hdots,B$ and some $w_{c2} \in [1:M_{c2}]$, $w_1 \in [1:M_1]$ and $t^B=(t_1,\hdots,t_B) \in [1:\hat{M}]^{B}$ such that $w_{c2} \neq 1$, i.e.,
\begin{align}
E_4 = \bigg\{ &\: \exists \: w_{c2} \in [1:M_{c2}],\: w_1 \in [1:M_1],\: t^B=(t_1,\hdots,t_B) \in [1:\hat{M}]^{B} \:\text{s.t.:} \: w_{c2} \neq 1,\nonumber\\
& \bigcap_{i=1}^{B} \Big\{\Big(\dv u_i(1), \dv x_{2,i}(1,w_{c2},t_{i-1}), \dv v_i(1,w_{c2},t_{i-1},t_i), \dv x_{1,i}(1,w_1), \dv y[i]\Big) \in \mc T_{\epsilon}^n(P_{U,X_2,V,X_1,Y})\Big\}\bigg\}.
\end{align}

To bound the probability of the event $E_4$, define the following event for given $w_{c2} \in [1:M_{c2}]$, $w_1 \in [1:M_1]$ and $(t_{i-1},t_i) \in [1:\hat{M}]^2$ such that $w_{c2} \neq 1$,
\begin{align*}
E_{4i}(w_{c2},t_{i-1},t_i,w_1) = \Big\{ &\:\Big(\dv u_i(1), \dv x_{2,i}(1,w_{c2},t_{i-1}), \dv v_i(1,w_{c2},t_{i-1},t_i), \dv x_{1,i}(1,w_1), \dv y[i]\Big) \in \mc T_{\epsilon}^n(P_{U,X_2,V,X_1,Y})\Big\}.
\end{align*}

For $w_{c2} \neq 1$ the vectors $\dv x_{2,i}(1,w_{c2},t_{i-1})$, $\dv v_i(1,w_{c2},t_{i-1},t_i)$ and $\dv x_{1,i}(1,w_1)$ are generated independently of $\dv y[i]$ conditionnally given $\dv u_i(1)$. Hence, by the joint typicality lemma \cite[Lecture Note 2]{GK11}, we get
\begin{align}
\text{Pr}\Big(E_{4i}(w_{c2},t_{i-1},t_i,w_1)|E_1^c,E_2^c,E_3^c\Big) & \leq 2^{-n[I(V,X_1,X_2;Y|U)-\epsilon]}.
\label{BoundingProbabilityIntermediaryEventE4_Step1}
\end{align}

Then, conditioned on the events $E_{1}^c$, $E_{2}^c$ and $E_{3}^c$, the probability of the event $E_4$ can be bounded as
\begin{align}
\text{Pr}(E_4|E_1^c,E_2^c,E_3^c) &= \text{Pr}\Big( \bigcup_{w_{c2} \neq 1} \bigcup_{w_1 \: \in \: [1:M_1]} \bigcup_{t^B \: \in \: [1:\hat{M}]^{B}} \bigcap_{i=1}^{B} E_{4i}(w_{c2},t_{i-1},t_i,w_1)|E_1^c,E_2^c,E_3^c\Big)\nonumber\\
&\stackrel{(a)}{\leq} \sum_{w_{c2} \neq 1} \sum_{w_1 \in [1:M_1]} \sum_{t^B \: \in \: [1:\hat{M}]^{B}}  \text{Pr}\Big(\bigcap_{i=1}^{B} E_{4i}(w_{c2},t_{i-1},t_i,w_1)|E_1^c,E_2^c,E_3^c\Big)\nonumber\\
&\stackrel{(b)}{=} \sum_{w_{c2} \neq 1} \sum_{w_1 \in [1:M_1]} \sum_{t^B \: \in \: [1:\hat{M}]^{B}}  \prod_{i=1}^{B} \text{Pr}\Big(E_{4i}(w_{c2},t_{i-1},t_i,w_1)|E_1^c,E_2^c,E_3^c\Big)\nonumber\\
&\leq \sum_{w_{c2} \neq 1} \sum_{w_1 \in [1:M_1]} \sum_{t^B \: \in \: [1:\hat{M}]^{B}}  \prod_{i=2}^{B} \text{Pr}\Big(E_{4i}(w_{c2},t_{i-1},t_i,w_1)|E_1^c,E_2^c,E_3^c\Big)\nonumber\\
&\stackrel{(c)}{\leq} \sum_{w_{c2} \neq 1} \sum_{w_1 \in [1:M_1]} \sum_{t^B \: \in \: [1:\hat{M}]^{B}} \prod_{i=2}^{B} 2^{-n\big[I(V,X_1,X_2;Y|U)-\epsilon\big]}\nonumber\\
&= \sum_{w_{c2} \neq 1} \sum_{w_1 \in [1:M_1]} \sum_{t_B \: \in \: [1:\hat{M}]} 2^{n(B-1)\big[\hat{R}+\hat{\eta}\epsilon\big]}2^{-n(B-1)\big[I(V,X_1,X_2;Y|U)-\epsilon\big]}\nonumber\\
&\leq M_{c2}M_1\hat{M}2^{-n(B-1)\big[I(V,X_1,X_2;Y|U)-\hat{R}-(\hat{\eta}+1)\epsilon\big]}\nonumber\\
&= 2^{-nB\big[\frac{B-1}{B}\big(I(V,X_1,X_2;Y|U)-\hat{R}\big)-(R_{c2}+R_1)-\frac{\hat{R}}{B}+\big(\eta_{c2}+\eta_1-\hat{\eta}-\frac{B-1}{B}\big)\epsilon\big]}
\label{BoundingProbabilityIntermediaryEventE4_Step2}
\end{align}
where: $(a)$ follows by the union bound; $(b)$ follows since the codebook is generated independently for each block $i \in [1:B]$ and the channel is memoryless; and $(c)$ follows by \eqref{BoundingProbabilityIntermediaryEventE4_Step1}.

The right hand side (RHS) of \eqref{BoundingProbabilityIntermediaryEventE4_Step2} tends to zero as $n \rightarrow \infty$ if
\begin{align}
  R_{c2}+R_1 \leq \frac{B-1}{B}\big(I(V,X_1,X_2;Y|U)-\hat{R}\big)-\frac{\hat{R}}{B}.
  \label{BoundingProbabilityIntermediaryEventE4_Step3}
\end{align}
Taking $B \rightarrow \infty$, we get $\text{Pr}(E_4|E^c_1,E^c_2,E^c_3) \rightarrow 0$ as long as
\begin{equation}
R_{c2}+R_1 \leq I(V,X_1,X_2;Y|U)-\hat{R}.
\label{condition2-on-sum-rate}
\end{equation}

\item For the decoding of the individual message $w_1=1$ at the receiver, let $E_5=\cup_{i=1}^{B} E_{5i}$ where $E_{5i}$ is the event that $\Big(\dv x_{2,i}(1,1,1)$, $\dv v_i(1,1,1,1)$, $x_{1,i}(1,1)$, $\dv y[i]\Big)$ is not jointly typical conditionnally given $\dv u_{i-1}(1)$, i.e.,
\begin{align}
E_5 &= \bigcup_{i=1}^{B} \Big\{\Big(\dv u_{i-1}(1), \dv x_{2,i}(1,1,1), \dv v_i(1,1,1,1), x_{1,i}(1,1), \dv y[i]\Big) \notin \mc T_{\epsilon}^n(P_{U,X_2,V,X_1,Y})\Big\}.
\end{align}

Conditioned on $E^c_{1i}$, the vectors $\dv s[i]$, $\dv u_{i-1}(1)$, $\dv x_{2,i}(1,1,1)$ and $\dv v_i(1,1,1,1)$ are jointly typical, and with $x_{1,i}(1,1)$. Then, conditioned on $E^c_{1i}$, the vectors $\dv s[i]$, $\dv u_{i-1}(1)$, $\dv x_{2,i}(1,1,1)$ and $\dv v_i(1,1,1,1)$, $x_{1,i}(1,1)$ and $\dv y[i]$ are jointly typical by the Markov lemma \cite[p. 436]{CT91}, i.e., $\text{Pr}(E_{2i}|E^c_{1i}) \rightarrow 0 \quad \text{as}\quad  n \rightarrow \infty$. Thus, by the union bound over the $B$ blocks, $\text{Pr}(E_5|E^c_1,E^c_2,E^c_3) \rightarrow 0 \quad \text{as}\quad  n \rightarrow \infty$.

\item For the decoding of the individual message $w_1=1$ at the receiver, let $E_6$ be the event that $\dv u_i(1)$, $\dv x_{2,i}(1,1,t_{i-1})$, $\dv v_i(1,1,t_{i-1},t_i)$, $x_{1,i}(1,w_1)$ and $\dv y[i]$ are jointly typical for all $i=1,\hdots,B$ and some $w_1 \in [1:M_1]$ and $t^B=(t_1,\hdots,t_B) \in [1:\hat{M}]^{B}$ such that $w_1 \neq 1$, i.e.,
\begin{align}
E_6 = \bigg\{ &\: \exists \: w_1 \in [1:M_1],\: t^B=(t_1,\hdots,t_B) \in [1:\hat{M}]^{B} \:\text{s.t.:} \: w_1 \neq 1,\nonumber\\
& \bigcap_{i=1}^{B} \Big\{\Big(\dv u_i(1), \dv x_{2,i}(1,1,t_{i-1}), \dv v_i(1,1,t_{i-1},t_i), \dv x_{1,i}(1,w_1), \dv y[i]\Big) \in \mc T_{\epsilon}^n(P_{U,X_2,V,X_1,Y})\Big\}\bigg\}.
\end{align}
To bound the probability of the event $E_6$, define the following event for given $w_1 \in [1:M_1]$ and $(t_{i-1},t_i) \in [1:\hat{M}]^2$,
\begin{align*}
E_{6i}(t_{i-1},t_i,w_1) = \Big\{ &\:\Big(\dv u_i(1), \dv x_{2,i}(1,1,t_{i-1}), \dv v_i(1,1,t_{i-1},t_i), \dv x_{1,i}(1,w_1), \dv y[i]\Big) \in \mc T_{\epsilon}^n(P_{U,X_2,V,X_1,Y})\Big\}.
\end{align*}
Then, we have
\begin{align}
\text{Pr}(E_6|E_1^c,E_2^c,E_3^c,E_4^c,E_5^c) &= \text{Pr}\Big(\bigcup_{w_1 \neq 1} \bigcup_{t^B \: \in \: [1:\hat{M}]^{B}} \bigcap_{i=1}^{B} E_{6i}(t_{i-1},t_i,w_1)|E_1^c,E_2^c,E_3^c,E^c_4,E_5^c,\Big)\nonumber\\
&\stackrel{(d)}{\leq} \sum_{w_1 \neq 1} \sum_{t^B \: \in \: [1:\hat{M}]^{B}} \text{Pr}\Big(\bigcap_{i=1}^{B} E_{6i}(t_{i-1},t_i,w_1)|E_1^c,E_2^c,E_3^c,E_4^c,5^c \Big)\nonumber\\
&\stackrel{(e)}{=} \sum_{w_1 \neq 1} \sum_{t^B \: \in \: [1:\hat{M}]^{B}} \prod_{i=1}^{B} \text{Pr}\Big(E_{6i}(t_{i-1},t_i,w_1)|E_1^c,E_2^c,E_3^c,E^c_4,5^c \Big)\nonumber\\
&\leq \sum_{w_1 \neq 1} \sum_{t^B \: \in \: [1:\hat{M}]^{B}} \prod_{i=2}^{B} \text{Pr}\Big(E_{6i}(t_{i-1},t_i,w_1)|E_1^c,E_2^c,E_3^c,E^c_4,5^c \Big)
\label{BoundingProbabilityIntermediaryEventE6_Step1}
\end{align}
where: $(d)$ follows by the union bound and $(e)$ follows since the codebook is generated independently for each block $i \in [1:B]$ and the channel is memoryless.

For $w_1 \neq 1$, the probability of the event $E_{6i}(t_{i-1},t_i,w_1)$ conditioned on $E_1^c,E_2^c,E_3^c,E_4^c,E_5^c$ can be bounded as follows, depending on the values of $t_{i-1}$ and $t_i$:

\begin{itemize}
\item[i)] if $t_{i-1} \neq 1$ then $\Big(\dv x_{2,i}(1,1,t_{i-1}), \dv v_i(1,1,t_{i-1},t_i), \dv x_{1,i}(1,w_1)\Big)$ is generated independently of the output vector $\dv y[i]$ conditionnally given $\dv u_i(1)$ irrespective to the value of $t_i$, and so, by the joint typicality lemma \cite[Lecture Note 2]{GK11}
\begin{align}
\text{Pr}\Big(E_{6i}(t_{i-1},t_i,w_1)|E_1^c,E_2^c,E_3^c,E_4^c,E_5^c\Big) & \leq 2^{-n[I(V,X_1,X_2;Y|U)-\epsilon]}.
\end{align}
\item[ii)] if $t_{i-1}=1$ and $t_i \neq 1$, then $\Big(\dv v_i(1,1,t_{i-1},t_i), \dv x_{1,i}(1,w_1)\Big)$ is generated independently of the output vector $\dv y[i]$ conditionnally given $\dv u_i(1)$ and $\dv x_{2,i}(1,1,t_{i-1})$; and, hence,
\begin{align}
\text{Pr}\Big(E_{6i}(t_{i-1},t_i,w_1)|E_1^c,E_2^c,E_3^c,E_4^c,E_5^c\Big) & \leq 2^{-n[I(V,X_1;Y|U,X_2)-\epsilon]}.
\end{align}
\item[iii)] if $t_{i-1}=1$ and $t_i=1$ then $\dv x_{1,i}(1,w_1)$ is generated independently of the output vector $\dv y[i]$ conditionnally given $\dv u_i(1)$, $\dv x_{2,i}(1,1,t_{i-1})$ and $\dv v_i(1,1,t_{i-1},t_i)$; and, hence,
\begin{align}
\text{Pr}\Big(E_{6i}(t_{i-1},t_i,w_1)|E_1^c,E_2^c,E_3^c,E_4^c,E_5^c\Big) & \leq 2^{-n[I(X_1;Y|U,V,X_2)-\epsilon]}.
\end{align}
\end{itemize}
Now, note that since $I(V,X_1;Y|U,X_2) \geq I(X_1;Y|U,V,X_2)$, if $w_1 \neq 1$ and $t_{i-1}=1$ the following holds irrespective to the value of $t_i$,
\begin{align}
\text{Pr}\Big(E_{6i}(t_{i-1},t_i,w_1)|E_1^c,E_2^c,E_3^c,E_4^c,E_5^c\Big) & \leq 2^{-n[I(X_1;Y|U,V,X_2)-\epsilon]}.
\end{align}
Let $I_1 := I(X_1;Y|U,V,X_2)$ and $I_2 := I(V,X_1,X_2;Y|U)$. If the sequence $(t_1,\hdots,t_{B-1})$ has $k$ ones, we have
\begin{align}
\prod_{i=2}^{B} \text{Pr}\Big(E_{6i}(t_{i-1},t_i,w_1)E_1^c,E_2^c,E_3^c,E_4^c,E_5^c\Big) & \leq 2^{-n[kI_1+(B-1-k)I_2-(B-1)\epsilon]}.
\end{align}
Continuing from \eqref{BoundingProbabilityIntermediaryEventE6_Step1}, we then bound the probability of the event $E_6$ as
\begin{align}
&\text{Pr}(E_6|E_1^c,E_2^c,E_3^c,E_4^c,E_5^c) \nonumber\\
&\leq \sum_{w_1 \neq 1} \sum_{t^B \: \in \: [1:\hat{M}]^{B}}  \prod_{i=2}^{B} \text{Pr}\Big(E_{6i}(t_{i-1},t_i,w_1)|E_1^c,E_2^c,E_3^c,E^c_4,5^c \Big)\nonumber\\
&= \sum_{w_1 \neq 1} \sum_{t_B \: \in \: [1:\hat{M}]}  \sum_{t^{B-1} \: \in \: [1:\hat{M}]^{B-1}} \prod_{i=2}^{B} \text{Pr}\Big(E_{6i}(t_{i-1},t_i,w_1)|E_1^c,E_2^c,E_3^c,E^c_4,5^c \Big)\nonumber\\
&\leq \sum_{w_1 \neq 1} \sum_{t_B \: \in \: [1:\hat{M}]} \sum_{k=0}^{B-1}\binom{B-1}{k}\: 2^{n(B-1-k)\big[\hat{R}+\hat{\eta}\epsilon\big]} 2^{-n\big[kI_1+(B-1-k)I_2-(B-1)\epsilon\big]}\nonumber\\
&= \sum_{w_1 \neq 1} \sum_{t_B \: \in \: [1:\hat{M}]} \sum_{j^{B-1} \: \in \: [1:J]^{B-1}} \sum_{k=0}^{B-1}\binom{B-1}{k}\: 2^{-n\big[kI_1+(B-1-k)(I_2-\hat{R})-(B-1-k)\hat{\eta}\epsilon-(B-1)\epsilon\big]}\nonumber\\
&= \sum_{w_1 \neq 1} \sum_{t_B \: \in \: [1:\hat{M}]} \sum_{k=0}^{B-1}\binom{B-1}{k}\: 2^{-n\big[kI_1+(B-1-k)(I_2-\hat{R})-(B-1)(\hat{\eta}+1)\epsilon\big]}\nonumber\\
&\leq \sum_{w_1 \neq 1} \sum_{t_B \: \in \: [1:\hat{M}]} \sum_{k=0}^{B-1}\binom{B-1}{k}\: 2^{-n\big[(B-1)\min(I_1,\: I_2-\hat{R})-(B-1)(\hat{\eta}+1)\epsilon\big]}\nonumber\\
&\leq M_1\hat{M} 2^{B} 2^{-n\big[(B-1)\min(I_1,\: I_2-\hat{R})-(B-1)(\hat{\eta}+1)\epsilon\big]}\nonumber\\
&= 2^{-nB\big[\frac{B-1}{B}\min(I_1,\: I_2-\hat{R})-R_1-\frac{\hat{R}}{B}-\frac{1}{n}+\big(\eta_1-\frac{\hat{\eta}}{B}-\frac{(B-1)(\hat{\eta}++1)}{B}\big)\epsilon\big]}\nonumber\\
&= 2^{-nB\big[\frac{B-1}{B}\min(I_1,\: I_2-\hat{R})-R_1-\frac{\hat{R}}{B}-\frac{1}{n}+\big(\eta_1-\hat{\eta}-\frac{B-1}{B}\big)\epsilon\big]}.
\label{BoundingProbabilityIntermediaryEventE6_Step4}
\end{align}

The right hand side (RHS) of \eqref{BoundingProbabilityIntermediaryEventE6_Step4} tends to zero as $n \rightarrow \infty$ if
\begin{align}
R_1 \leq \frac{B-1}{B}\big(\min(I_1,\: I_2-\hat{R})-\frac{\hat{R}}{B}.
\label{BoundingProbabilityIntermediaryEventE6_Step5}
\end{align}
Taking $B \rightarrow \infty$, we get $\text{Pr}(E_6|E^c_1,E^c_2,E^c_3),E^c_4,E^c_5 \rightarrow 0$ as long as
\begin{align}
R_1 &\leq I(X_1;Y|U,V,X_2)\\
R_1 &\leq I(V,X_1,X_2;Y|U)-\hat{R}.
\label{condition-on-individual-rate}
\end{align}

\end{itemize}
\textbf{Summarizing:} From the above, we get that the error probability is small provided that $n$ and $B$ are large and
 \begin{subequations}
 \begin{align}
 R_1 &\leq I(X_1;Y|U,V,X_2)\\
 R_1 &\leq I(V,X_1,X_2;Y|U)-\hat{R}\\
 R_{c2} + R_1 &\leq I(V,X_1,X_2;Y|U)-\hat{R}\\
 R_c + R_1 &\leq I(U,V,X_1,X_2;Y)-\hat{R}.
 \end{align}
 \label{Theorem2RateRegion__Step1}
 \end{subequations}
 
Finally, using Fourier-Motzkin Elimination to successively project out $R_{c2}$ and $\hat{R}$ from \eqref{Theorem2RateRegion__Step1}, we get 
\begin{subequations}
\begin{align}
R_1 &\leq I(X_1;Y|U,V,X_2)\\
R_1 &\leq I(V,X_1,X_2;Y|U)-I(V;S|U,X_2)\\
R_c + R_1 &\leq I(U,V,X_1,X_2;Y)-I(V;S|U,X_2).
\end{align}
\label{Theorem2RateRegion__Step3}
\end{subequations}

This completes the proof of Theorem~\ref{theorem-inner-bound-asymmetric-strictly-causal-states-setting}.

\renewcommand{\theequation}{K-\arabic{equation}}
\setcounter{equation}{0}  
\subsection{Proof of Theorem~\ref{theorem-capacity-informed-helper}}\label{appendix-proof-theorem-capacity-informed-helper}

\subsubsection{Direct Part} Recall the inner bound of Theorem~\ref{theorem-inner-bound-asymmetric-strictly-causal-states-setting}. Setting $R_c=0$, we obtain
\begin{subequations}
\begin{align}
\label{inner-bound-informed-helper-step1-ineq1}
R_1 \: &\leq \: I(X_1;Y|U,V,X_2)\\
R_1 \: &\leq \: I(V,X_1,X_2;Y|U)-I(V;S|U,X_2)
\label{inner-bound-informed-helper-step1-ineq2}
\end{align}
\label{inner-bound-informed-helper-step1}
\end{subequations}
for some measure 
\begin{equation}
P_{S,U,V,X_1,X_2,Y} = Q_SP_{U}P_{X_2|U}P_{X_1|U}P_{V|S,U,X_2}W_{Y|S,X_1,X_2}.
\label{measure-inner-bound-informed-helper}
\end{equation}
(Note that the bound on the sum rate is redundant).

\noindent Setting $V=S$ and $U=\emptyset$ in \eqref{inner-bound-informed-helper-step1-ineq1}, we obtain the first term of the minimum in the capacity expression \eqref{capacity-informed-helper}. Similarly, setting $V=S$ and $U=\emptyset$ in \eqref{inner-bound-informed-helper-step1-ineq2}, we obtain
\begin{align}
R_1 &\leq I(V,X_1,X_2;Y|U) - I(V;S|U,X_2)\\
    &= I(X_1,X_2;Y|U) - I(V;S|U,X_1,X_2,Y)\\
    &= I(X_1,X_2;Y) - H(S|X_1,X_2,Y)\\
    &= I(X_1,X_2,Y)
\end{align}
where the last equality holds since the state $S$ is a deterministic function of $(X_1,X_2,Y)$. 

\subsubsection{Converse Part} The converse proof also follows in a manner that is similar to that of Proposition~\ref{proposition-alternative-outer-bound-strictly-causal-states-setting}, by noticing that in this case the channel inputs are independent.

\renewcommand{\theequation}{L-\arabic{equation}}
\setcounter{equation}{0}  
\subsection{Proof of Theorem~\ref{theorem-capacity-region-causal-states-setting}}\label{appendix-proof-theorem-capacity-region-causal-states-setting}

To see that the knowledge of the states strictly causally at the encoders does not increase the sum-rate capacity, observe that we can bound the sum rate as follows.

\subsubsection{Direct Part} The achievability follows straightforwardly by using Shannon strategies, without Block-Markov coding.

\subsubsection{Converse Part} The converse proof also follows through straightforward steps. More specifically, let us define $V_i=(W_c,Y^{i-1})$ and $U_i=(W_1,V_i)$, $i=1,\hdots,n$.

We can bound the sum rate $(R_c+R_1)$ as follows.

{\allowdisplaybreaks
\begin{align}
 n(R_c+R_1)  &\leq H(W_c,W_1)\nonumber \\
      &= I(W_c,W_1;Y^n)+H(W_c,W_1|Y^n)\nonumber\\
      &\leq I(W_c,W_1;Y^n)+n\epsilon_n\nonumber\\
      &= \sum_{i=1}^{n} I(W_c,W_1;Y_i|Y^{i-1})+n\epsilon_n\nonumber\\
      &\leq \sum_{i=1}^{n} I(W_c,W_1,Y^{i-1};Y_i)+n\epsilon_n\nonumber\\
      &\stackrel{(a)}{\leq} \sum_{i=1}^{n} I(V_i,U_i;Y_i)+n\epsilon_n
\end{align}
where $(a)$ follows by substituting using the definitions of $U_i$ and $V_i$.

Similarly, we can bound the individual rate $R_1$ as follows

{\allowdisplaybreaks
\begin{align}
 nR_1  &\leq H(W_1) \nonumber\\
       &=H(W_1|W_c) \nonumber\\
      &= I(W_1;Y^n|W_c)+H(W_1|W_c,Y^n)\nonumber\\
      &\leq I(W_1;Y^n|W_c)+n\epsilon_n\nonumber\\
      &= \sum_{i=1}^{n} I(W_1;Y_i|W_c,Y^{i-1})+n\epsilon_n\nonumber\\
      &\leq \sum_{i=1}^{n} I(W_1,W_c,Y^{i-1};Y_i|W_c,Y^{i-1})+n\epsilon_n\nonumber\\
      &\stackrel{(b)}{\leq} \sum_{i=1}^{n} I(U_i;Y_i|V_i)+n\epsilon_n
\end{align}
where $(b)$ follows by substituting using the definitions of $U_i$ and $V_i$.

\noindent The rest of the proof of Theorem~\ref{theorem-capacity-region-causal-states-setting} follows by standard single-letterization.

\bibliographystyle{IEEEtran}
\bibliography{manuscript}

\begin{thebibliography}{10}
\providecommand{\url}[1]{#1}
\csname url@rmstyle\endcsname
\providecommand{\newblock}{\relax}
\providecommand{\bibinfo}[2]{#2}
\providecommand\BIBentrySTDinterwordspacing{\spaceskip=0pt\relax}
\providecommand\BIBentryALTinterwordstretchfactor{4}
\providecommand\BIBentryALTinterwordspacing{\spaceskip=\fontdimen2\font plus
\BIBentryALTinterwordstretchfactor\fontdimen3\font minus
  \fontdimen4\font\relax}
\providecommand\BIBforeignlanguage[2]{{%
\expandafter\ifx\csname l@#1\endcsname\relax
\typeout{** WARNING: IEEEtran.bst: No hyphenation pattern has been}%
\typeout{** loaded for the language `#1'. Using the pattern for}%
\typeout{** the default language instead.}%
\else
\language=\csname l@#1\endcsname
\fi
#2}}

\bibitem{Sh58}
C.~E. Shannon, ``Channels with side information at the transmitter,'' \emph{IBM
  journal of Research and Development}, vol.~2, pp. 289--293, Oct. 1958.

\bibitem{LS13a}
A.~Lapidoth and Y.~Steinberg, ``The multiple access channel with causal side
  information: common state,'' \emph{IEEE Trans. Inf. Theory}, vol.~59, pp.
  32--50, Jan. 2013.

\bibitem{LS13b}
------, ``The multiple access channel with causal side information: double
  state,'' \emph{IEEE Trans. Inf. Theory}, vol.~99, pp. 1379--1393, 2013.

\bibitem{LSY13}
M.~Li, O.~Simeone, and A.~Yener, ``Multiple access channels with states
  causally known at the transmitters,'' \emph{IEEE Trans. Inf. Theory},
  vol.~99, pp. 1394--1404, 2013.

\bibitem{ZPS13}
A.~Zaidi, P.~Piantanida, and S.~{Shamai (Shitz)}, ``Capacity region of
  cooperative multiple access channel with states,'' \emph{IEEE Trans. Inf.
  Theory}, vol.~59, pp. 6153--6174, Oct. 2013.

\bibitem{Sh56}
C.~E. Shannon, ``The zero error capacity of a noisy channel,'' \emph{IRE Trans.
  on Inf. Theory}, vol.~2, pp. 8--19, 1956.

\bibitem{ZPS11a}
A.~Zaidi, P.~Piantanida, and S.~{Shamai (Shitz)}, ``Multiple access channel
  with states known noncausally at one encoder and only strictly causally at
  the other encoder,'' in \emph{Proc. {IEEE} Int. Symp. Information Theory},
  Saint-Petersburg, Russia, 2011, pp. 2801--2805.

\bibitem{ZPS12a}
------, ``Wyner-ziv type versus noisy network coding for a state-dependent
  mac,'' in \emph{Proc. {IEEE} Int. Symp. Information Theory}, Cambridge, USA,
  Jul. 2012, pp. 1682--1686.

\bibitem{M-AT12}
M.~A. Maddah-Ali and D.~Tse, ``Completely stale transmitter channel state
  information is still very useful,'' \emph{{IEEE} Trans. Inf. Theory},
  vol.~58, no.~7, pp. 4418--4431, Jul. 2012.

\bibitem{ZT03}
L.~Zheng and D.~Tse, ``Diversity and multiplexing: {A} fundamental tradeoff in
  multiple-antenna channels,'' \emph{{IEEE} Trans. Inf. Theory}, vol.~49,
  no.~5, pp. 1073--1096, 2003.

\bibitem{J10}
S.~A. Jafar, ``Interference alignment --- {A} new look at signal dimensions in
  a communication network,'' \emph{Foundations and Trends in Communications and
  Information Theory}, vol.~7, no.~1, pp. 1--134, 2010.

\bibitem{VV11a}
\BIBentryALTinterwordspacing
C.~S. Vaze and M.~K. Varanasi, ``The degrees of freedom region of the two-user
  and certain three-user {MIMO} broadcast channel with delayed {CSI},'' 2011.
  [Online]. Available: \url{http://arxiv.org/abs/1101.0306}
\BIBentrySTDinterwordspacing

\bibitem{AGK11}
M.~J. Abdoli, A.~Ghasemi, and A.~K. Khandani, ``On the degrees of freedom of
  three-user {MIMO} broadcast channel with delayed {CSIT},'' in \emph{IEEE Int.
  Sym. on Info. Theory}, St. Petersburg, Russia, Aug. 2011, pp. 209--213.

\bibitem{XAJ12}
J.~Xu, J.~G. Andrews, and S.~A. Jafar, ``Broadcast channels with delayed
  finite-rate feedback: Predict or observe ?'' \emph{IEEE Trans. on Wireless
  Comm.}, vol.~11, pp. 1456--1467, Apr. 2012.

\bibitem{VV11b}
\BIBentryALTinterwordspacing
C.~S. Vaze and M.~K. Varanasi, ``The degrees of freedom region and interference
  alignment for the {MIMO} interference channel with delayed {CSI},'' 2011.
  [Online]. Available: \url{http://arxiv.org/abs/1101.5809}
\BIBentrySTDinterwordspacing

\bibitem{VM-AA13}
A.~Vahid, M.~A. Maddah-Ali, and A.~S. Avestimehr, ``Capacity results for binary
  fading interference channels with delayed {CSIT},'' \emph{IEEE Trans. Inf.
  Theory, submitted for publication. Available at
  url{http://arxiv.org/abs/1301.5309}}, 2013.

\bibitem{ZASV13a}
A.~Zaidi, Z.~Awan, S.~{Shamai (Shitz)}, and L.~Vandendorpe, ``Secure degrees of
  freedom of {MIMO} {X}-channels with output feedback and delayed {CSI},''
  \emph{IEEE Transactions on Information Forensics and Security, submitted for
  publication}, 2013.

\bibitem{GMK11}
A.~Ghasemi, A.~S. Motahari, and A.~K. Khandani, ``On the degrees of freedom of
  {X} channel with delayed {CSIT},'' in \emph{Proc. {IEEE} Int. Symp.
  Information Theory}, Jul. 2011.

\bibitem{YZ12}
L.~Yang and W.~Zhang, ``On achievable degrees of freedom for {MIMO} {X}
  channels,'' \emph{{IEEE} Trans. Inf. Theory, submitted for publication,
  available at \url{arxiv.org/abs/1208.2900}}, Aug. 2012.

\bibitem{SCCK05}
A.~Sutivong, M.~C. T.~M. Cover, and Y.-H. Kim, ``Channel capacity and state
  estimation for state-dependent gaussian channels,'' \emph{{IEEE} Trans. Inf.
  Theory}, vol.~51, pp. 1486--1495, 2005.

\bibitem{CH-KM12}
C.~Choudhuri, Y.-H. Kim, and U.~Mitra, ``Causal state amplification,''
  \emph{Available in \url{http://arxiv.org/abs/1203.6027}}, 2012.

\bibitem{CSW12}
Y.-K. Chia, R.~Soundararajan, and T.~Weissman, ``Estimation with a helper who
  knows the interference,'' \emph{Available in
  \url{http://arxiv.org/abs/1203.4311}}, 2012.

\bibitem{ZKLV10}
A.~Zaidi, S.~Kotagiri, J.~N. Laneman, and L.~Vandendorpe, ``Cooperative
  relaying with state available non-causally at the relay,'' \emph{IEEE Trans.
  Inf. Theory}, vol.~56, pp. 2272--2298, May 2010.

\bibitem{ZKLV08a}
------, ``Cooperative relaying with state at the relay,'' in \emph{Proc. {IEEE}
  Information Theory Workshop}, Porto, Portugal, May 2008, pp. 139--143.

\bibitem{ZSPV10a}
A.~Zaidi, S.~{Shamai (Shitz)}, P.~Piantanida, and L.~Vandendorpe, ``Bounds on
  the capacity of the relay channel with noncausal state information at
  source,'' in \emph{Proc. {IEEE} Int. Symp. Information Theory}, Austin, TX,
  USA, 2010, pp. 639--643.

\bibitem{ZSPV10b}
------, ``Bounds on the capacity of the relay channel with noncausal state at
  source,'' \emph{IEEE Trans. Inf. Theory}, vol.~5, pp. 2639--2672, May 2013.

\bibitem{ZV07b}
A.~Zaidi and L.~Vandendorpe, ``Rate regions for the partially-cooperative
  relay-broadcast channel with non-causal side information,'' in \emph{Proc.
  {IEEE} Int. Symp. Information Theory}, Nice, France, Jun. 2007, pp.
  1246--1250.

\bibitem{ZV09b}
------, ``Lower bounds on the capacity of the relay channel with states at the
  source,'' \emph{EURASIP Journal on Wireless Commnunications and Networking},
  vol. Article ID 634296. doi:10.1155/2009/634296, 2009.

\bibitem{AMA09}
B.~Akhbari, M.~Mirmohseni, and M.~R. Aref, ``Compress-and-forward strategy for
  the relay channel with non-causal state information,'' in \emph{Proc. {IEEE}
  Int. Symp. Information Theory}, Seoul, Korea, Jun.-Jul. 2009, pp. 1169--1173.

\bibitem{KE-GS13}
M.~N. Khormuji, A.~{El Gamal}, and M.~Skoglund, ``State-dependent relay
  channel: Achievable rate and capacity of a semideterministic class,''
  \emph{IEEE Trans. Inf. Theory}, vol.~5, pp. 2629--2638, May 2013.

\bibitem{LSY11}
M.~Li, O.~Simeone, and A.~Yener, ``Message and state cooperation in a relay
  channel when only the relay knows the state,'' \emph{Submitted for
  publication in IEEE Trans. Inf. Theory. Available at
  \url{http://arxiv.org/abs/1102.0768}}, 2011.

\bibitem{LW13}
A.~Lapidoth and L.~Wang, ``The state-dependent semideterministic broadcast
  channel,'' \emph{IEEE Trans. Inf. Theory, submitted for publication.
  Available in \url{http://arxiv.org/abs/1111.1144}}, 2012.

\bibitem{OS13}
T.~J. Oechtering and M.~Skoglund, ``Bidirectional broadcast channel with random
  states noncausally known at the encoder,'' \emph{IEEE Trans. Inf. Theory},
  vol.~59, pp. 64--75, 2013.

\bibitem{K-FM11a}
R.~Khosravi-Farsani and F.~Marvasti, ``Capacity bounds for multiuser channels
  with non-causal channel state information at the transmitters,'' in
  \emph{Proc. {IEEE} Information Theory Workshop}, Paraty, Brasil, Oct. 2011,
  pp. 195 -- 199.

\bibitem{SCYA11a}
N.~\c{S}en, G.~Como, S.~Y{\"u}ksel, and F.~Alajaji, ``On the capacity of
  memoryless finite-state multiple access channels with asymmetric noisy state
  information at the encoders,'' in \emph{Proc. of 49th Annual Conf. on
  Communication, Control, and Computing (Allerton)}, IL, USA, Sep. 2011.

\bibitem{ZVD07a}
A.~Zaidi, L.~Vandendorpe, and P.~Duhamel, ``Lower bounds on the capacity
  regions of the multi-relay channel and the multi-relay broadcast channel with
  non-causal side-information,'' in \emph{Proc. {IEEE} Int. Conf. on
  Communications, ICC}, Glasgow, UK, Jun. 2007, pp. 6005--6011.

\bibitem{ZS13a}
A.~Zaidi and S.~{Shamai (Shitz)}, ``On multiple access channels with delayed
  {CSI},'' in \emph{Proc. {IEEE} Int. Symp. Information Theory}, Istanbul,
  Turkey, Jun.-Jul. 2013, pp. 982--986.

\bibitem{ZS14a}
------, ``Asymmetric cooperative multiple access channels with delayed {CSI},''
  in \emph{Proc. {IEEE} Int. Symp. Information Theory}, Honolulu, Hawai, USA,
  Jun.-Jul. 2014.

\bibitem{ZPD05b}
A.~Zaidi, P.~Piantanida, and P.~Duhamel, ``Broadcast- and {MAC}-aware coding
  strategies for multiple user information embedding,'' \emph{{IEEE}
  Transactions on Signal Processing}, vol.~55, no.~6, pp. 2974--2992, Jun.
  2007.

\bibitem{ZV09a}
A.~Zaidi and L.~Vandendorpe, ``Coding schemes for relay-assisted information
  embedding,'' \emph{{IEEE} Transactions on Information Forensics and
  Security}, vol.~4, no.~1, pp. 70--85, Jan. 2009.

\bibitem{H-LKGC11}
S.~H. Lim, Y.-H. Kim, A.~E. Gamal, and S.-Y. Chung, ``Noisy network coding,''
  \emph{IEEE Trans. Inf. Theory}, vol.~57, pp. 3132--3152, May 2011.

\bibitem{ADT11}
A.~Avestimehr, S.~Diggavi, and D.~Tse, ``Wireless network information flow: a
  determenistic approach,'' \emph{IEEE Trans. Inf. Theory}, vol.~57, pp.
  1872--1905, April 2011.

\bibitem{G68}
R.~G. Gallager, \emph{Information Theory and Reliable Communication}.\hskip 1em
  plus 0.5em minus 0.4em\relax New York: John Willey, 1968.

\bibitem{GK11}
A.~E. Gamal and Y.-H. Kim, \emph{Network information theory}.\hskip 1em plus
  0.5em minus 0.4em\relax Cambridge University Press, 2011.

\bibitem{BLW08}
S.~I. Bross, A.~Lapidoth, and M.~A. Wigger, ``The gaussian mac with
  conferencing encoders,,'' in \emph{Proc. of IEEE Int. Symp. Information
  Theory}, Toronto, ON, Jul. 2008, pp. 2702--2706.

\bibitem{W83}
F.~M.~J. Willems, ``The discrete memoryless multiple channel with partially
  cooperating encoders,'' \emph{IEEE Trans. Inf. Theory}, vol.~29, pp.
  441--445, May 1983.

\bibitem{C83}
M.~H.~M. Costa, ``Writing on dirty paper,'' \emph{{IEEE} Trans. Inf. Theory},
  vol.~29, pp. 439--441, May 1983.

\bibitem{ZKLV09a}
A.~Zaidi, S.~Kotagiri, J.~N. Laneman, and L.~Vandendorpe, ``Multiaccess
  channels with state known to one encoder: Another case of degraded message
  sets,'' in \emph{Proc. {IEEE} Int. Symp. Information Theory}, Seoul, Korea,
  Jun.-Jul. 2009, pp. 2376--2380.

\bibitem{DLKS13}
R.~duan, Y.~Liang, A.~Khisti, and S.~S. (Shitz), ``State-dependent {G}aussian
  {Z}-channel with mismatched side-information and interference,'' in
  \emph{Proc. of IEEE Information Theory Workshop}, Sevilla, Spain, Sep. 2013,
  pp. 1--5.

\bibitem{SBSV07a}
A.~Somekh-Baruch, S.~{Shamai (Shitz)}, and S.~Verd\`u, ``Cooperative multiple
  access encoding with states available at one transmitter,'' \emph{IEEE Trans.
  Inf. Theory}, vol.~54, pp. 4448--4469, Oct. 2008.

\bibitem{KL07}
S.~Kotagiri and J.~Laneman, ``Multiaccess channels with state known to one
  encoder: A case of degraded message sets,'' in \emph{Proc. {IEEE} Int. Symp.
  Information Theory}, Nice, France, Jun. 2007, pp. 1566--1570.

\bibitem{KL07a}
S.~Kotagiri and J.~N. Laneman, ``Multiaccess channels with state known to some
  encoders and independent messages,'' \emph{EURASIP Journal on Wireless
  Commnunications and Networking}, vol. Article ID 450680.
  doi:10.1155/2008/450680, 2008.

\bibitem{SK05}
S.~Sigurjonsson and Y.~H. Kim, ``On multiple user channels with state
  information at the transmitters,'' in \emph{Proc. {IEEE} Int. Symp.
  Information Theory}, Sep. 2005.

\bibitem{K-FM10}
R.~Khosravi-Farsani and F.~Marvasti, ``Multiple access channels with
  cooperative encoders and channel state information,'' \emph{Available in
  \url{http://arxiv.org/abs/1009.6008}}, 2010.

\bibitem{CT91}
T.~M. Cover and J.~A. Thomas, \emph{Elements of Information Theory}.\hskip 1em
  plus 0.5em minus 0.4em\relax New York: John Willey \& Sons INC., 1991.

\bibitem{CK78}
{I. Csisz\'ar} and {J. K\"orner}, ``Broadcast channels with confidential
  messages,'' \emph{IEEE Trans. Inf. Theory}, vol.~24, pp. 339--348, 1978.

\bibitem{W82}
F.~M.~J. Willems, \emph{Informationtheoretical Results for the Discrete
  Memoryless Multiple Access Channel}.\hskip 1em plus 0.5em minus 0.4em\relax
  Leuven, Belgium: Doctor in de Wetenschappen Proefschrift dissertation, Oct.
  1982.

\bibitem{CK81}
{I. Csisz\'ar} and {J. K\"orner}, \emph{Information Theory: Coding Theorems for
  Discrete Memoryless Systems}.\hskip 1em plus 0.5em minus 0.4em\relax London,
  U. K.: Academic Press, 1981.

\end{thebibliography}

\end{document}